\documentclass[iop, twocolumn, twocolappendix]{emulateapj}

\usepackage{xspace}

\newcommand\cxo{\textit{Chandra}\xspace}
\newcommand\xmm{\textit{XMM-Newton}\xspace}
\newcommand\zp{$\zeta$~Pup\xspace}
\newcommand\lya{Ly$\alpha$\xspace}
\newcommand\lyb{Ly$\beta$\xspace}
\newcommand\taustar{\ensuremath{\tau_*}\xspace}
\newcommand\ro{\ensuremath{R_0}\xspace}
\newcommand\rstar{\ensuremath{R_*}\xspace}
\newcommand\rmax{\ensuremath{R_{\mathrm{max}}\xspace}}
\newcommand\Msunyr{$M_\odot \mathrm{yr}^{-1}$\xspace}
\newcommand\kms{$\mathrm{km\, s^{-1}}$\xspace}
\newcommand\vinf{\ensuremath{v_\infty}\xspace}
\newcommand\hinf{\ensuremath{h_\infty}\xspace}
\newcommand\gax{\>\vcenter{\hbox{$>$\hskip-.75em\lower1.0ex\hbox{$\sim$}}}\>}
\newcommand\lax{\>\vcenter{\hbox{$<$\hskip-.75em\lower1.0ex\hbox{$\sim$}}}\>}
\def\profilePlotPanelSize{0.33}
\def\appendixPlotPanelSize{0.33}

\begin{document}

\submitted{Accepted by ApJ}

\title{Constraints on porosity and mass loss in O-star winds
  from modeling of X-ray emission line profile shapes}

\author{Maurice A. Leutenegger\altaffilmark{1,}\altaffilmark{2} David
  H. Cohen\altaffilmark{3}, Jon O. Sundqvist\altaffilmark{4, 5}, and Stanley
  P. Owocki\altaffilmark{4}}

\altaffiltext{1}{CRESST and X-ray Astrophysics Laboratory NASA/Goddard
  Space Flight Center, Greenbelt, MD 20771, USA;
  \email{Maurice.A.Leutenegger@nasa.gov}}

\altaffiltext{2}{Department of Physics, University of Maryland,
  Baltimore County, 1000 Hilltop Circle, Baltimore, MD 21250, USA}

\altaffiltext{3}{Department of Physics and Astronomy, Swarthmore
  College, Swarthmore, PA 19081, USA}

\altaffiltext{4}{Bartol Research Institute, University of Delaware,
  Newark, DE 19716, USA} 

\altaffiltext{5}{Universit\"{a}tssternwarte M\"{u}nchen, Scheinerstr. 1, D-81679 M\"{u}nchen, Germany}

\shorttitle{Line profiles from porous winds}
\shortauthors{Leutenegger et al.}

\begin{abstract}
  We fit X-ray emission line profiles in high resolution \xmm and \cxo
  grating spectra of the early O supergiant \zp with models that
  include the effects of porosity in the stellar wind. We explore the
  effects of porosity due to both spherical and flattened clumps. We
  find that porosity models with flattened clumps oriented parallel to
  the photosphere provide poor fits to observed line shapes. However,
  porosity models with isotropic clumps can provide acceptable fits to
  observed line shapes, but only if the porosity effect is
  moderate. We quantify the degeneracy between porosity effects from
  isotropic clumps and the mass-loss rate inferred from the X-ray line
  shapes, and we show that only modest increases in the mass-loss rate
  ($\lax 40\%$) are allowed if moderate porosity effects ($\hinf \lax
  \rstar$) are assumed to be important. Large porosity lengths, and
  thus strong porosity effects, are ruled out regardless of
  assumptions about clump shape. Thus, X-ray mass-loss rate estimates
  are relatively insensitive to both optically thin and optically
  thick clumping. This supports the use of X-ray spectroscopy as a
  mass-loss rate calibration for bright, nearby O stars.
\end{abstract}

 --- stars: mass-loss --- stars: winds,
outflows --- X-rays: stars

\section{Introduction}
\label{sec:intro}

Radiatively driven winds in massive stars have been known to be
structured and inhomogeneous virtually since their discovery
\citep{Morton67a}, both from an observational and theoretical
standpoint \citep{LS70, 1971MNRAS.153..435B}.  The same instability in
the radiative driving force that is responsible for the inhomogeneous
nature of these winds is also responsible for generating shocks
resulting in X-ray emission \citep{LW80, OCR88, FPP97}. Since these
phenomena are so intimately related, it is no surprise that the
question of the effects of wind inhomogeneities on X-ray radiative
transfer has arisen as well.

A good understanding of X-ray line profile formation in massive star
winds is of great importance, as the degree of asymmetry in X-ray line
profiles can provide an independent diagnostic of stellar mass loss
\citep{Cohen2010}. From an observational perspective, the traditional
mass-loss rate diagnostics in the optical, radio, and UV bands are
easier to use, but all these diagnostics are subject to modeling
uncertainties resulting from wind inhomogeneities. H$\alpha$
recombination lines and radio free-free emission are both the result
of collisional processes that scale with the square of density, and
thus are sensitive to the degree of clumping in the wind. H$\alpha$
emission originates near the stellar photosphere, while radio
free-free emission originates farther out in the wind, and each of
these measurements depends on the local degree of inhomogeneity
\citep[e.g.,][]{Pel06}. UV P Cygni absorption features probe the ion
column density in the wind, but the derivation of a mass-loss rate
from an ion column density requires accurate knowledge of the
ionization balance \citep{MFSH03, FMP06}. Furthermore, inhomogeneities
in the distribution of wind material in line-of-sight velocity
complicate the construction of model absorption profiles
\citep{2008cihw.conf..121O, 2007A&A...476.1331O, 2010A&A...510A..11S,
  2011A&A...528A..64S}. If X-ray line profiles are demonstrated to be
relatively free of systematic effects from wind inhomogeneities, then
X-ray spectroscopy may be used to calibrate the effects of
inhomogeneities on traditional mass-loss rate diagnostics.

Because X-ray radiative transfer in stellar winds is generally only
affected by continuum photoelectric absorption in the cool bulk of the
wind, the only possible way for inhomogeneities to have an effect is
if the clumps are optically thick in the continuum and have a large
interclump mean free path, leading to {\it porosity}. On the other
hand, optically thin clumps will only affect diagnostics with
opacities that scale with the {\it square} of the density, and so do
not affect X-ray absorption.  The goal of this paper is to empirically
constrain the potential effect of porosity on X-ray line diagnostics.

\citet{FOH03} and \citet{OFH04} have developed a formalism for X-ray
radiative transfer in a porous stellar wind, focusing on the case of
flat clumps oriented parallel to the stellar surface. They find that
porous winds can produce significantly more symmetric line profiles
for a given mass-loss rate. Since the assumed mass-loss rate also
influences the degree of profile asymmetry, the mass-loss rate and
degree of porosity might be expected to become degenerate parameters
of line profile fitting. This would reduce the utility of X-ray line
shapes as mass-loss rate diagnostics, unless some other observational
constraints could be placed on porosity lengths.

In \citet{OC06} and \citet[][hereafter Paper I]{2012MNRAS.420.1553S},
we have developed a generalized formalism for X-ray radiative transfer
in a clumpy stellar wind. In particular, Paper I gives a unified
treatment of spherical and flat clumps, and explores the effects of
a distribution of clump scales. Paper I shows that porosity in models
with either flat or spherical clumps tends to produce more symmetric
line shapes for a given mass-loss rate. However, flat clumps result in
profiles with a distinct shape, with a bump arising near line center
(the ``Venetian blind effect''; e.g., columns 3 and 4 of Figure 3 in
Paper I).

\citet{OFH06} have applied their X-ray radiative transfer formalism to
the calculation of emission line profiles, and have compared their
model to the observed line profiles of four O stars observed with
\cxo. They found that a model including porosity from flat clumps
could qualitatively reproduce the observed profile shapes. However,
they have not formally evaluated the quality of their fits, and have
not quantitatively explored the tradeoffs between porosity and mass
loss rate.

In this paper, we aim to confront models of X-ray emission line
profiles with high quality observational data to evaluate how well
porous and non-porous models can explain the X-ray emission of O
stars, to evaluate the effects of both flat and spherical clumps, and
to quantify the tradeoffs between porosity and mass-loss rate in
modeling the observations. We focus on the high quality archival
observations of the bright, well-studied O star \zp, which has been
observed by \cxo, as well as by \xmm for nearly one million
seconds. In addition to the high quality of the available data, \zp
has the advantage of being a single, non-magnetic O star, and it has
visibly asymmetric X-ray line profiles, indicating that absorption of
X-rays by the wind is significant and measurable. 

This paper is organized as follows.  Section~\ref{sec:model} briefly
recapitulates our X-ray emission line profile models. In
\S~\ref{sec:data} we describe the observations, and our data reduction
and fitting procedures. Section~\ref{sec:results} gives the results of our
line profile modeling. In Section~\ref{sec:comparison} we compare our
results to previous work. In Section~\ref{sec:discussion} we discuss our
results and give our conclusions.

\section{Model}
\label{sec:model}

The models used in this paper are described in \citet{OC01}, with
elaborations for porosity in \citet{OC06} and Paper I. Here we give a
brief recapitulation of these models and a description of their
implementation.

The emergent X-ray spectrum from the wind of a massive star is
calculated assuming the X-rays are formed in a small fraction of the
wind, while the cool bulk of the wind absorbs the X-rays as they
propagate through it. The emergent luminosity at a given wavelength
$\lambda$ is
\begin{equation}
  L_\lambda = 4\pi \int dV \eta_\lambda e^{-\tau}\, ,
  \label{eq:windintegral}
\end{equation}
where the integral is taken over the X-ray emitting volume of the
wind, $\eta_\lambda$ gives the X-ray emissivity, and $\tau$ is the
X-ray optical depth due to bound-free transitions in the cool bulk of
the wind, evaluated by an integral along the line of sight. The X-ray
emitting volume is typically taken to have a lower radial cutoff
\ro. In most of this work, we assume that there is no upper
radial cutoff \rmax, but in Section~\ref{sec:comparison} we instead
assume a finite value for \rmax in order to compare our work with
that of \citet{OFH06}.

It is possible to incorporate porosity within the framework of
\citet{OC01} by making appropriate modifications to the optical
depth term in Equation~\ref{eq:windintegral}. In this paper, we use the
implementations of isotropic and anisotropic porosity described in Paper
I. The bound-free absorption of an X-ray photon emitted at $z_{\rm e}$
in direction $\hat{z}$ is given by the optical depth
\begin{equation}
  \tau = \int_{z_{\rm e}}^\infty \chi_{\rm eff} \, dz,  
  \label{eq:tauinte}
\end{equation}  
where the \textit{effective} opacity $\chi_{\rm eff}$ accounts for any
porosity. $\chi_{\rm eff}$ is calculated using the ``inverse'' (or
Rosseland) law for bridging the optically thin and optically thick regimes
\begin{equation}
 \frac{\chi_{\rm eff}}{\chi}= \frac{1}{1+\tau_{\rm cl}}, 
	\label{Eq:chieff}
\end{equation} 
where $\chi$ is the atomic opacity per unit length, proportional to
the mass-loss rate and here characterized by the fiducial optical
depth $\tau_\star = \dot{M} \kappa/(4 \pi R_\star \vinf)$, with
mass absorption coefficient $\kappa$ and wind terminal speed
\vinf. The clump optical depth is
\begin{equation}
  \tau_{\rm cl} = \chi h/|\mu|,  
	\label{Eq:taucl}
\end{equation}
where $\mu$ is the directional cosine for a photon impacting a clump;
isotropic porosity is recovered by replacing $\mu$ with unity (Paper
I).  Note from Equation~(\ref{Eq:chieff}) that the effective opacity
approaches the atomic opacity when $\tau_{\rm cl} \ll 1$. For the
porosity length $h$ we assume the ``velocity stretch'' form
$h(r)/\hinf = v(r)/\vinf$, where the velocity field is given by
the standard ``$\beta$-law'' $v(r)/\vinf = (1-R_\star/r)^\beta$.
Thus, to evaluate X-ray line profiles accounting for porosity, it is
only necessary to specify one additional parameter beyond \taustar and
$R_0$: the terminal porosity length \hinf.

It is also possible to account for the effects of UV photoexcitation
of metastable levels in He-like ions and resonance scattering of
strong X-ray lines within this framework \citep{LPKC06,
  LOKP07}. Because the focus of this paper is to derive observational
constraints on porosity and its effects on mass-loss rate
determinations, we do not use fits to He-like triplets for this
purpose, since they are blended. However, in our exploration of the
effects of a finite upper radial cutoff \rmax in
Section~\ref{sec:comparison}, we fit He-like triplet complexes to obtain
constraints on \rmax. Also, with the exception of our fits to
He-like complexes, we also do not account for resonance scattering,
which might have a moderate effect on derived mass-loss rates, but
will not affect our conclusions about the suitability of porous models
or the quantitative tradeoff between porosity and mass-loss rate.

All of the models are implemented as local models in the X-ray
spectral fitting package XSPEC \citep{1996ASPC..101...17A}. The code
is freely available for download\footnote{
\url{http://heasarc.nasa.gov/xanadu/xspec/models/windprof.html}} and is
issued under the General Public License.

\section{Observations, data reduction, and model fitting}
\label{sec:data}

\begin{figure*}[p]
\includegraphics[angle=90,scale=0.12,width=170mm]{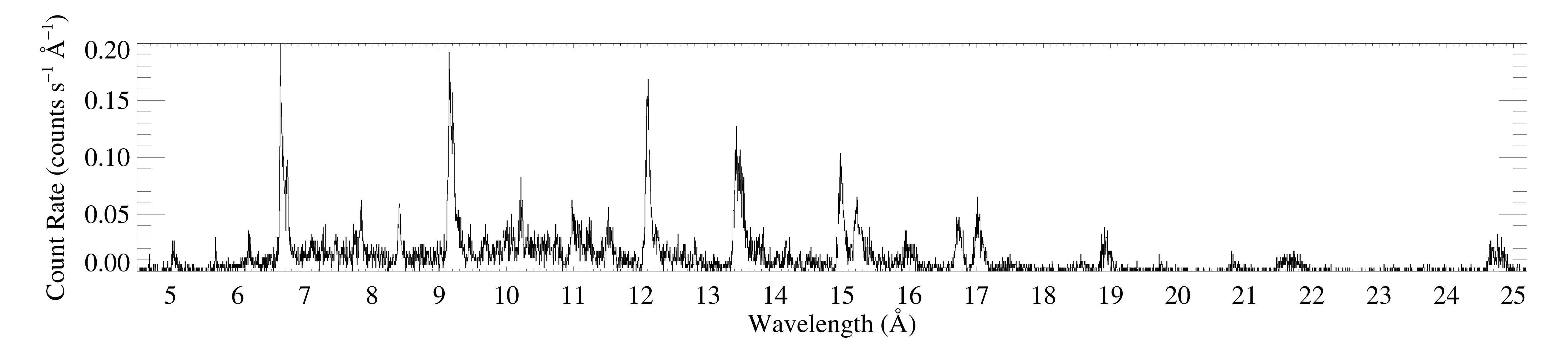}
\includegraphics[angle=90,scale=0.12,width=170mm]{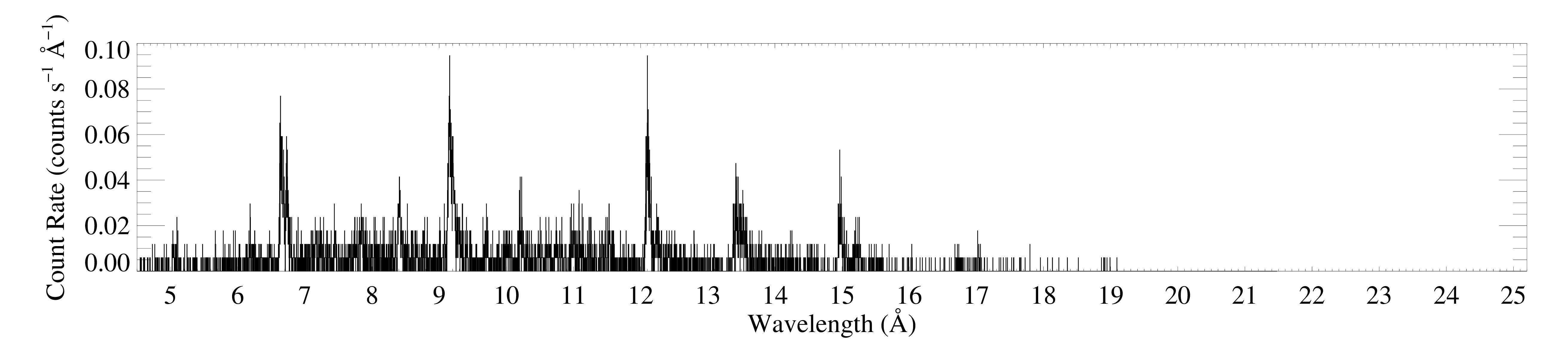}
\includegraphics[angle=90,scale=0.12,width=170mm]{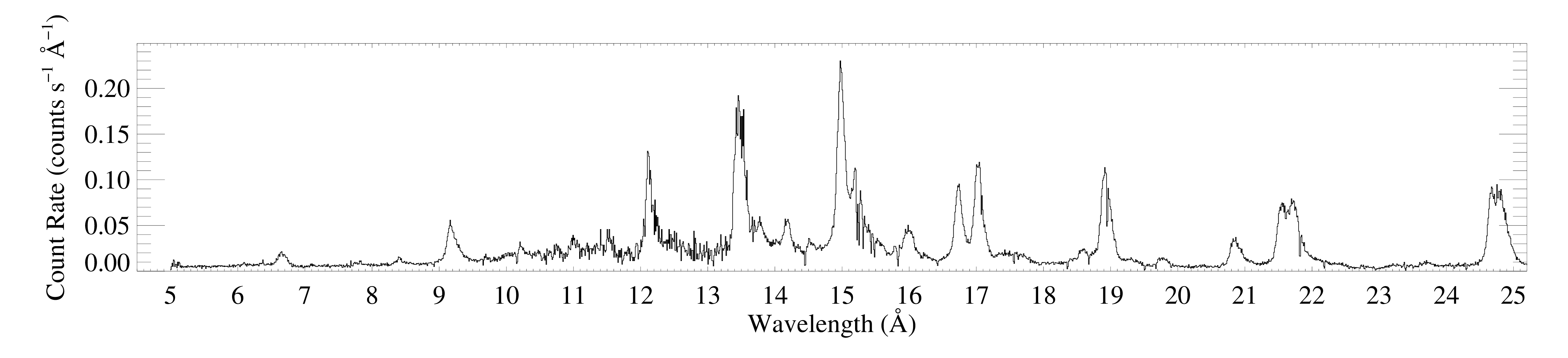}
\includegraphics[angle=90,scale=0.12,width=170mm]{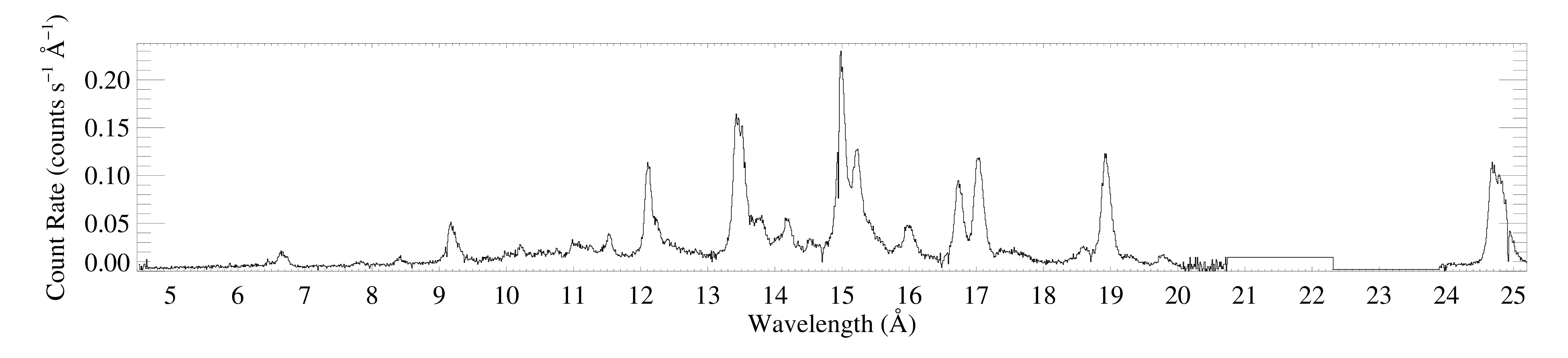}
  \caption{The extracted coadded positive and negative first order MEG
    (top panel) and HEG (second panel) spectra along with the negative
    first order RGS1 (third panel) and RGS2 (bottom panel) spectra
    (coadded 18 pointings).}
\label{fig:entire_spectrum}
\end{figure*}

\zp was observed by \cxo on 2000 March 28-29 using the High Energy
Transmission Grating Spectrometer (HETGS; ObsID 640, 67.7 ks
effective exposure time). \zp was also observed several times in the
last $\sim$ 10 years by \xmm, with total effective exposure time in
excess of 0.8 Ms; the list of observations used is given in
Table~\ref{tab:obslog}. The \cxo Medium Energy Grating (MEG), High
Energy Grating (HEG), and \xmm Reflection Grating Spectrometer (RGS)
spectra are presented in Figure~\ref{fig:entire_spectrum}.

The HETGS data were reprocessed using CIAO 4.2 and CALDB 4.3.1 in
order to apply optimized contemporary charge transfer inefficiency
(CTI) calibrations. Standard procedures were used to produce spectra,
response matrices, and ancillary response files.

The RGS \citep{dHel01} data were processed using the XMM SAS
11.0.0. {\tt rgsproc} was used with the following significant
modifications from default settings: pixel-by-pixel corrections to CTI
based on contemporary diagnostic exposures were applied; periods of
high background were rejected; events on CCD node interfaces, near the
edges of CCDs, and next to bad pixels were not rejected; ``cool''
pixels that have been flagged as having consistently high CTI were
rejected. The resulting spectra were then coadded using {\tt
  rgscombine}.

RGS spectra are affected by small, systematic shifts in the wavelength
scale. The shifts are different for each observation, but constant
within each observation. The shifts have an rms deviation of $\sigma
= 8 {\mathrm {m\AA}}$ \citep{dHel01}. The most recent calibration
implements a correction to the RGS boresight which zeroes out the mean
systematic wavelength scale shift in the data used for calibration
\citep{tn0080}. Since we have combined a large number of observations
of \zp, the overall systematic wavelength scale shift is reduced, and
the combined data set is broadened. For $N$ independent observations
of equal length, the systematic shift in the combined data set is
expected to be of order $8 \mathrm{m\AA} / \sqrt{N}$. Since we are
using 18 observations of roughly equal length, we estimate a net
systematic uncertainty in the wavelength scale of $\sim$2 m\AA.

Based on the fitting results we report in Section~\ref{sec:results}, we
believe that the corrections to the RGS boresight implemented in the
most recent version of the SAS still result in relative differences
between the RGS1 and RGS2 wavelength scales of $\sim 3-5$ m\AA. Thus,
any quantitative measurement of line profile asymmetry using RGS data
will still be dominated by the systematic wavelength scale
uncertainty, and joint fits to data from both RGS instruments will
have poor fit statistics which are dominated by the systematic
disagreement in the relative calibration of these instruments. 

Regardless of these systematic uncertainties in the absolute and
relative wavelength scales of the RGS instruments, the data are of
extremely high statistical quality and are well suited to answering
our two key questions: first, do the data show a bump near line center
indicating that anisotropic porosity is important?; and second, what
is the quantitative tradeoff between porosity effects
and mass-loss rate reductions?

We fit the strong emission lines in the HETGS and RGS spectra with the
line profile models described in Section~\ref{sec:model} using XSPEC 12.6
\citep{1996ASPC..101...17A}. Each line was fit separately, rather than
performing a global fit. The continuum was modeled locally for each
line using a power law in energy with index 2, which is a good
approximation to the shape of the continuum for a small range in
energy. For the \cxo data the continuum strength was determined for a
given spectral region by direct fitting while excluding all lines. For
the RGS data this is not possible because of the extended wings of the
instrumental line spread function (LSF), so the continuum strength was fit
simultaneously with the line profiles.  For lines where data from both
RGS and \cxo are usable, we fit them separately and compare
results. For lines where data from both RGS instruments is usable,
they are fit jointly, using only negative first order data. In all
fits to HETGS data, we used only first order data, fitting positive
and negative orders simultaneously. We fit MEG and HEG simultaneously,
when HEG data were available.  When possible, we fit the line over a
wavelength range about 20\% bigger than the Doppler shift associated
with the wind terminal velocity of \vinf = 2250 \kms
\citep{Haser95PHD}.  When there is a nearby line which could be
blended with the line in question, we eliminate part of the wing of
the line from the fit, a procedure which has been shown to have little
effect on the fit and the derived parameters \citep{Cohen2010}. We
explicitly note these cases in our discussion of individual lines at
the end of this section.

Closely spaced doublets in the \lya lines are fit with a single
profile model centered at the emissivity-weighted wavelength of the
two components.

We assess the goodness of our model fitting and assign confidence
limits using the C statistic \citep{C79} for the low-background,
low-count-per-bin \cxo spectra, and use the chi-square statistic with
Churazov weighting \citep{1996ApJ...471..673C} for the higher
background and higher count RGS data.  Confidence limits are assigned
using the $\Delta \chi^2$ formalism described in Chapter 15.6 of
Numerical Recipes \citep{NR2007}.  We report formal 68.3\% confidence
limits throughout this paper, unless otherwise noted. 

Because of the systematic wavelength offset between RGS1 and RGS2, the
fits to lines where data were available for both RGS instruments
usually had $\chi^2_\nu > 2$. This also occurred in some cases where
models provided very poor fits to the data. XSPEC deems a model with
$\chi^2_\nu > 2$ a poor fit and does not provide confidence intervals
for model parameters, so we do not give them in our results for the
two lines where even the best-fit models do not provide formally good
fits. For models with large porosity lengths, many other lines also do
not have formal confidence limits quoted.

Although it is possible to extract meaningful constraints on model
parameters from fits to blended lines, we have chosen to focus on a
few strong lines with minimal blending in order to emphasize
constraints on porosity and quantitative tradeoffs between porosity
and mass-loss rate. Thus, we do not include fits to He-like triplets
in our study of the effects of porosity, nor do we include the \ion{N}{7}
Ly$\alpha$ line, which is blended with \ion{N}{6} He$\beta$.

Below, we give notes on several of the lines which we fit. Also, note
that for the RGS there are several lines which show variations in the
effective area due to hot pixels (visible as narrow ``dips'' in both
the spectrum and model). These variations are correctly accounted for
by the standard RGS calibration data and pipeline processing. We also
tested the effect of excluding these bins from our fits, and
found that our results did not change significantly.

{\it \ion{Mg}{12} Ly$\alpha$}: The RGS has relatively poor resolving
power at this wavelength and does not provide strong constraints on
profile shape, so we only fit the HETGS data.

{\it \ion{Ne}{10} Ly$\alpha$}: The \ion{Ne}{10} Ly$\alpha$ doublet is
blended with a nearby Fe XVII line at 12.124 \AA, which has a
comparable temperature-dependent emissivity.  It is so close in
wavelength to the Ne X lines, which are themselves very close
together, that the feature can safely be modeled as a single
profile. In contrast, there are two additional iron lines --- another
Fe XVII line at 12.266 \AA\ and an Fe XXI line at 12.284 \AA, which
forms at higher temperatures --- that may be blended with the red wing
of the Ne X line. We therefore exclude the longest wavelength portion
of this line from our fitting for both the \cxo and \xmm data.  

{\it \ion{Fe}{17} 15.014\AA}: The RGS data for this line are
moderately affected by blending from the wings of neighboring lines,
especially the strong \ion{Fe}{17} line at 15.261\AA, so we have
excluded the red wing of this line for the RGS data sets. There are
also a number of weak lines from the Rydberg series of \ion{O}{8} in
this spectral region, as well as satellite lines from
\ion{Fe}{16}. Neither of these are strong enough to have a significant
effect on the profile shape.

{\it \ion{Fe}{17} 16.780\AA}: This line is slightly contaminated by
the Rydberg series of \ion{O}{7}, but this does not have a significant
effect on the line profile shape.  We use only the MEG data and not the
RGS data, because of contamination from the extended wings of the RGS
LSF from the nearby lines at 17.051 and 17.096 \AA.

{\it \ion{O}{8} Ly$\alpha$}: There is slight blending with the Rydberg
series of \ion{N}{6}, but these lines are sufficiently weak that they
do not have a significant effect on the \ion{O}{8} profile shape. Most
of the MEG counts come from the negative first order, which falls on
one of the back illuminated ACIS CCD chips, which has much higher
detection efficiency.

{\it \ion{N}{7} Ly$\beta$}: This line is somewhat weak, but valuable
because it is unblended. The \cxo data are too weak to give meaningful
constraints. Most of the RGS observations put part of this line on a
gap between CCD chips of RGS1, but enough observations exist with
small pointing offsets to give acceptable exposure over the whole
line. This line falls on a failed CCD of RGS2, so no data are available
from that spectrometer.

\begin{deluxetable}{cccc}
\tablecaption{List of \xmm Observations of \zp with Net Exposure Times. \label{tab:obslog}}
\tablehead{
  \colhead{ObsID} &
  \colhead{Date} &
  \multicolumn{2}{c}{Exposure Time (ks)} \\
  \colhead{} &
  \colhead{} & 
  \colhead{RGS1} &
  \colhead{RGS2}
}
\startdata
  0095810301 & 2000 Jun 8 & 52.6 & 51.0 \\
  0095810401 & 2000 Oct 15 & 39.9 & 38.5 \\
  0157160401 & 2002 Nov 10 & 41.6 & 40.2 \\
  0157160501 & 2002 Nov 17 & 38.7 & 38.7 \\
  0157160901 & 2002 Nov 24 & 43.5 & 43.5 \\
  0157161101 & 2002 Dec 15 & 27.8 & 27.8 \\
  0159360101 & 2003 May 30 & 66.3 & 66.3 \\
  0163360101 & 2003 Dec 6 & 41.5 & 41.5 \\
  0159360301 & 2004 Apr 12 & 27.7 & 27.7 \\
  0159360401 & 2004 Nov 14 & 57.3 & 57.3 \\
  0159360501 & 2005 Apr 16 & 34.6 & 34.6 \\
  0159360701 & 2005 Oct 15 & 23.5 & 23.4 \\
  0159360901 & 2005 Dec 3 & 48.3 & 48.2 \\
  0159361101 & 2006 Apr 17 & 42.5 & 42.4 \\
  0414400101 & 2007 Apr 9 & 58.5 & 58.5 \\
  0159361301 & 2008 Oct 13 & 61.2 & 61.2 \\
  0561380101 & 2009 Nov 3 & 64.1 & 64.2 \\
  0561380201 & 2010 Oct 7 & 76.7 & 76.7 \\
  \hline
  Total & & 846.2 & 841.8 \\
\enddata
\end{deluxetable}

\section{Results}
\label{sec:results}

We fit each line in the \xmm and \cxo grating spectra according to
the methodology described in Section~\ref{sec:data}. As discussed in that
section, we exclude from our analysis those lines that are too weak or
too blended to provide strong constraints on the parameters of the
line profile model, described in Section~\ref{sec:model}. For those lines
that do provide useful results, Tables~\ref{tab:chandra_nonporous} and
\ref{tab:rgs_nonporous} summarize the results of the line profile fits
that do not include any
porosity. Tables~\ref{tab:chandra_aniso}--\ref{tab:rgs_iso} summarize
the results of line profile fits including porosity.

In this section, we give single parameter confidence limits for all
parameters of interest. These single parameter confidence limits are
determined with all other parameters of interest free. They are
evaluated for a change in the fit statistic of 1.0 (68.3\% confidence,
corresponding to 1-$\sigma$). These confidence limits are based only
on the change in fit statistic under variation of model parameters and
do not reflect systematic uncertainties. 

\begin{deluxetable*}{ccccccc}
  \tablecaption{Wind profile model without porosity: fit results to
    the \cxo HETGS spectra} 
  \tablehead{ 
    \colhead{ion} & 
    \colhead{$\lambda$} &
    \colhead{\taustar} & 
    \colhead{\ro} & 
    \colhead{Normalization} & 
    \colhead{$C$} & 
    \colhead{$N_{\rm bins}$} \\
    \colhead{}  & 
    \colhead{(\AA)} &
    \colhead{} & 
    \colhead{(\rstar)} & 
    \colhead{($10^{-4}$ photons cm$^{-2}$ s$^{-1}$)} & 
    \colhead{} & 
    \colhead{}
  }
  \startdata
  \ion{Mg}{12} \lya & 8.4210 & $1.22_{-.44}^{+.78}$ & $1.34_{-.21}^{+.17}$ & $0.294_{-.022}^{+.024}$ & 186.5 & 188  \\
  \ion{Ne}{10} \lya & 12.1339 & $2.01_{-.24}^{+.27}$ & $1.45_{-.08}^{+.13}$ & $2.71_{-.09}^{+.09}$ & 191.4 & 176  \\
  \ion{Fe}{17} & 15.014 & $1.94_{-.33}^{+.32}$ & $1.55_{-.12}^{+.12}$ & $5.24_{-.17}^{+.24}$ & 280.8 & 308  \\
  \ion{Fe}{17} & 16.780 & $3.01_{-.70}^{+.32}$ & $1.01_{-.01}^{+.59}$ & $2.45_{-.17}^{+.13}$ & 174.9 & 308  \\
  \ion{O}{8} \lya & 18.969 & $3.00_{-.54}^{+.54}$ & $1.22_{-.21}^{+.37}$ & $3.70_{-.35}^{+.29}$ & 150.9 & 130  \\
  \enddata
  \label{tab:chandra_nonporous}
\end{deluxetable*}

\begin{deluxetable*}{ccccccc}
  \tablecaption{Wind profile model without porosity: fit results to the \xmm RGS spectra}
  \tablehead{ 
    \colhead{Ion} & 
    \colhead{$\lambda$} &
    \colhead{\taustar} & 
    \colhead{\ro} & 
    \colhead{Normalization} & 
    \colhead{$\chi^2$} & 
    \colhead{$N_{\rm bins}$} \\
    \colhead{}  & 
    \colhead{(\AA)} &
    \colhead{} & 
    \colhead{(\rstar)} & 
    \colhead{($10^{-4}$ photons cm$^{-2}$ s$^{-1}$)} & 
    \colhead{} & 
    \colhead{}
  }
  \startdata
  \ion{Ne}{10} \lya  & 12.1339 & $1.81_{-.22}^{+.25}$ & $1.61_{-.18}^{+.15}$ & $3.10_{-.08}^{+.08}$ & 24.0 & 19  \\
  \ion{Fe}{17} & 15.014 & $1.77$ & $1.57$ & $6.39$ & 129.5 & 52  \\
  \ion{O}{8} \lya  & 18.969 & $3.14$ & $1.01$ & $4.66$ & 204.3 & 72  \\
  \ion{N}{7} \lyb  & 20.910 & $4.93_{-.97}^{+.66}$ & $1.41_{-.40}^{+.62}$ & $1.66_{-.10}^{+.07}$ & 40.5 & 36  \\
  \enddata
  \label{tab:rgs_nonporous}
\end{deluxetable*}

\subsection{Comparison of RGS and \cxo Results \label{sec:rgs_chandra}}

\begin{figure*}
\includegraphics[angle=90,scale=\profilePlotPanelSize]{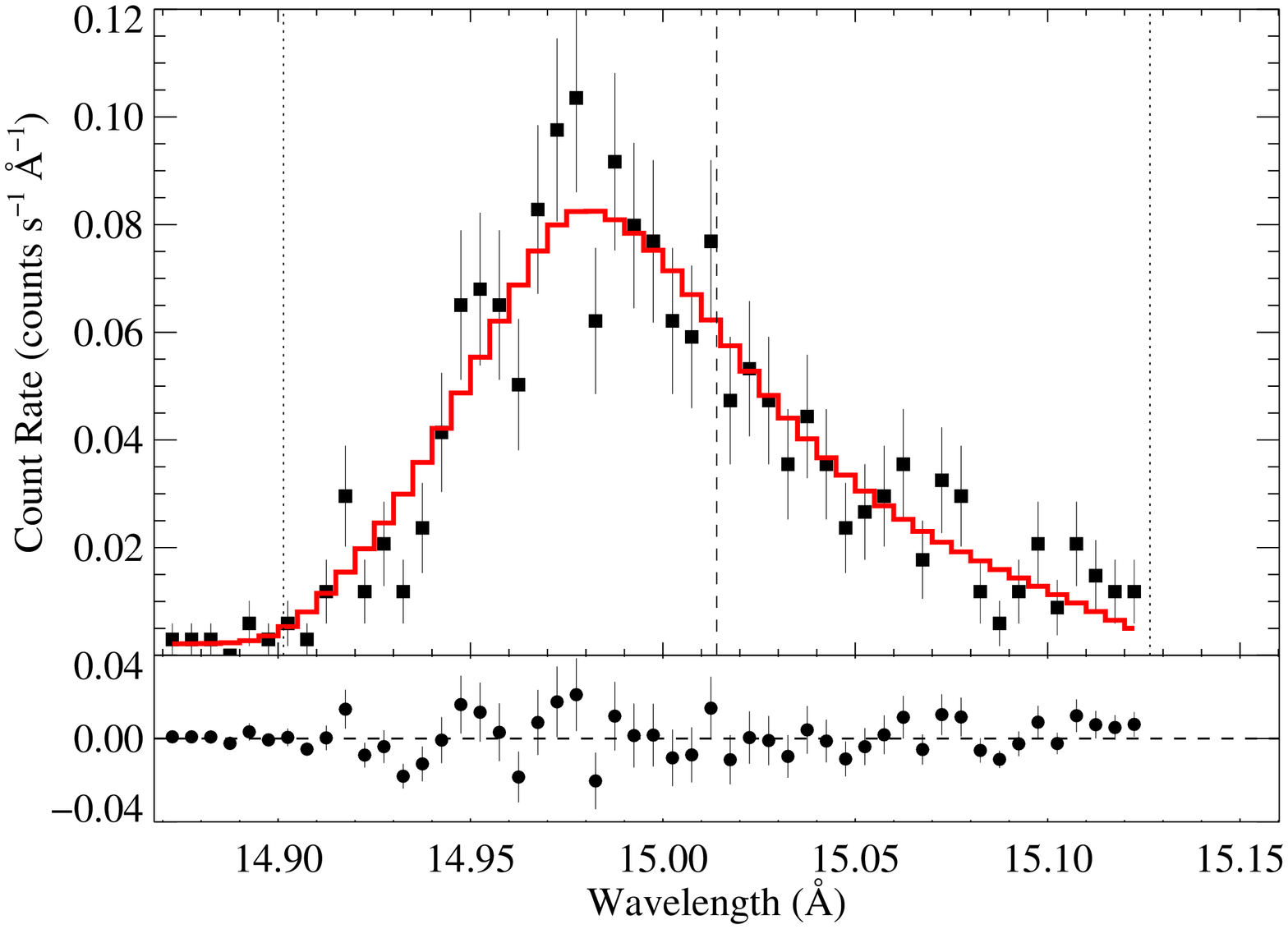}
\includegraphics[angle=90,scale=\profilePlotPanelSize]{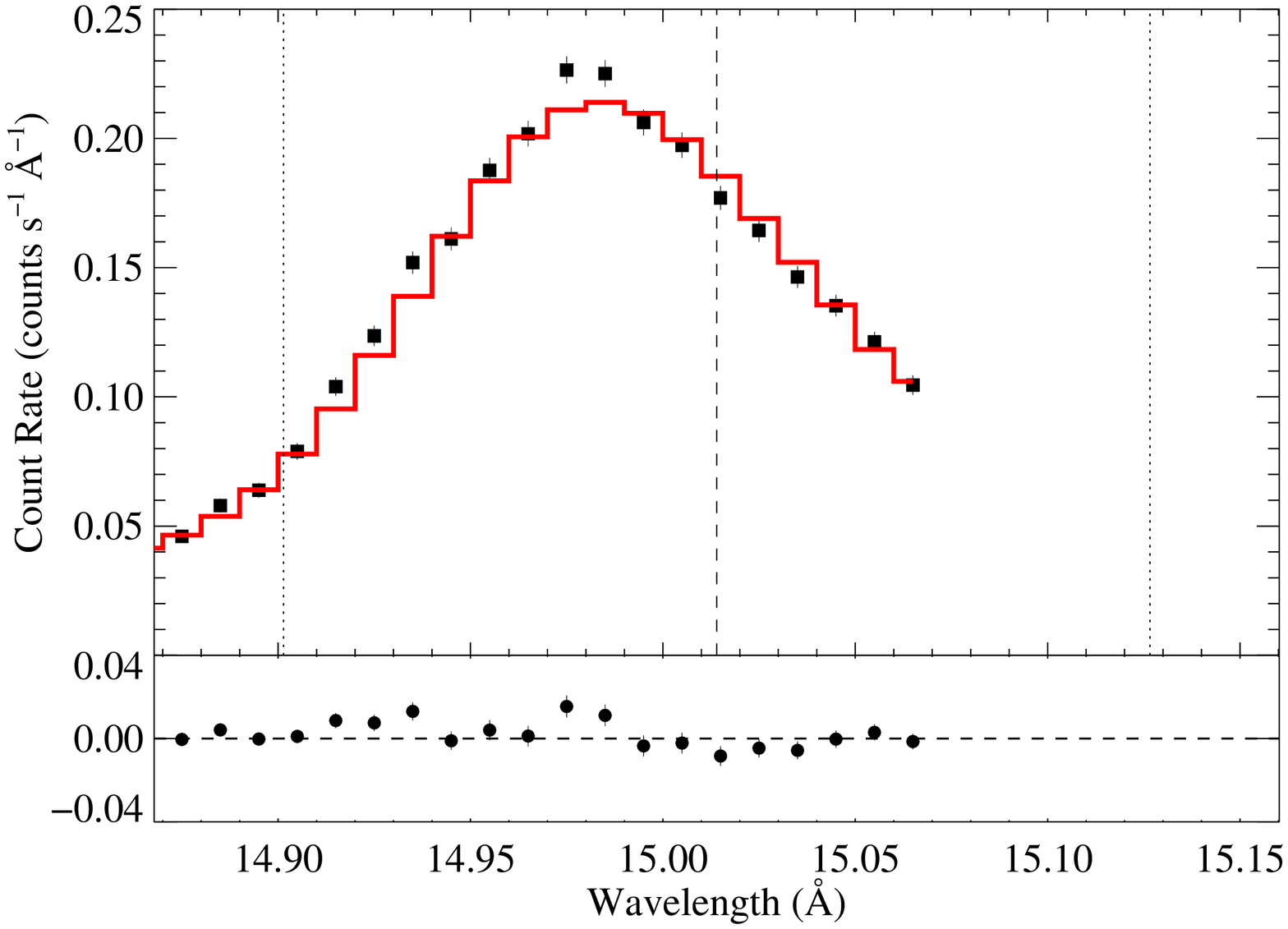}
\caption{The \cxo MEG (left panel) and \xmm RGS1 (right panel) data
  for the \ion{Fe}{17} line at 15.014 \AA, along with the respective
  best-fit models. The vertical dashed line in each figure represents
  the laboratory rest wavelength of the line. The vertical dotted
  lines denote the Doppler shift associated with the wind terminal
  velocity of $\vinf = 2250\, \mathrm{km\, s^{-1}}$. Although the left
  panel shows only the MEG data, the HEG data were fit simultaneously.
  The right panel shows only the RGS1 data; however, both RGS datasets
  were fit simultaneously. (A color version of this figure is
  available in the online journal.)}
\label{fig:15.014_nonporous}
\end{figure*}

We first fit each line without including the effects of porosity.  As
can be seen by inspecting the \taustar and \ro parameters and their
confidence limits in Tables \ref{tab:chandra_nonporous} and
\ref{tab:rgs_nonporous}, we find very good agreement for the three
lines common to both RGS and \cxo datasets.  As an illustrative case,
Figure~\ref{fig:15.014_nonporous} shows the fits to the \ion{Fe}{17}
line at 15.014 \AA\ in the RGS and HETGS spectra.  The fitted \taustar
values are consistent with a mass-loss rate of $\sim 3 \times 10^{-6}$
\Msunyr (or a correspondingly lower mass-loss rate if higher wind
opacity is adopted), as derived from the ensemble of \cxo HETGS lines
in \citet{Cohen2010}, and the fitted \ro values of $\sim 1.5$ \rstar
are consistent with the predictions of line-driven instability (LDI)
simulations of wind shock formation \citep{FPP97,RO02}. The overall
agreement between the \taustar values found in lines from these two
datasets has recently been confirmed by \citet{Naze2012}.


\subsection{Constraints on Porosity Models from Individual Lines
\label{sec:individual_porosity}}

The fits presented in the previous subsection show that the simple
line-profile model without porosity provides good fits to both the
\cxo and \xmm data. The derived parameters are physically reasonable
in the context of embedded wind shocks. However, it is possible that
models that include the effects of porosity could also fit the data.
We now turn our attention to line profile models that include
anisotropic or isotropic porosity.

\begin{figure*}
  \begin{tabular}{cc}
  \includegraphics[angle=90,scale=\profilePlotPanelSize]{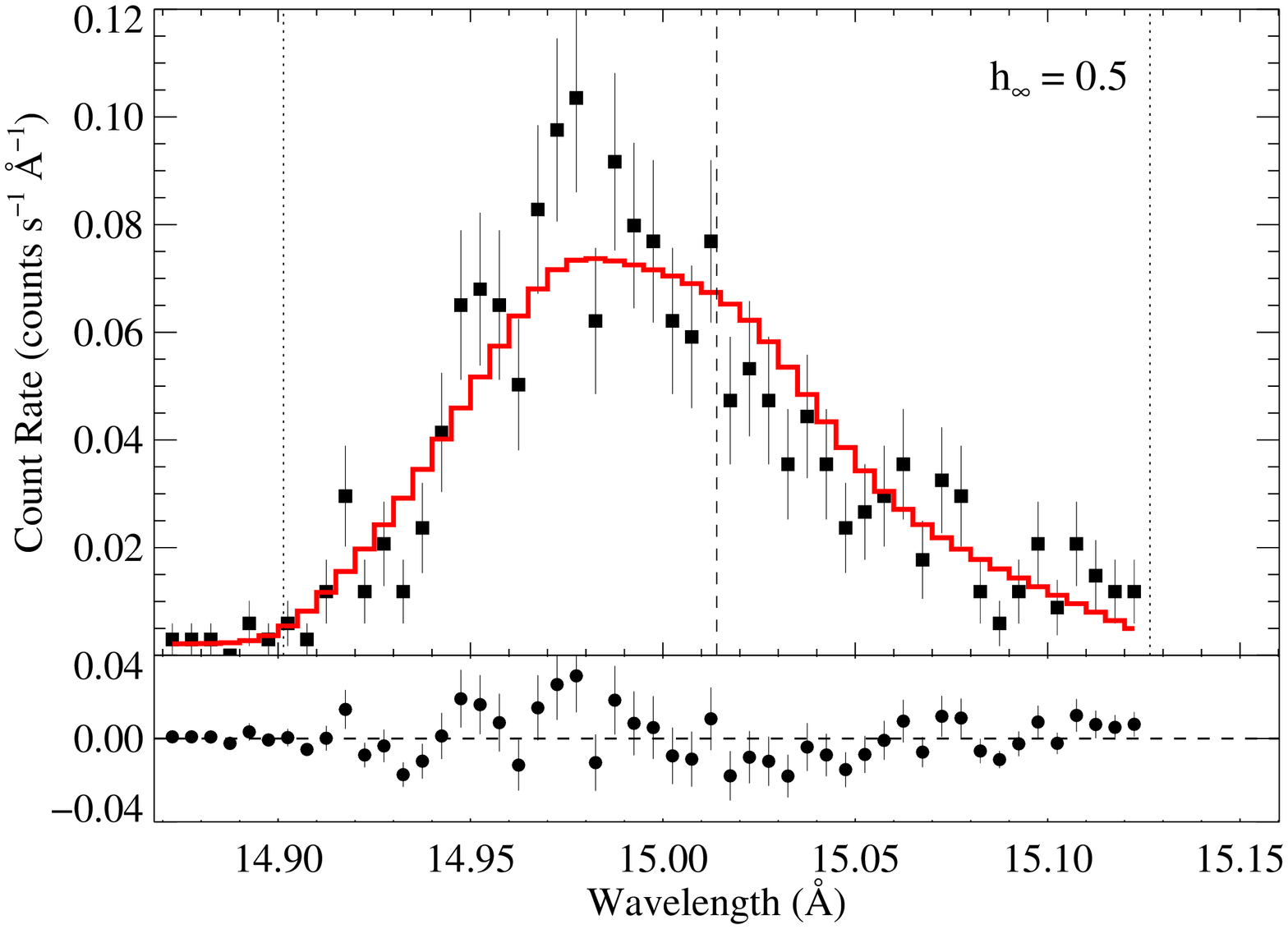} &
  \includegraphics[angle=90,scale=\profilePlotPanelSize]{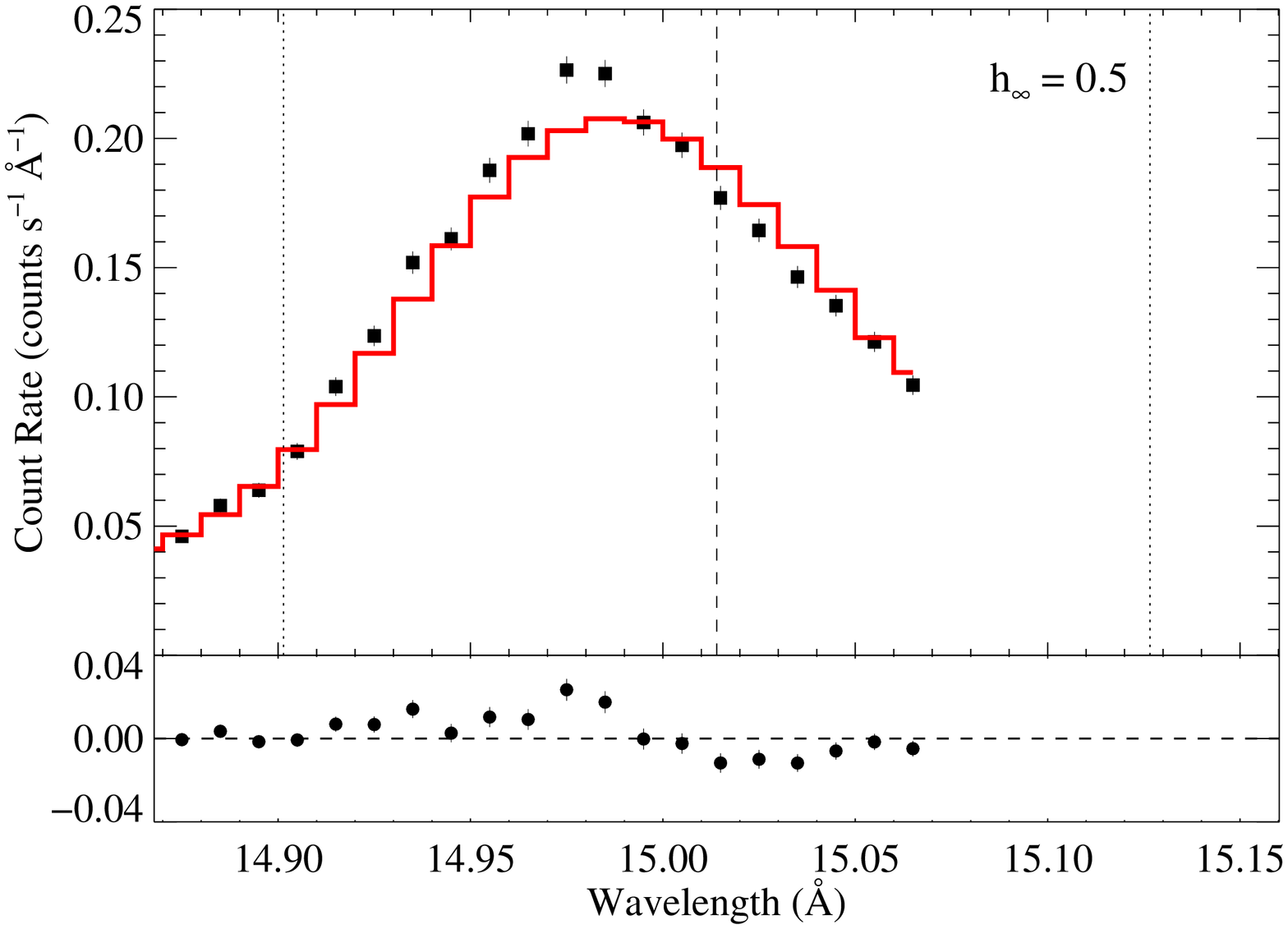} \\
  \includegraphics[angle=90,scale=\profilePlotPanelSize]{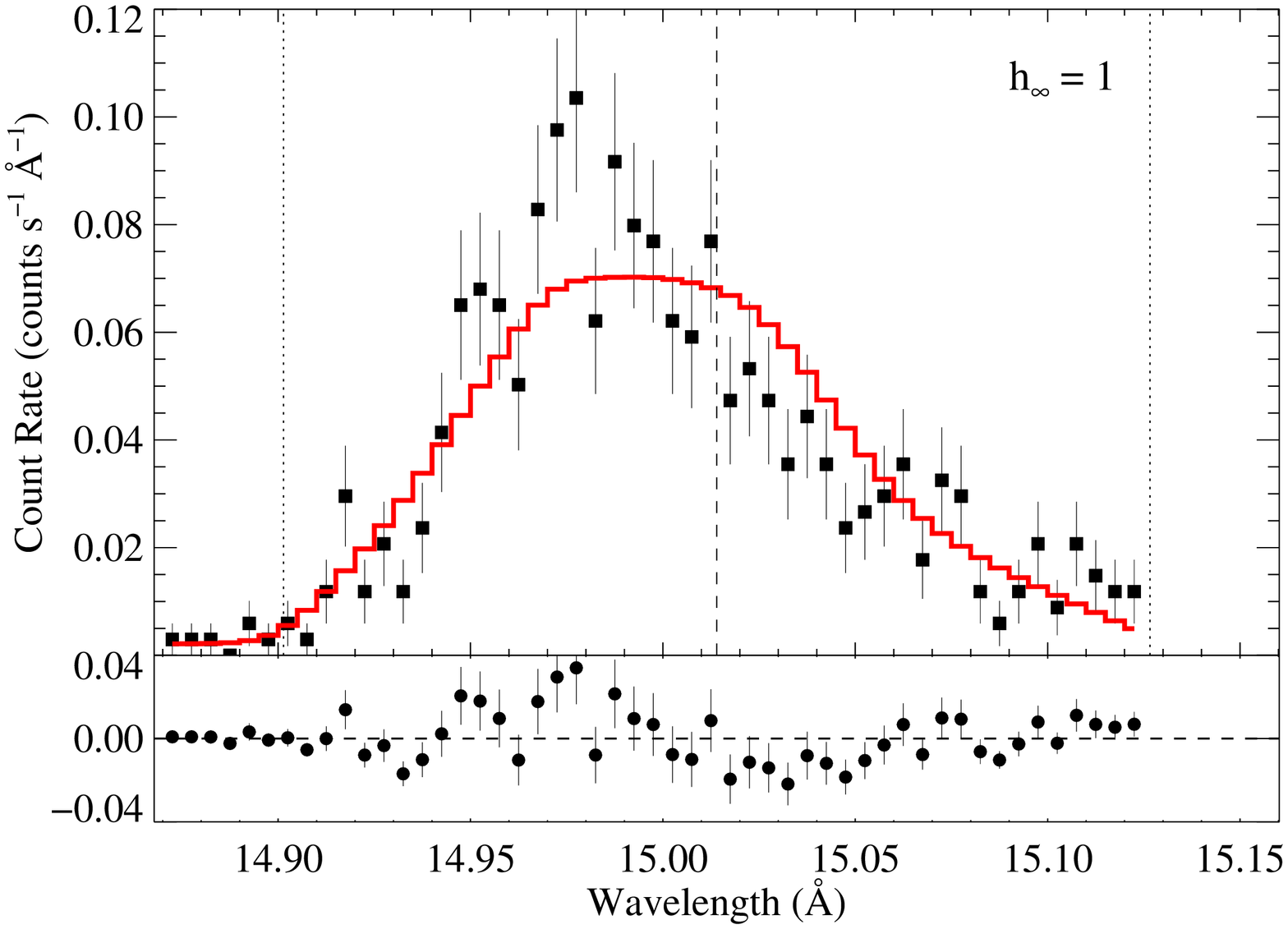} &
  \includegraphics[angle=90,scale=\profilePlotPanelSize]{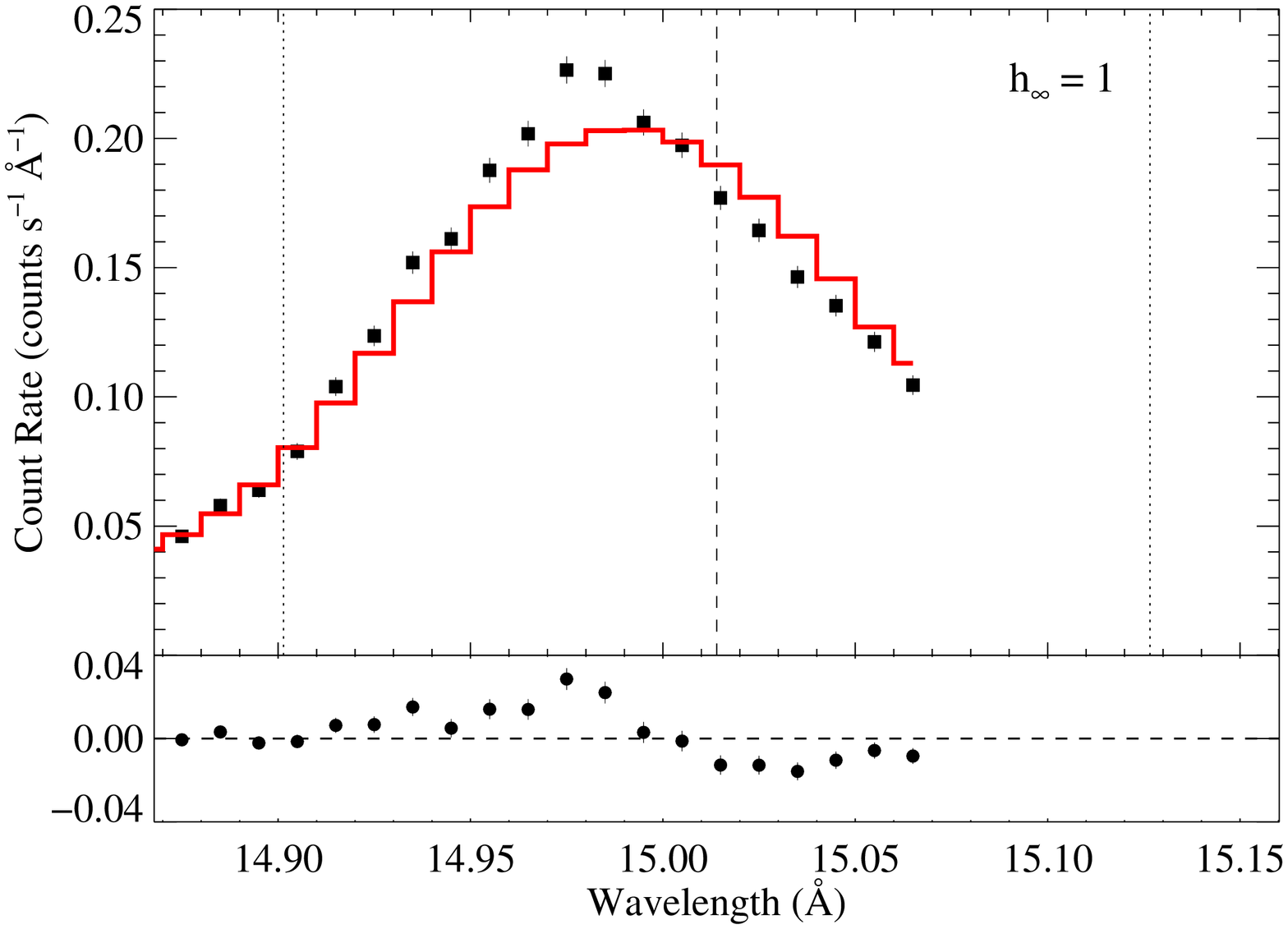} \\
  \includegraphics[angle=90,scale=\profilePlotPanelSize]{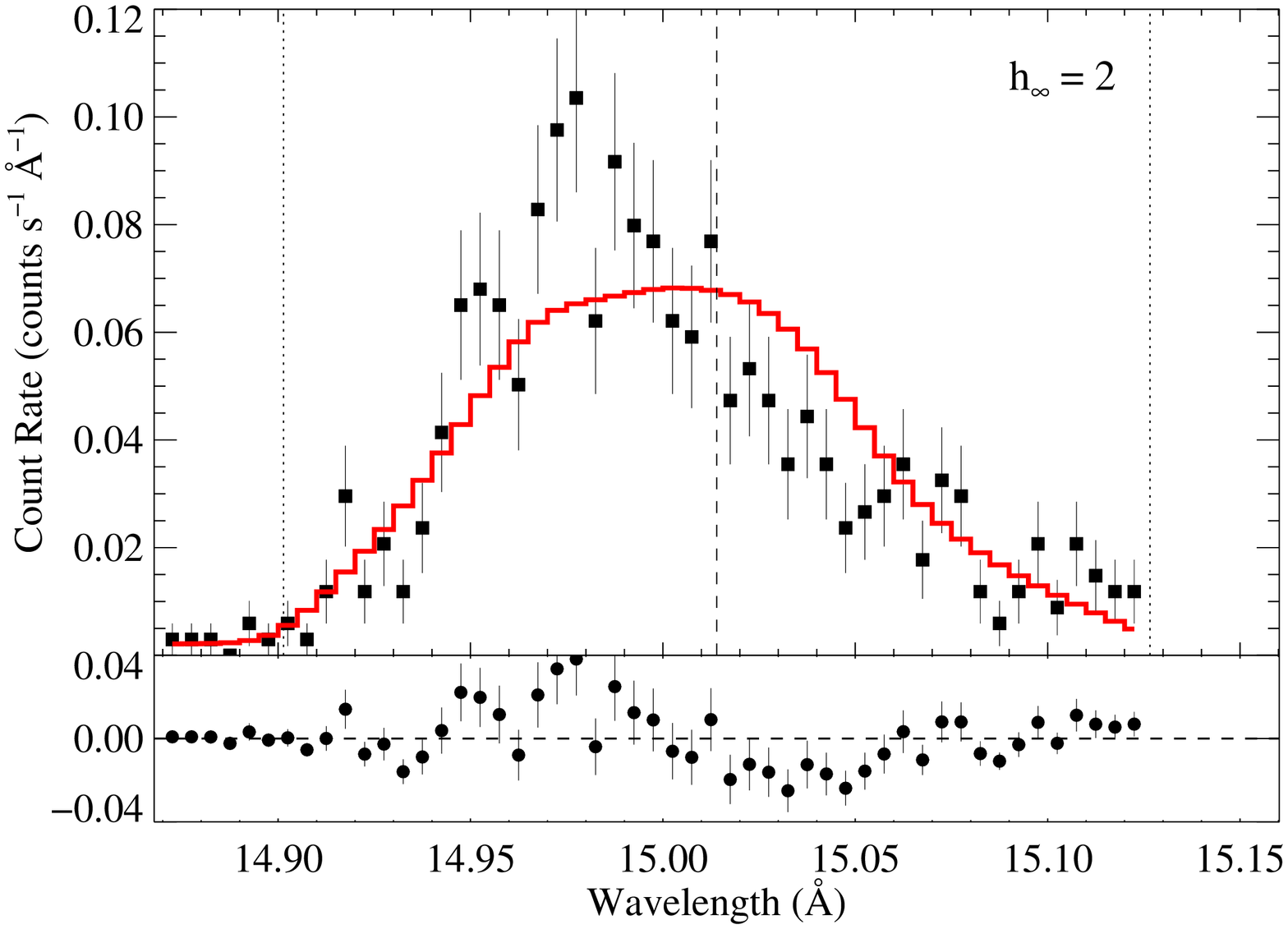} &
  \includegraphics[angle=90,scale=\profilePlotPanelSize]{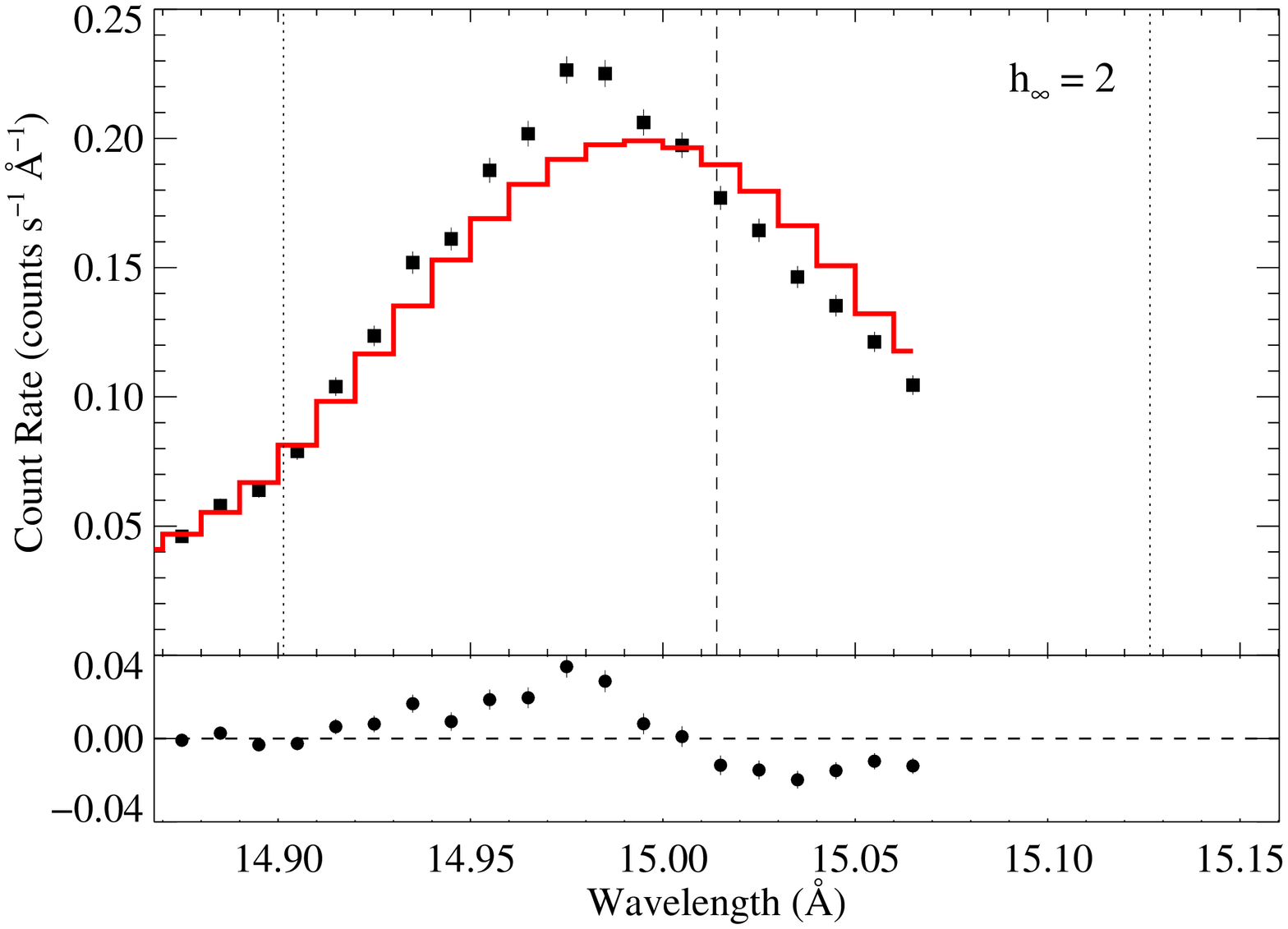} \\
  \includegraphics[angle=90,scale=\profilePlotPanelSize]{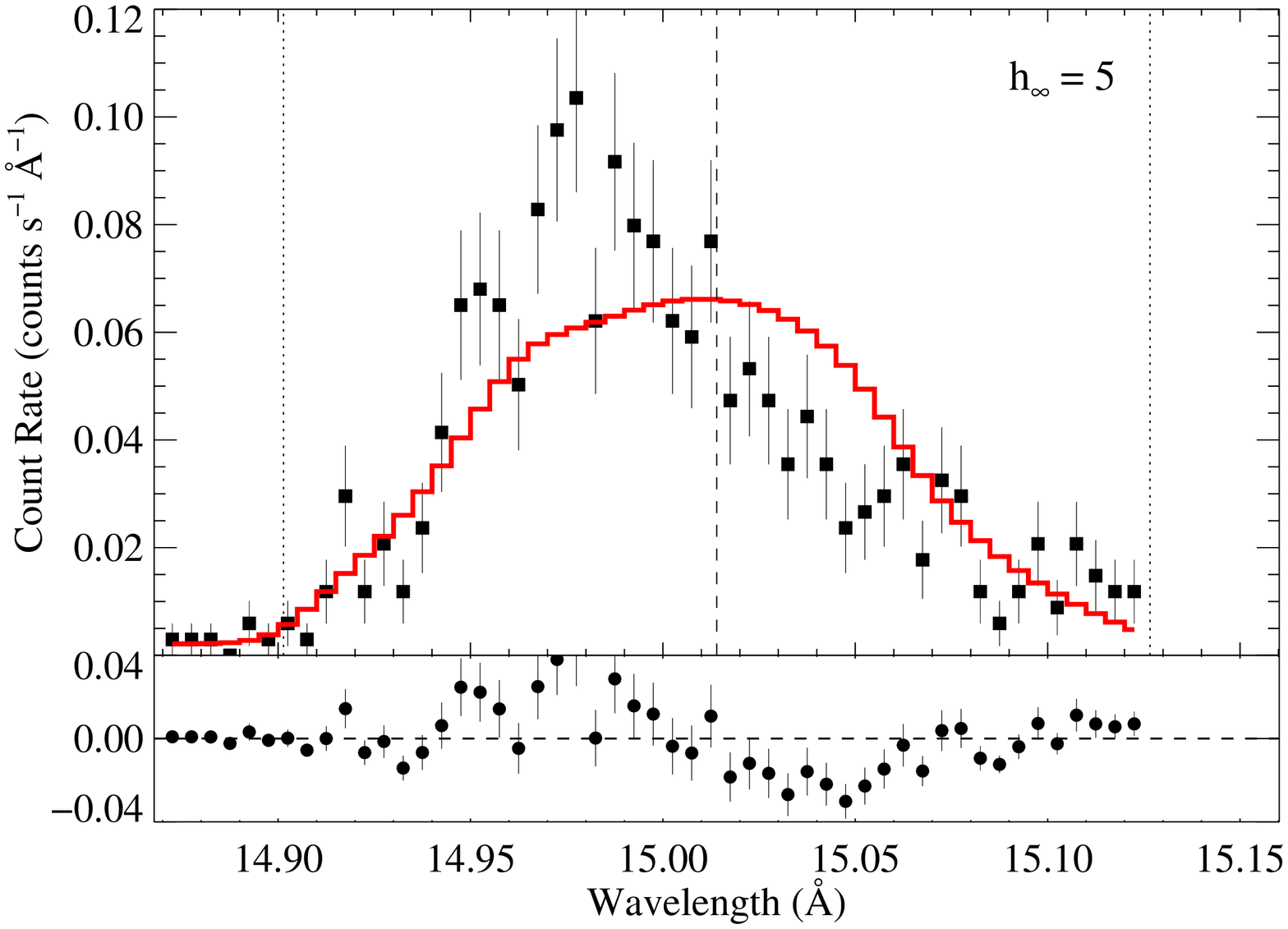} &
  \includegraphics[angle=90,scale=\profilePlotPanelSize]{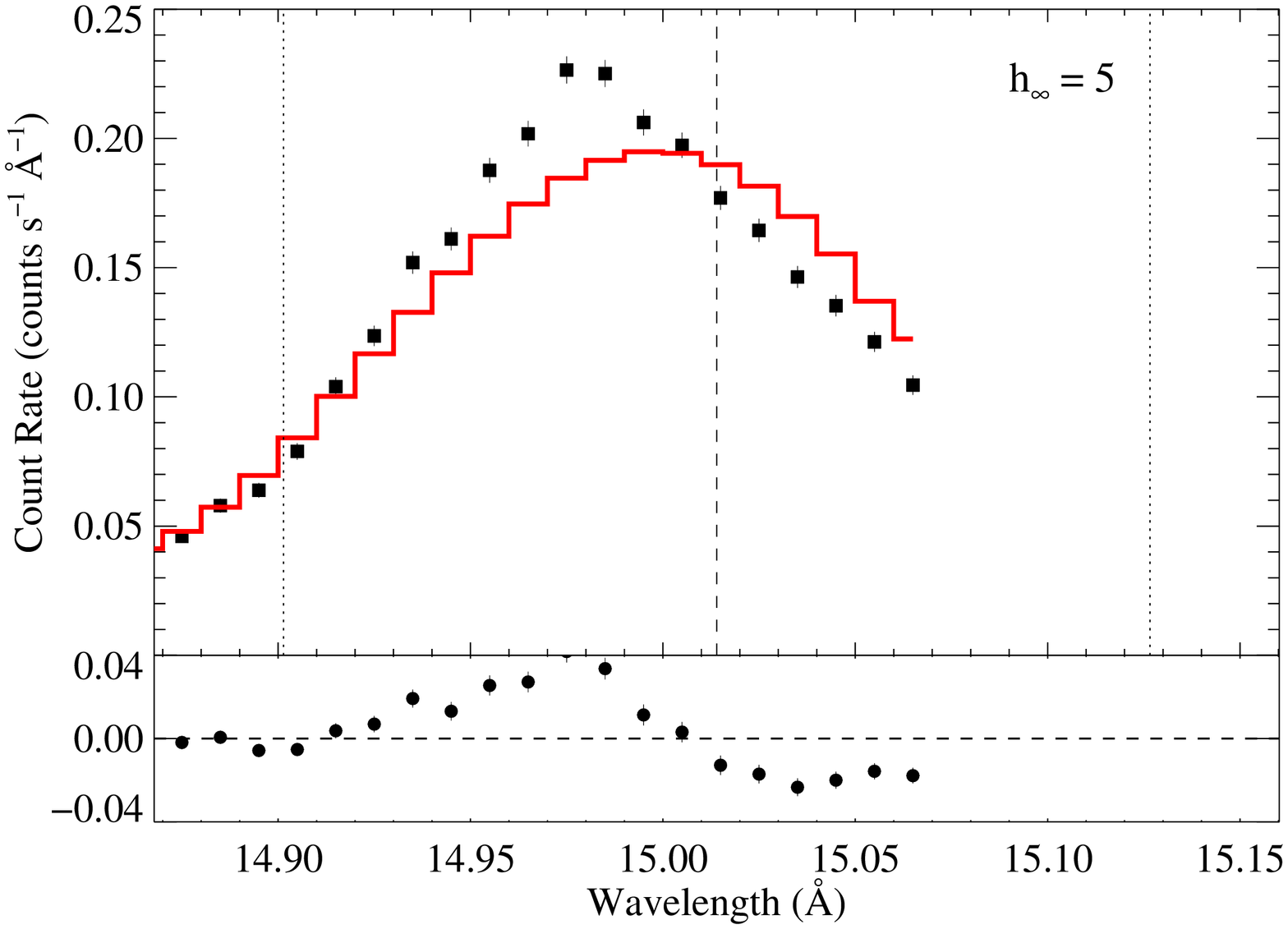} \\
  \end{tabular}
  \caption{The \cxo MEG (left column) and \xmm RGS (right column)
    measurements of the \ion{Fe}{17} line at 15.014 \AA, identical to
    the data shown in Figure~\ref{fig:15.014_nonporous}, but with
    best-fit anisotropic porosity models superimposed.  The
    characteristic ``Venetian blind'' bump at line center (see Figure
    4 of \citet{OFH04} or Figure 3 of Paper I) can be seen in all
    these models, with the feature getting stronger with increasing
    \hinf.  The effect of porosity on the profile shapes can be seen
    qualitatively even for the models with the smallest porosity
    lengths, and even with the moderate resolution of the RGS data.
    These anisotropic porosity profile models do not provide a good
    fit to the data. (A color version of this figure is available in
    the online journal.)}
  \label{fig:15.014_aniso}
\end{figure*}

First, we report on the fitting of models that assume {\it
  anisotropic} porosity, from radially oriented, flattened clumps.  We
first performed fits that allowed the terminal porosity length to be a
free parameter.  These fits generally prefer $\hinf = 0$ (i.e., no
porosity, recovering the same results as in
Section~\ref{sec:rgs_chandra}).  To further explore the effects of
anisotropic porosity on the line profiles, we next fit models with
fixed values of the terminal porosity length, $\hinf$ = 0.5, 1, 2, and
5 \rstar. We find increasingly poor fits as the value of the terminal
porosity length increases.  To illustrate these trends,
Figure~\ref{fig:15.014_aniso} shows the \ion{Fe}{17} line at 15.014
\AA, with the best-fit models assuming each of the four values of
$\hinf$. Plots of fits to the other lines in the spectrum of \zp using
anisotropic porosity models are shown in the Appendix.

Tables~\ref{tab:chandra_aniso} and \ref{tab:rgs_aniso} summarize the
results of fitting anisotropic porosity models to all of the lines. To
facilitate comparison with the non-porous model fitting results, the
table repeats those results in the rows with $\hinf = 0$.  For
models that provide very poor fits (i.e.\ rejected with high
probability compared to the best-fitting $\hinf = 0$ models), we do
not list formal uncertainties on the free model parameters, as the
formalism for assigning these uncertainties assumes that the
best-fitting models do indeed provide formally good fits.

One clear result that can readily be seen in these tables is that in
no case is an anisotropic porous model statistically preferred over
the non-porous model. Even for those lines with a lower
signal-to-noise ratio, the mildest anisotropic porosity models (with
$\hinf = 0.5$ \rstar) can be rejected at the 68.3\% confidence level.

\begin{deluxetable*}{cccccccc}
  \tablecaption{Wind profile model with anisotropic porosity: fit
    results to the \cxo HETGS spectra}
  \tablehead{
    \colhead{ion} &
    \colhead{$\lambda$}&
    \colhead{$h_\infty$} &
    \colhead{\taustar} &
    \colhead{\ro} &
    \colhead{Normalization} & 
    \colhead{$C$} &
    \colhead{$N_{\rm bins}$} \\
    \colhead{} & 
    \colhead{(\AA)} & 
    \colhead{(\rstar)} &
    \colhead{} & 
    \colhead{(\rstar)} & 
    \colhead{($10^{-4}$ photons cm$^{-2}$ s$^{-1}$)} & 
    \colhead{} & 
    \colhead{} 
  }
  \startdata
  \ion{Mg}{12} \lya & 8.4210 & 0  & $1.22_{-.44}^{+.78}$ & $1.34_{-.21}^{+.17}$ & $0.294_{-.022}^{+.024}$ & 186.5 & 188  \\
& & 0.5 & $1.72_{-.81}^{+.77}$ & $1.37_{-.14}^{+.17}$ & $0.281_{-.008}^{+.050}$ & 189.8 & 188  \\
& & 1 & $1.46_{-.60}^{+1.58}$ & $1.45_{-.16}^{+.14}$ & $0.305_{-.033}^{+.018}$ & 191.5 & 188  \\
& & 2 & $1.83_{-.95}^{+2.08}$ & $1.49_{-.12}^{+.13}$ & $0.298_{-.024}^{+.023}$ & 194.1 & 188  \\
& & 5 & $2.37_{-1.48}^{+6.19}$ & $1.54_{-.12}^{+.15}$ & $0.297_{-.023}^{+.024}$ & 198.6 & 188  \\

  \ion{Ne}{10} \lya & 12.1339 & 0 & $2.01_{-.24}^{+.27}$ & $1.45_{-.08}^{+.13}$ & $2.71_{-.09}^{+.09}$ & 191.4 & 176  \\
  & & 0.5 & $2.92_{-.41}^{+.44}$ & $1.49_{-.09}^{+.09}$ & $2.72_{-.09}^{+.09}$ & 208.8 & 176  \\
  & & 1 & $3.60_{-.61}^{+.64}$ & $1.56_{-.08}^{+.08}$ & $2.71_{-.08}^{+.10}$ & 223.8 & 176  \\
  & & 2 & $5.00_{-1.01}^{+1.37}$ & $1.65_{-.09}^{+.07}$ & $2.73_{-.09}^{+.10}$ & 245.0 & 176  \\
  & & 5 & $11.32_{-3.72}^{+5.74}$ & $1.77_{-0.10}^{+0.11}$ & $2.76_{-0.09}^{+0.10}$ & 277.6 & 176  \\

  \ion{Fe}{17} & 15.014 & 0 & $1.94_{-.33}^{+.32}$ & $1.55_{-.12}^{+.12}$ & $5.24_{-.17}^{+.24}$ & 280.8 & 308  \\
   &  & 0.5 & $2.51_{-.40}^{+.68}$ & $1.61_{-.14}^{+.08}$ & $5.23_{-.18}^{+.23}$ & 293.9 & 308  \\
   &  & 1 & $3.02_{-.65}^{+.84}$ & $1.65_{-.10}^{+.10}$ & $5.22_{-.18}^{+.23}$ & 303.8 & 308  \\
   &  & 2 & $3.76_{-1.00}^{+1.56}$ & $1.74_{-.11}^{+.09}$ & $5.22_{-.18}^{+.23}$ & 317.4 & 308  \\
   &  & 5 & $6.95_{-3.01}^{+5.32}$ & $1.84_{-.11}^{+.11}$ & $5.21_{-.18}^{+.23}$ & 338.4 & 308  \\

  \ion{Fe}{17} & 16.780 & 0 & $3.01_{-.70}^{+.32}$ & $1.01_{-.01}^{+.59}$ & $2.45_{-.17}^{+.13}$ & 174.9 & 308  \\
 & & 0.5 & $4.12_{-.90}^{+.93}$ & $1.40_{-.19}^{+.22}$ & $2.40_{-.11}^{+.19}$ & 180.7 & 308  \\
 & & 1 & $5.45_{-1.33}^{+1.46}$ & $1.43_{-.09}^{+.25}$ & $2.47_{-.18}^{+.13}$ & 185.6 & 308  \\
 & & 2 & $8.77_{-2.60}^{+3.39}$ & $1.58_{-.15}^{+.17}$ & $2.41_{-.12}^{+.19}$ & 193.2 & 308  \\
 & & 5 & $28.99_{-11.96}^{+15.55}$ & $1.64_{-.16}^{+.17}$ & $2.38_{-.09}^{+.23}$ & 206.3 & 308  \\

  \ion{O}{8} \lya & 18.969 & 0 & $3.00_{-.54}^{+.54}$ & $1.22_{-.21}^{+.37}$ & $3.70_{-.35}^{+.29}$ & 150.9 & 130  \\
  & & 0.5 & $4.26_{-1.16}^{+1.18}$ & $1.43_{-.21}^{+.31}$ & $3.70_{-.26}^{+.28}$ & 152.9 & 130  \\
  & & 1 & $5.22_{-1.61}^{+2.17}$ & $1.57_{-.23}^{+.26}$ & $3.70_{-.27}^{+.30}$ & 155.1 & 130  \\
  & & 2 & $7.83_{-2.93}^{+3.92}$ & $1.66_{-.20}^{+.25}$ & $3.62_{-.19}^{+.38}$ & 158.3 & 130  \\
  & & 5 & $21.10_{-10.50}^{+15.92}$ & $1.74_{-.20}^{+.23}$ & $3.57_{-.15}^{+.42}$ & 165.0 & 130  \\
  \enddata
  \label{tab:chandra_aniso}
\end{deluxetable*}

\begin{deluxetable*}{cccccccc}
  \tablecaption{Wind profile model with anisotropic porosity: fit results to the \xmm RGS spectra}
  \tablehead{
    \colhead{ion} & 
    \colhead{$\lambda$} &
    \colhead{$h_\infty$} & 
    \colhead{\taustar} & 
    \colhead{\ro} & 
    \colhead{Normalization} & 
    \colhead{$\chi^2$} &
    \colhead{$N_{\rm bins}$} \\
    \colhead{} & 
    \colhead{(\AA)} & 
    \colhead {(\rstar)} & 
    \colhead{} & 
    \colhead{(\rstar)} & 
    \colhead{($10^{-4}$ photons cm$^{-2}$ s$^{-1}$)} & 
    \colhead{} & 
    \colhead{}
  }
  \startdata
  \ion{Ne}{10} \lya & 12.1339 & 0 & $1.81_{-.22}^{+.25}$ & $1.61_{-.18}^{+.15}$ & $3.10_{-.08}^{+.08}$ & 24.0 & 19  \\
  & & 0.5 & $3.92_{-.57}^{+.83}$ & $1.35_{-.19}^{+.21}$ & $3.18_{-.08}^{+.08}$ & 27.3 & 19  \\
  & & 1 & $5.71_{-1.00}^{+1.23}$ & $1.41_{-.14}^{+.15}$ & $3.22_{-.09}^{+.08}$ & 34.2 & 19  \\
  & & 2 & $12.61$ & $1.46$ & $3.31$ & 47.2 & 19  \\
  & & 5 & $64.82$ & $1.35$ & $3.44$ & 66.9 & 19  \\

  \ion{Fe}{17} & 15.014 & 0 & $1.77$ & $1.57$ & $6.39$ & 129.5 & 52  \\
  & & 0.5 & $3.29$ & $1.44$ & $6.47$ & 180.4 & 52  \\
  & & 1 & $4.66$ & $1.50$ & $6.52$ & 249.6 & 52  \\
  & & 2 & $8.24$ & $1.57$ & $6.56$ & 370.1 & 52  \\
  & & 5 & $35.96$ & $1.54$ & $6.66$ & 577.2 & 52  \\

  \ion{Fe}{17} & 16.780 & 0 & $3.38_{-.45}^{+.31}$ & $1.54_{-.39}^{+.33}$ & $3.01_{-.06}^{+.07}$ & 30.3 & 32  \\
  & & 0.5 & $4.01_{-.83}^{+.83}$ & $1.99_{-.25}^{+.26}$ & $3.20_{-.07}^{+.07}$ & 42.2 & 32  \\
  & & 1 & $4.17_{-1.03}^{+1.26}$ & $2.24_{-.25}^{+.24}$ & $3.29_{-.07}^{+.08}$ & 48.0 & 32  \\
  & & 2 & $4.47_{-1.64}^{+1.87}$ & $2.49_{-.24}^{+.23}$ & $3.41_{-.08}^{+.09}$ & 54.5 & 32  \\
  & & 5 & $4.28$ & $2.89$ & $3.59$ & 62.3 & 32  \\

  \ion{O}{8} \lya  & 18.969 & 0 & $3.14$ & $1.01$ & $4.66$ & 204.3 & 72  \\
  &  & 0.5 & $5.30$ & $1.35$ & $4.76$ & 275.2 & 72  \\
  &  & 1 & $8.14$ & $1.42$ & $4.81$ & 355.1 & 72  \\
  &  & 2 & $16.81$ & $1.46$ & $4.87$ & 478.8 & 72  \\
  &  & 5 & $73.81$ & $1.37$ & $4.92$ & 645.8 & 72  \\

  \ion{N}{7} \lyb & 20.910 & 0 & $4.93_{-1.03}^{+.66}$ & $1.41_{-.40}^{+.62}$ & $1.66_{-.10}^{+.07}$ & 40.5 & 36  \\
  & & 0.5 & $7.96_{-1.81}^{+2.20}$ & $1.60_{-.36}^{+.41}$ & $1.69_{-.10}^{+.11}$ & 52.9 & 36  \\
  & & 1 & $12.95_{-3.34}^{+5.01}$ & $1.70_{-.33}^{+.44}$ & $1.75_{-.12}^{+.11}$ & 62.4 & 36  \\
  & & 2 & $34.11$ & $1.62$ & $1.86$ & 73.5 & 36  \\
  & & 5 & $> 100$ & $1.73$ & $1.91$ & 84.4 & 36  \\
  
  \enddata

  \label{tab:rgs_aniso}
\end{deluxetable*}

Let us next consider models with {\it isotropic} porosity. In this
case, there is some degeneracy between porous profiles with lower
optical depths and non-porous profiles with higher optical depths, as
shown in Figure 3 in Paper I. We find that models with isotropic
porosity provide better fits than the corresponding anisotropic
porosity profile models.  However, when we allow $\hinf$ to be a
free parameter, non-porous ($\hinf = 0$) models are almost always
preferred, and in no case is $\hinf = 0$ rejected at greater than
90\% confidence for any one line.

\begin{figure*}
  \begin{tabular}{cc}
  \includegraphics[angle=90,scale=\profilePlotPanelSize]{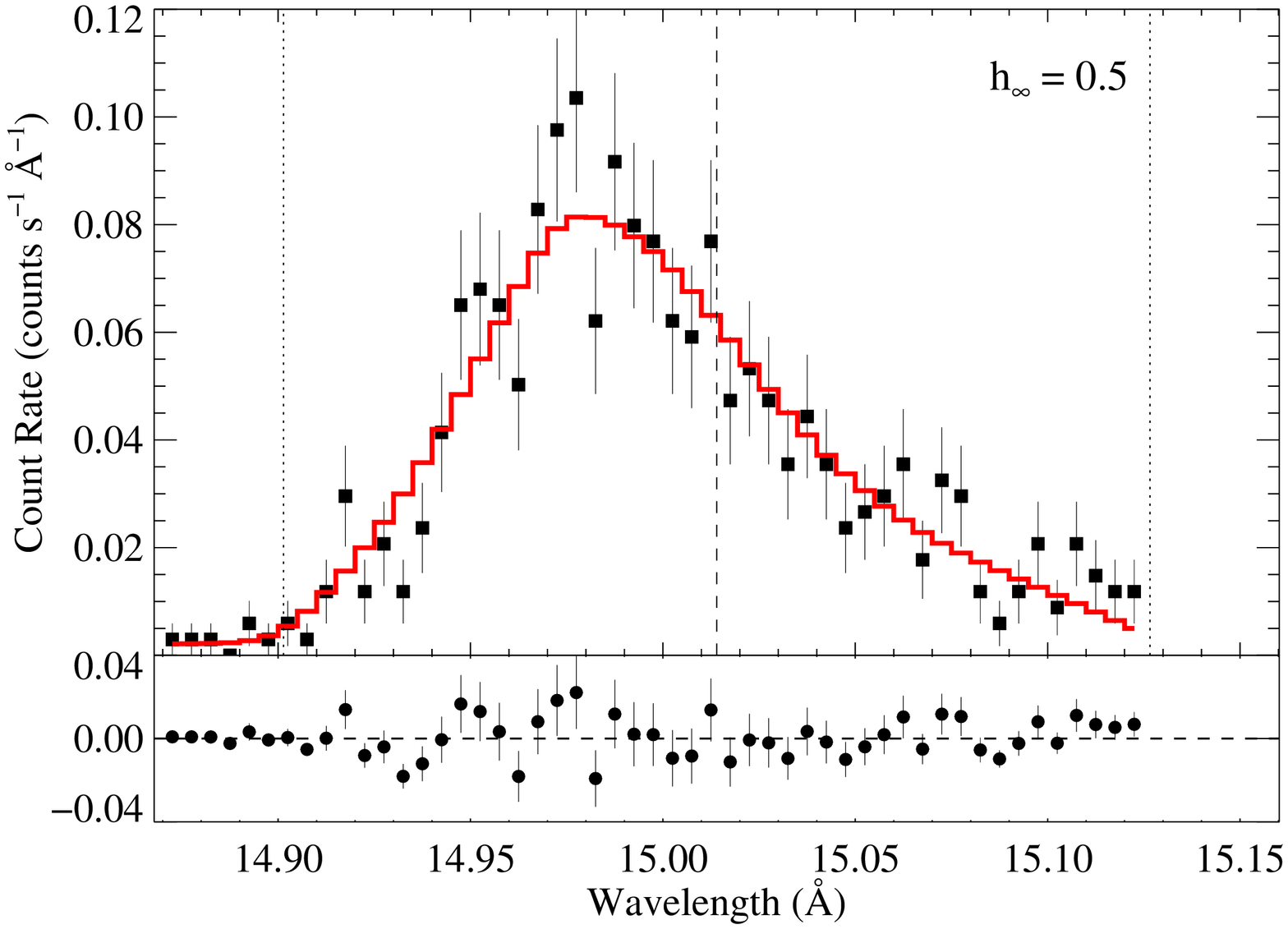} &
  \includegraphics[angle=90,scale=\profilePlotPanelSize]{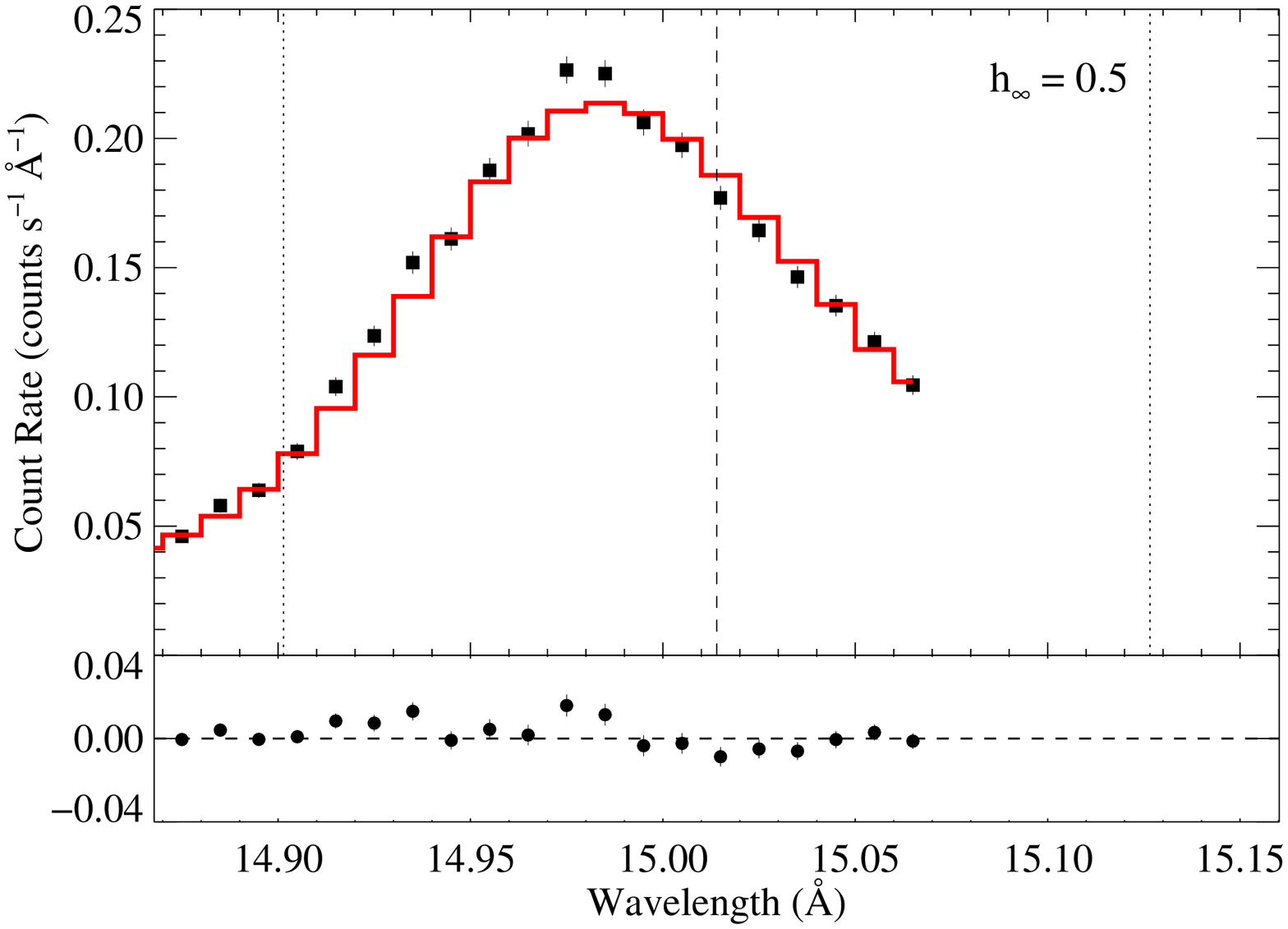} \\
  \includegraphics[angle=90,scale=\profilePlotPanelSize]{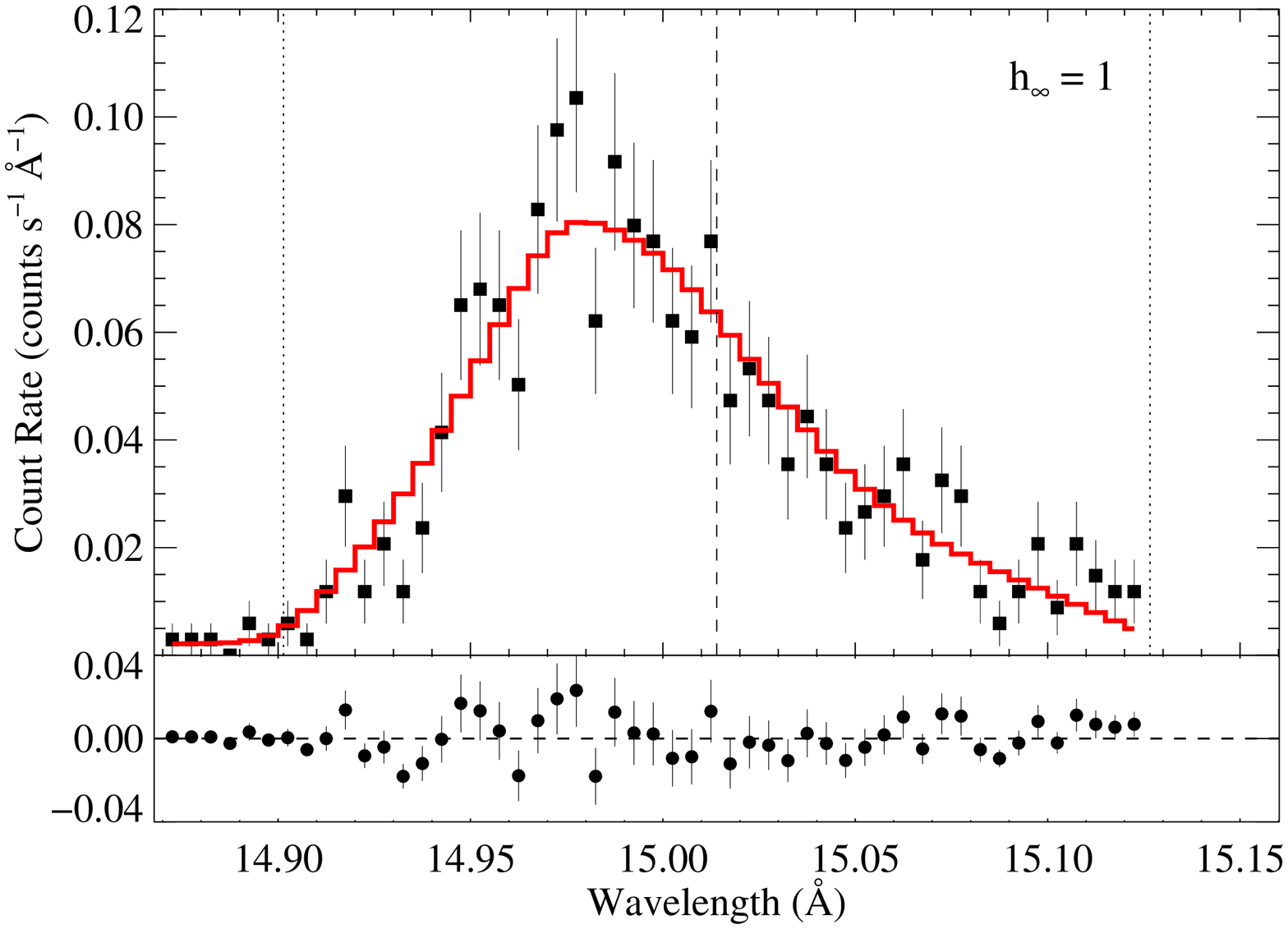} &
  \includegraphics[angle=90,scale=\profilePlotPanelSize]{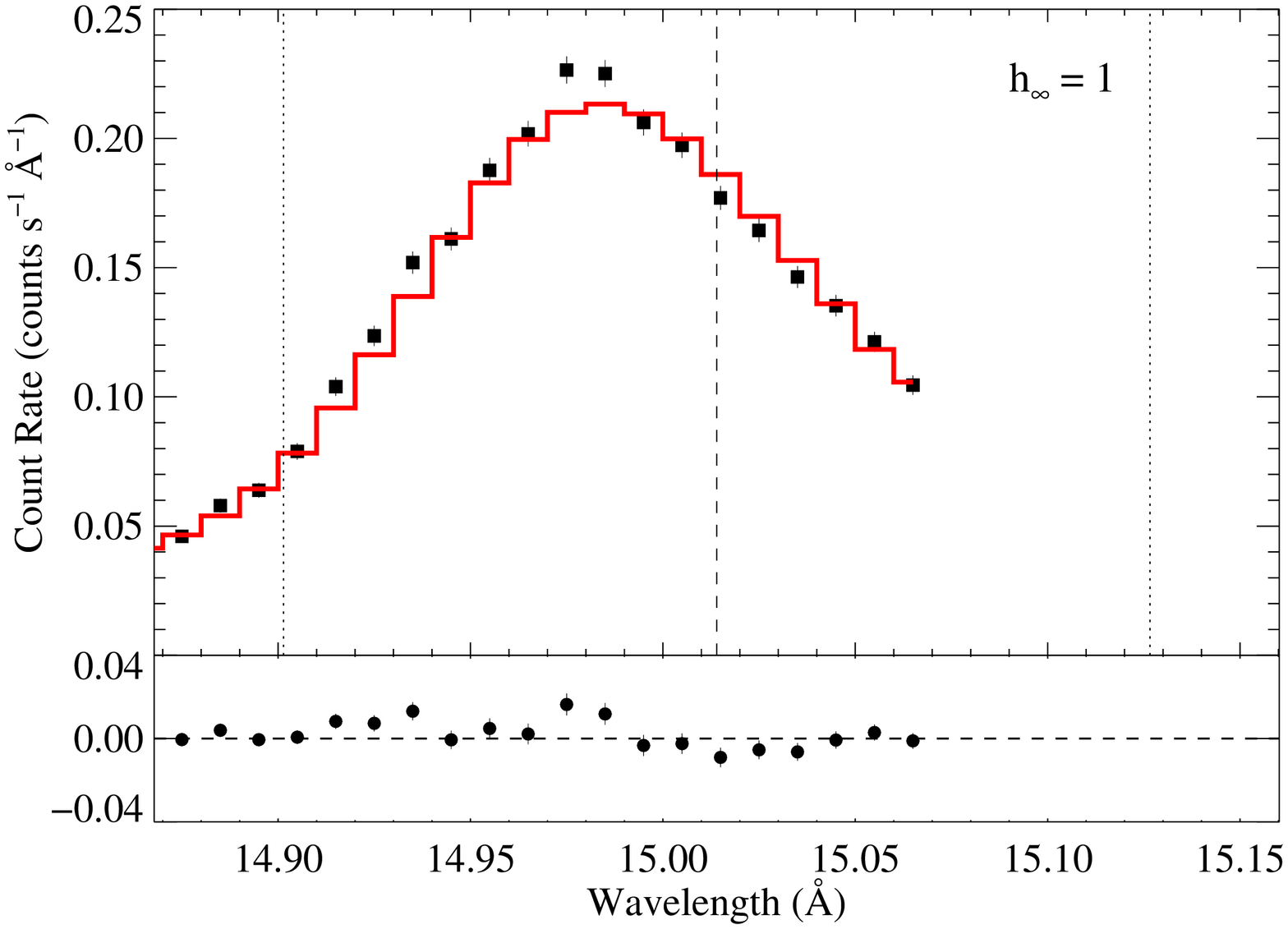} \\
  \includegraphics[angle=90,scale=\profilePlotPanelSize]{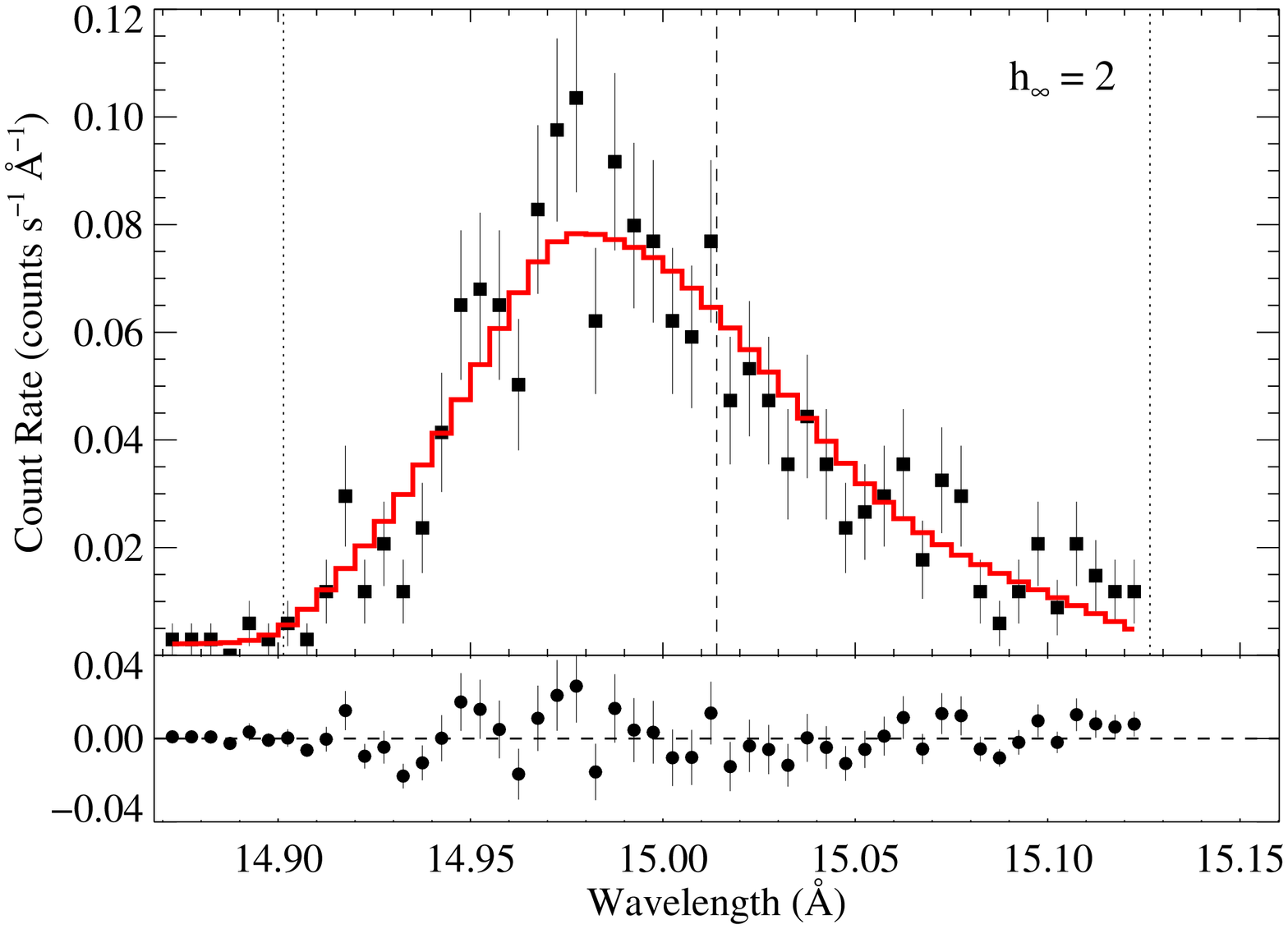} &
  \includegraphics[angle=90,scale=\profilePlotPanelSize]{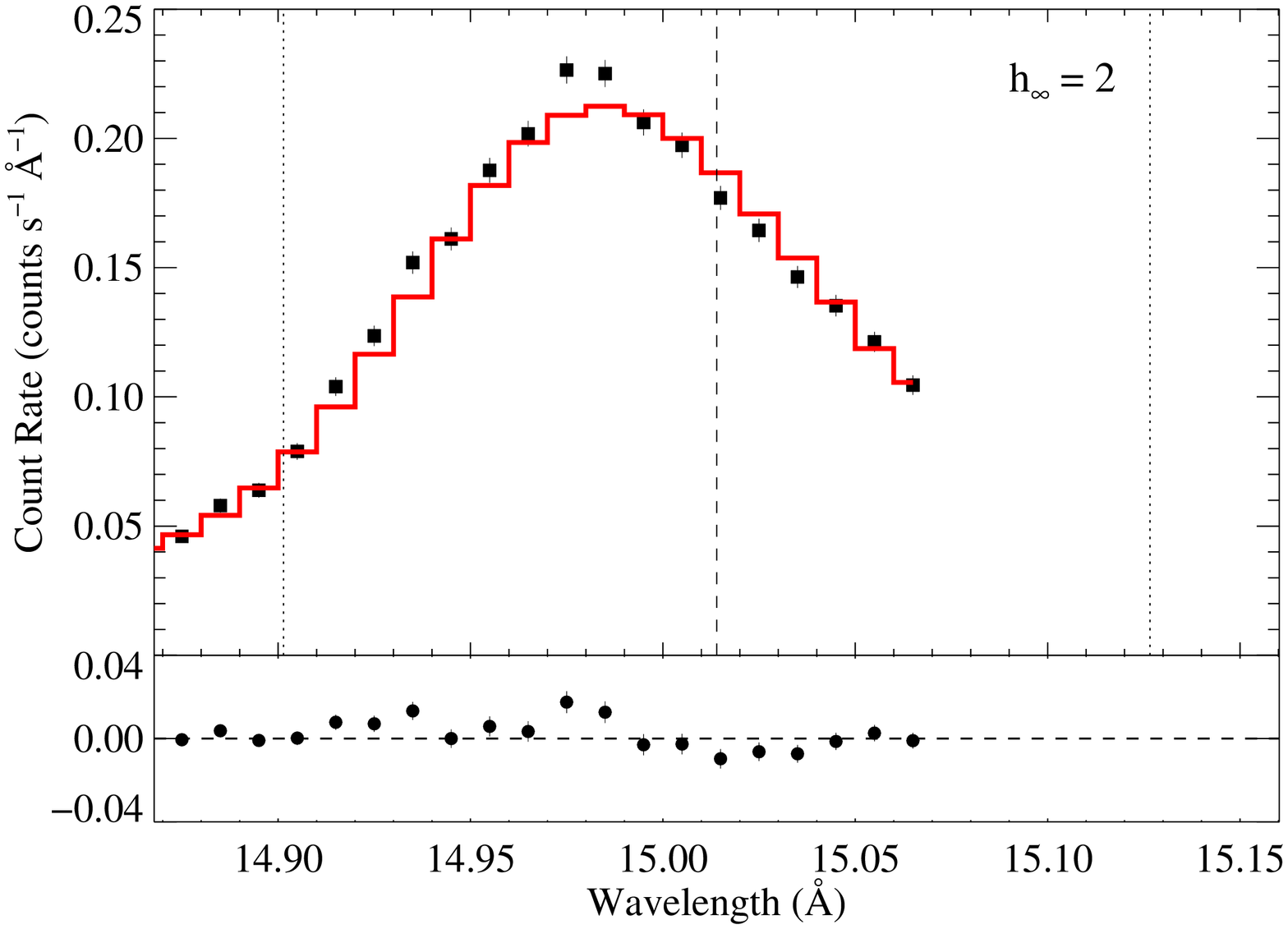} \\
  \includegraphics[angle=90,scale=\profilePlotPanelSize]{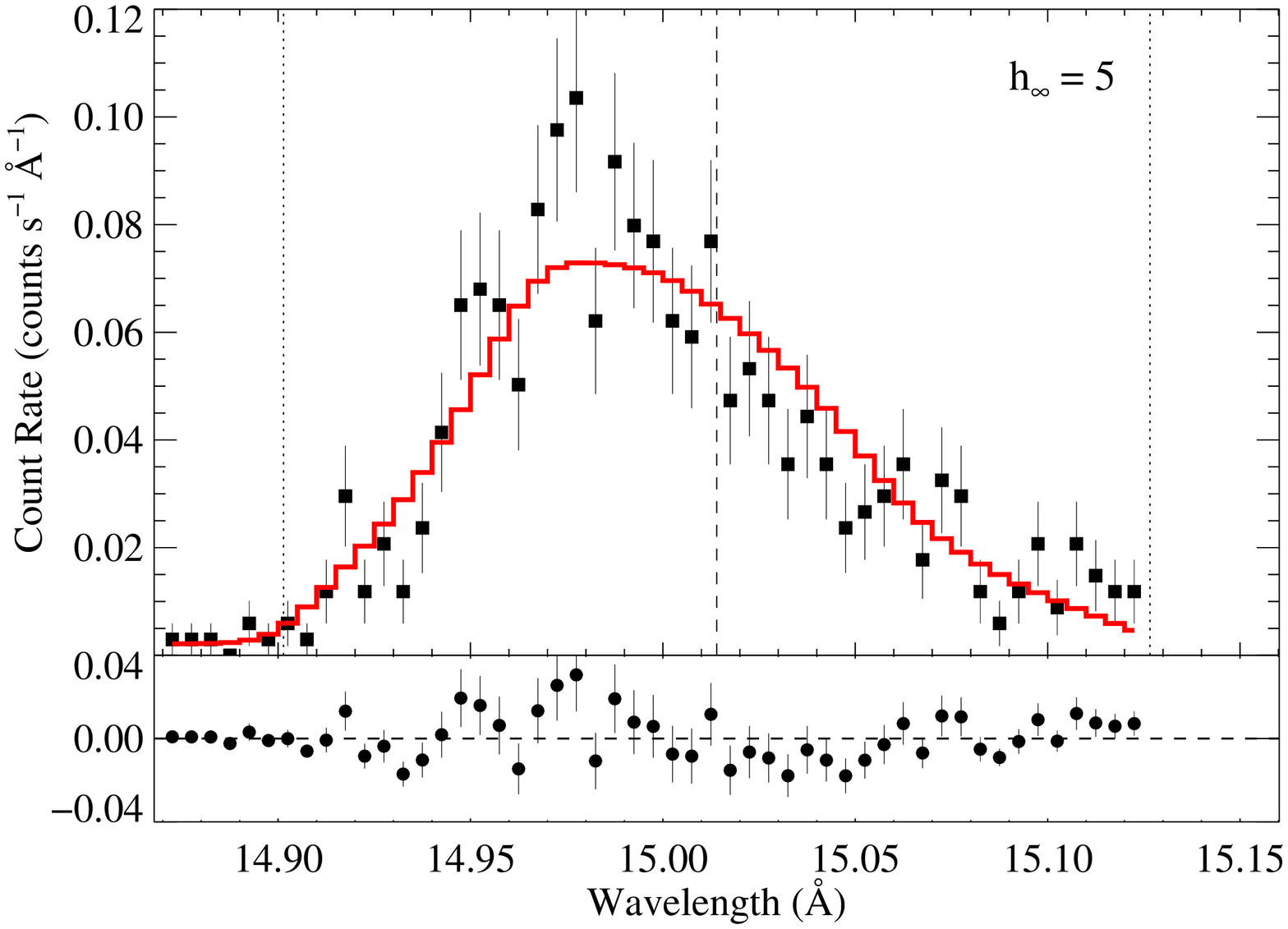} &
  \includegraphics[angle=90,scale=\profilePlotPanelSize]{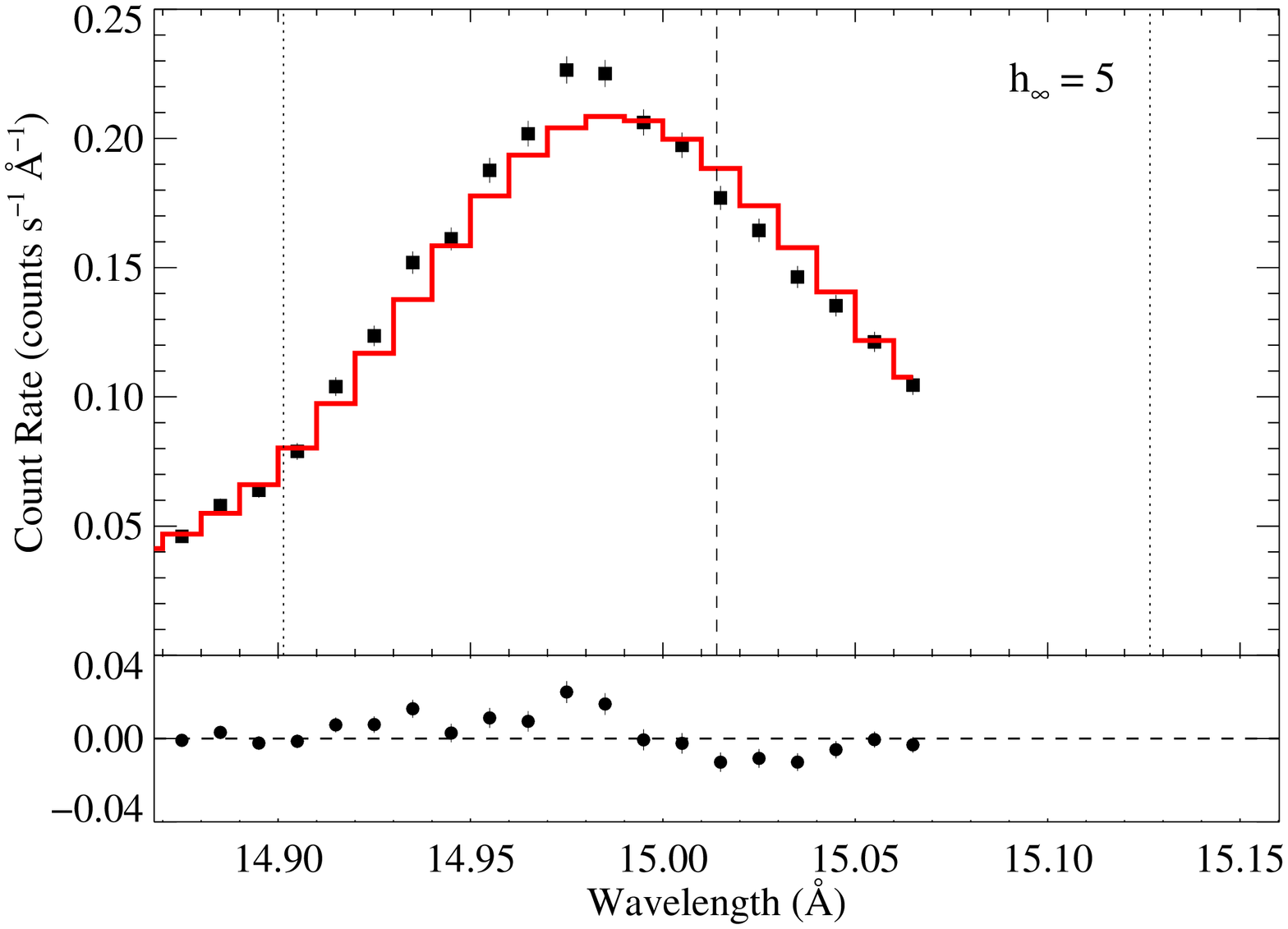} \\
  \end{tabular}
  \caption{The \cxo MEG (left column) and \xmm RGS (right column)
    measurements of the \ion{Fe}{17} line at 15.014 \AA; identical to
    the data shown in Figures~\ref{fig:15.014_nonporous} and
    \ref{fig:15.014_aniso}, but with best-fit isotropic porosity
    models superimposed. (A color version of this figure is available
    in the online journal.)}
  \label{fig:15.014_iso}
\end{figure*}

We then fit models with fixed values of terminal porosity length,
$\hinf = 0.5$, 1, 2, and 5 \rstar. Figure~\ref{fig:15.014_iso}
shows fits of the isotropic porosity models to the \ion{Fe}{17} line
at 15.014 \AA, and Tables~\ref{tab:chandra_iso} and \ref{tab:rgs_iso}
present fits to all the lines. Plots of fits to the other lines in the
spectrum of \zp using isotropic porosity models are shown in the Appendix.

Figure~\ref{fig:15.014_contour} shows confidence limit contours for the
fits to the 15.014 \AA\ line in \taustar--$\hinf$ parameter space,
illustrating the tradeoff between porosity and wind optical
depth. Modest porosity ($\hinf < \rstar$) is not strongly ruled
out, but it does not have a strong effect on \taustar, and thus on
mass-loss rate estimates from X-ray line profiles.

Figure~\ref{fig:15.014_model} shows the best fit models for the
\ion{Fe}{17} 15.014 \AA\, line in the \cxo data without accounting for
the instrument response. This allows evaluation of the level of
degeneracy of \hinf and \taustar in producing roughly comparable
line profiles. Even for a small porosity length of $\hinf = 0.5
\rstar$, the anisotropic porosity model shows a significant change in
shape near line center, with the bump becoming more pronounced for
larger porosity lengths. On the other hand, the isotropic porosity
models with small \hinf do not look very different from the
non-porous model. However, the shape does become increasingly
different as porosity length increases. The changes in profile shape
support the same conclusions as Figure~\ref{fig:15.014_contour}: for
isotropic porosity, small porosity lengths can produce similar line
shapes that adequately fit the data, while large porosity lengths are
ruled out; and for anisotropic porosity, even small porosity lengths
change the profile shape so that it does not fit the data.


\begin{figure*}
  \includegraphics[angle=90,scale=\profilePlotPanelSize]{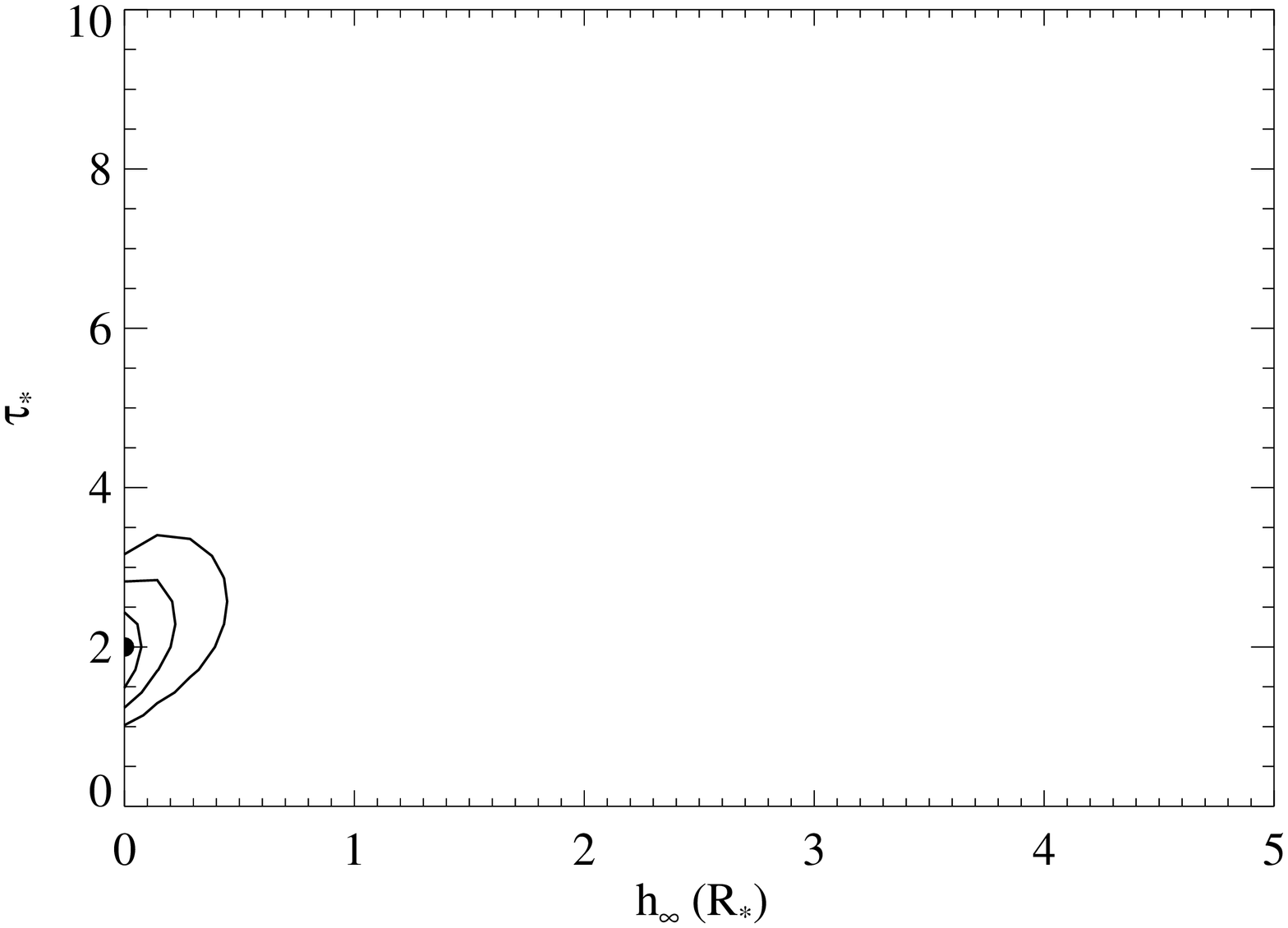}
  \includegraphics[angle=90,scale=\profilePlotPanelSize]{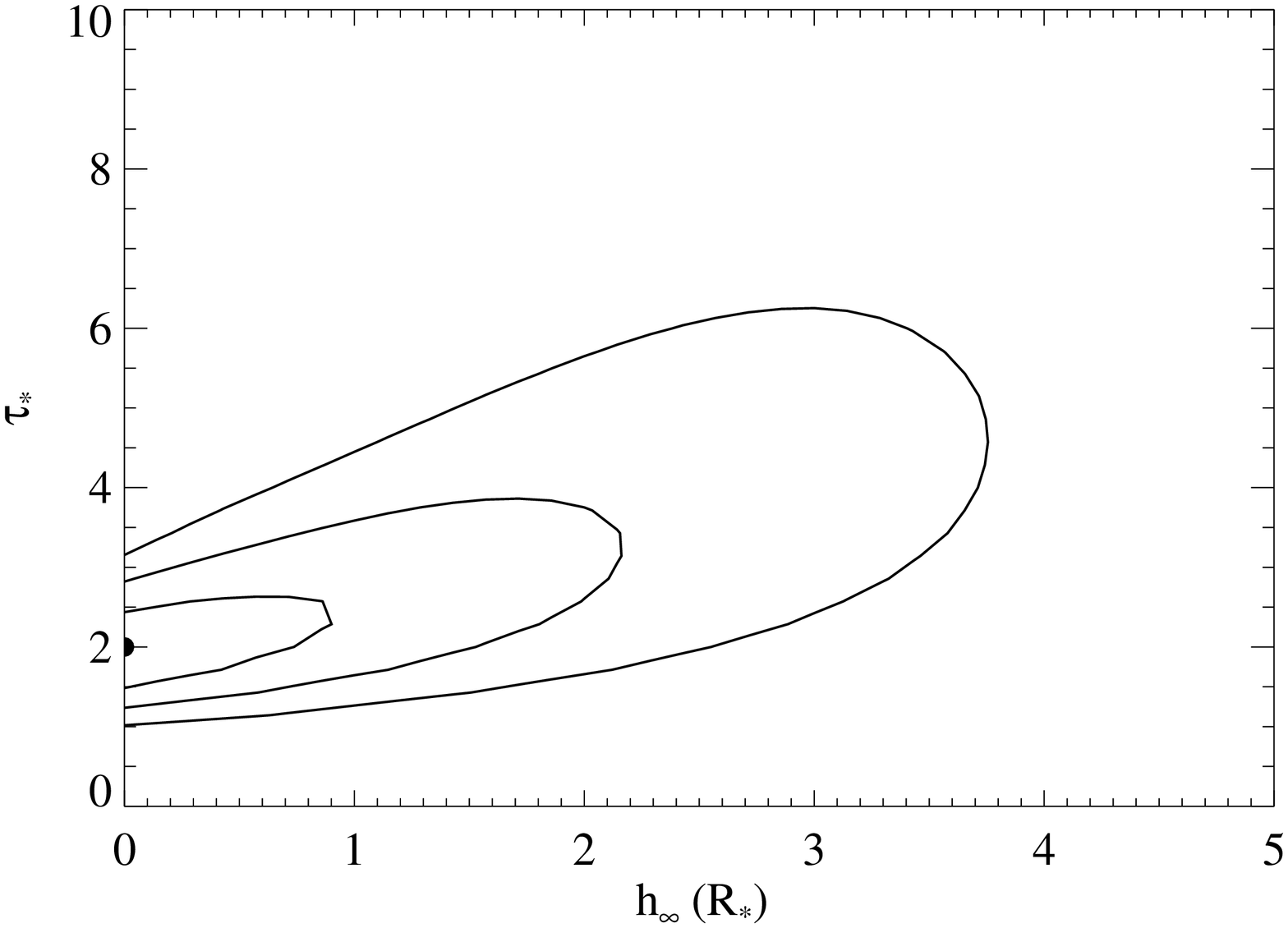}
  \caption{The 68.3\%, 95.4\%, and 99.7\% confidence limit contours in
    \taustar--\hinf parameter space for the anisotropic (left panel)
    and isotropic (right panel) porosity model fits to the
    \ion{Fe}{17} line at 15.014 \AA\ in the \cxo data.  The best fit
    model is denoted by the filled circle. }
  \label{fig:15.014_contour}
\end{figure*}

\begin{figure*}
  \begin{tabular}{cc}
  \includegraphics[angle=0,scale=\profilePlotPanelSize]{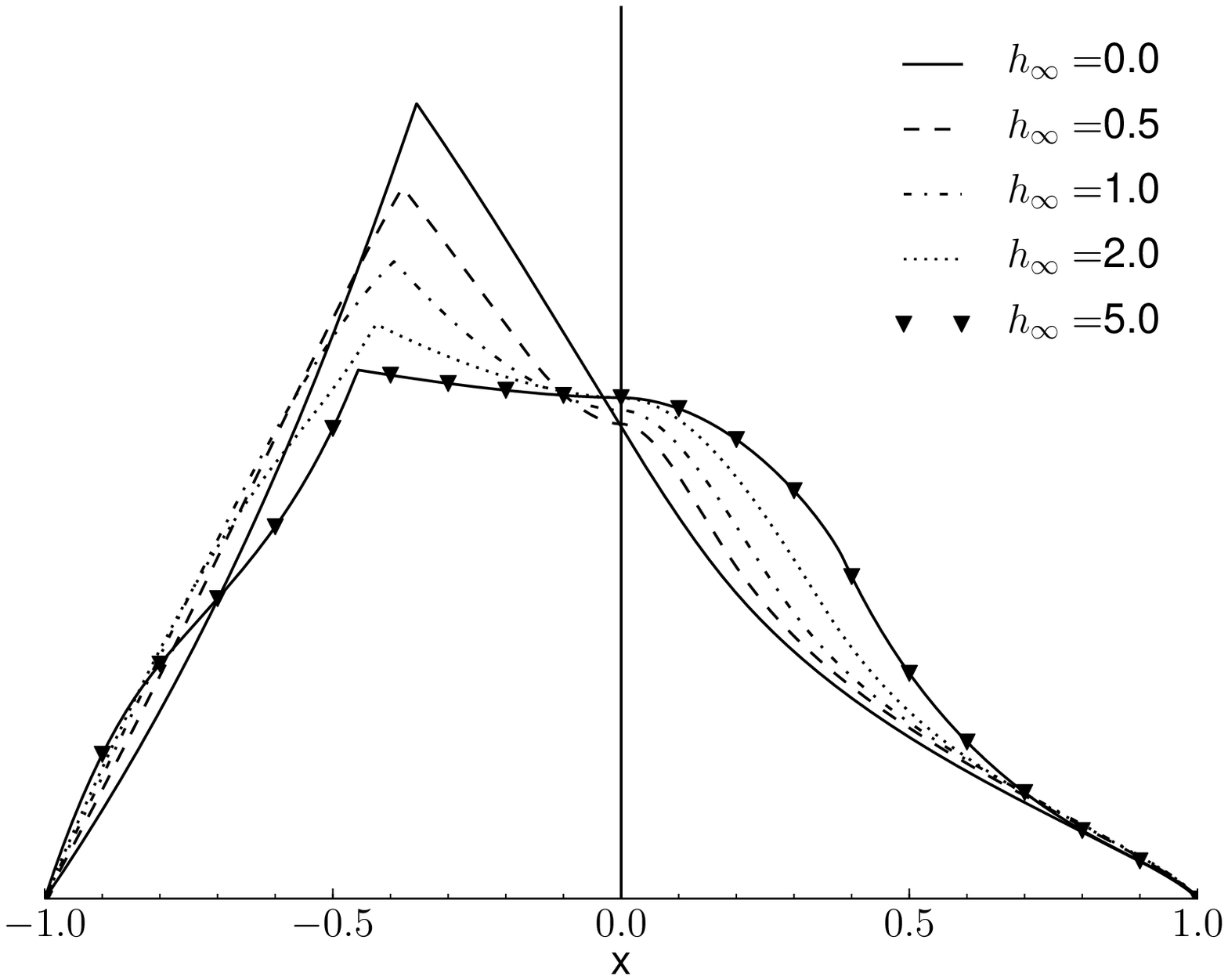} &
  \includegraphics[angle=0,scale=\profilePlotPanelSize]{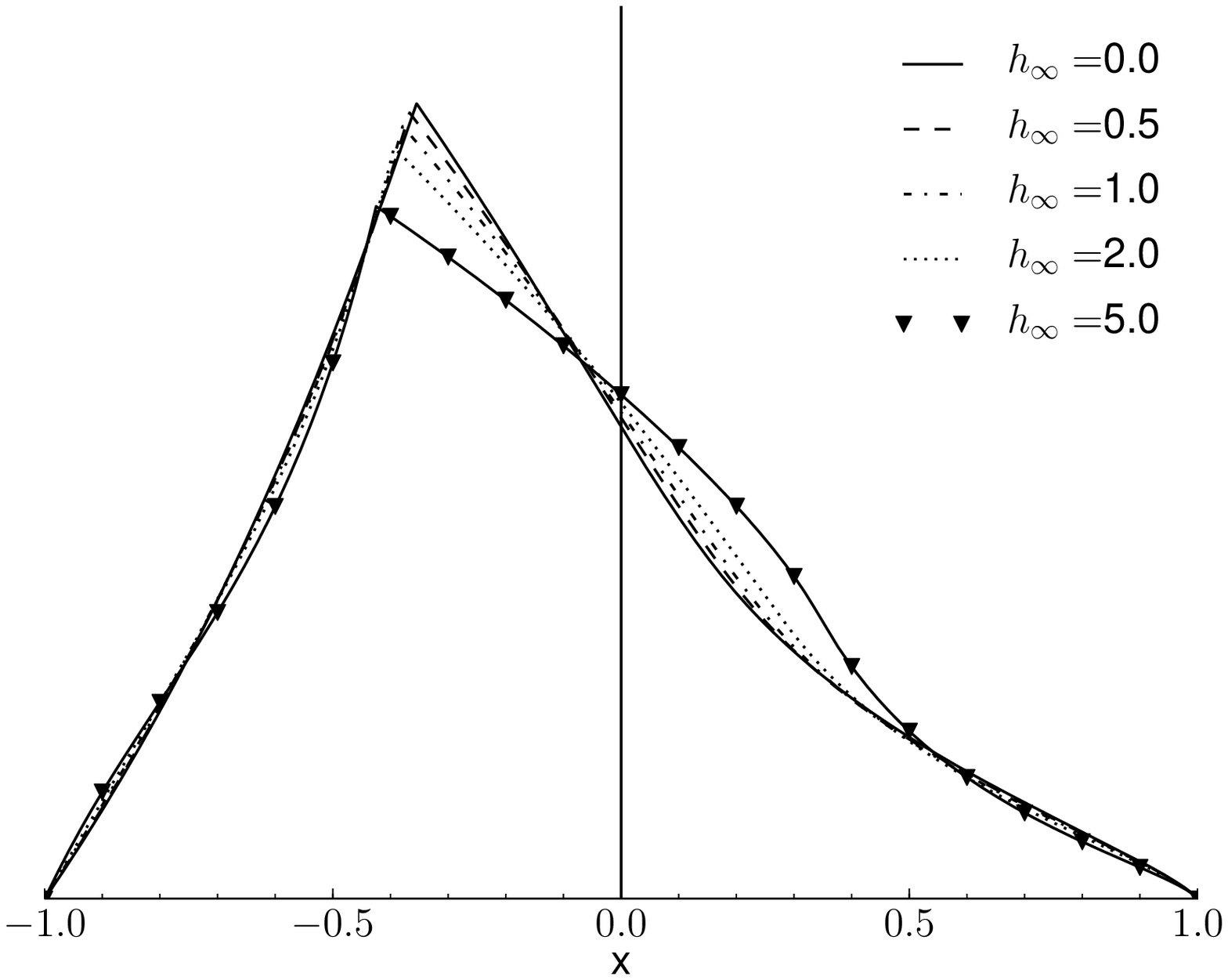} \\
  \end{tabular}
  \caption{Comparison of the best fit models for the \ion{Fe}{17}
    15.014 \AA\, line in the \cxo data. The model profile flux is
    shown as a function of the scaled wavelength $x \equiv (\delta
    \lambda / \lambda_0) (c / \vinf)$.  Anisotropic porosity is shown
    in the left panel, while isotropic is shown in the right
    panel. Note that the best fit anisotropic porosity models have a
    very different shape from the best fit nonporous model because of
    the enhancement near line center from the ``Venetian blind
    effect''.}
  \label{fig:15.014_model}
\end{figure*}

\begin{deluxetable*}{cccccccc}
  \tablecaption{Wind profile model with isotropic porosity: fit results to the \cxo HETGS spectra}
  \tablehead{
    \colhead{Ion} &
    \colhead{$\lambda$} &
    \colhead{$h_\infty$} &
    \colhead{\taustar} &
    \colhead{\ro} &
    \colhead{Normalization} & 
    \colhead{$C$} &
    \colhead{$N_{\rm bins}$} \\
    \colhead{} & 
    \colhead{(\AA)} & 
    \colhead{(\rstar)} &
    \colhead{} & 
    \colhead{(\rstar)} & 
    \colhead{($10^{-4}$ photons cm$^{-2}$ s$^{-1}$)} & 
    \colhead{} & 
    \colhead{} 
  }
  \startdata
  \ion{Mg}{12} \lya & 8.4210 & 0  & $1.22_{-.44}^{+.78}$ & $1.34_{-.21}^{+.17}$ & $0.294_{-.022}^{+.024}$ & 186.5 & 188  \\
  & & 0.5 & $1.31_{-.48}^{+.74}$ & $1.38_{-.18}^{+.14}$ & $0.300_{-.029}^{+.019}$ & 186.8 & 188  \\
  & & 1 & $1.46_{-.58}^{+.93}$ & $1.39_{-.15}^{+.14}$ & $0.300_{-.028}^{+.019}$ & 187.1 & 188  \\
  & & 2 & $1.79_{-.80}^{+1.42}$ & $1.41_{-.12}^{+.14}$ & $0.300_{-.028}^{+.020}$ & 187.7 & 188  \\
  & & 5 & $3.19_{-1.84}^{+4.33}$ & $1.47_{-.10}^{+.12}$ & $0.301_{-.028}^{+.020}$ & 189.7 & 188  \\
  \ion{Ne}{10} \lya & 12.1339 & 0 & $2.01_{-.24}^{+.27}$ & $1.45_{-.08}^{+.13}$ & $2.71_{-.09}^{+.09}$ & 191.4 & 176  \\
  & & 0.5 & $2.39_{-.35}^{+.33}$ & $1.49_{-.08}^{+.10}$ & $2.69_{-.07}^{+.11}$ & 191.8 & 176  \\
  & & 1 & $2.80_{-.44}^{+.44}$ & $1.51_{-.07}^{+.09}$ & $2.69_{-.08}^{+.10}$ & 192.5 & 176  \\
  & & 2 & $3.86_{-.69}^{+.75}$ & $1.54_{-.08}^{+.07}$ & $2.69_{-.08}^{+.10}$ & 195.0 & 176  \\
  & & 5 & $9.24_{-2.15}^{+2.82}$ & $1.61_{-.06}^{+.08}$ & $2.69_{-.09}^{+.09}$ & 208.5 & 176  \\
  \ion{Fe}{17} & 15.014 & 0 & $1.94_{-.33}^{+.32}$ & $1.55_{-.12}^{+.12}$ & $5.24_{-.17}^{+.24}$ & 280.8 & 308  \\
   &  & 0.5 & $2.18_{-.39}^{+.42}$ & $1.58_{-.10}^{+.11}$ & $5.24_{-.17}^{+.24}$ & 282.0 & 308  \\
   &  & 1 & $2.46_{-.47}^{+.54}$ & $1.61_{-.13}^{+.08}$ & $5.24_{-.17}^{+.24}$ & 283.4 & 308  \\
   &  & 2 & $3.09_{-.66}^{+.89}$ & $1.65_{-.09}^{+.08}$ & $5.24_{-.18}^{+.24}$ & 286.4 & 308  \\
   &  & 5 & $5.70_{-1.66}^{+2.75}$ & $1.74_{-.09}^{+.07}$ & $5.23_{-.18}^{+.23}$ & 297.2 & 308  \\
  \ion{Fe}{17} & 16.780 & 0 & $3.01_{-.70}^{+.32}$ & $1.01_{-.01}^{+.59}$ & $2.45_{-.17}^{+.13}$ & 174.9 & 308  \\
 & & 0.5 & $3.59_{-.87}^{+.93}$ & $1.35_{-.31}^{+.28}$ & $2.39_{-.11}^{+.19}$ & 174.1 & 308  \\
 & & 1 & $4.38_{-1.06}^{+1.11}$ & $1.42_{-.17}^{+.21}$ & $2.43_{-.15}^{+.15}$ & 173.3 & 308  \\
 & & 2 & $6.52_{-1.75}^{+1.99}$ & $1.48_{-.11}^{+.17}$ & $2.40_{-.12}^{+.19}$ & 172.6 & 308  \\
 & & 5 & $17.88_{-6.00}^{+6.52}$ & $1.60_{-.11}^{+.14}$ & $2.43_{-.14}^{+.17}$ & 176.3 & 308  \\
  \ion{O}{8} \lya & 18.969 & 0 & $3.00_{-.54}^{+.54}$ & $1.22_{-.21}^{+.37}$ & $3.70_{-.35}^{+.29}$ & 150.9 & 130  \\
  & & 0.5 & $3.85_{-1.10}^{+1.05}$ & $1.32_{-.20}^{+.38}$ & $3.70_{-.28}^{+.29}$ & 150.9 & 130  \\
  & & 1 & $4.42_{-1.47}^{+1.74}$ & $1.45_{-.23}^{+.34}$ & $3.71_{-.29}^{+.28}$ & 151.0 & 130  \\
  & & 2 & $5.85_{-2.19}^{+2.88}$ & $1.56_{-.19}^{+.28}$ & $3.62_{-.19}^{+.37}$ & 151.1 & 130  \\
  & & 5 & $13.23_{-5.80}^{+8.74}$ & $1.69_{-.15}^{+.21}$ & $3.60_{-.17}^{+.39}$ & 152.1 & 130  \\
  \enddata
  \label{tab:chandra_iso}
\end{deluxetable*}

\begin{deluxetable*}{cccccccc}
  \tablecaption{Wind profile model with isotropic porosity: fit results to the \xmm RGS spectra}
  \tablehead{
    \colhead{Ion} & 
    \colhead{$\lambda$} &
    \colhead{$h_\infty$} & 
    \colhead{\taustar} & 
    \colhead{\ro} & 
    \colhead{Normalization} & 
    \colhead{$\chi^2$} & 
    \colhead{$N_{\rm bins}$} \\
    \colhead{} & 
    \colhead{(\AA)} & 
    \colhead {(\rstar)} & 
    \colhead{} & 
    \colhead{(\rstar)} & 
    \colhead{($10^{-4}$ photons cm$^{-2}$ s$^{-1}$)} & 
    \colhead{} & 
    \colhead{} 
  }
  \startdata
  \ion{Ne}{10} \lya & 12.1339 & 0 & $1.81_{-.22}^{+.25}$ & $1.61_{-.18}^{+.15}$ & $3.10_{-.08}^{+.08}$ & 24.0 & 19  \\
  & & 0.5 & $2.07_{-.27}^{+.34}$ & $1.61_{-.16}^{+.15}$ & $3.10_{-.08}^{+.08}$ & 24.0 & 19  \\
  & & 1 & $2.41_{-.34}^{+.45}$ & $1.61_{-.15}^{+.15}$ & $3.10_{-.08}^{+.09}$ & 24.1 & 19  \\
  & & 2 & $3.42_{-.61}^{+.87}$ & $1.60_{-.13}^{+.14}$ & $3.11_{-.09}^{+.08}$ & 24.2 & 19  \\
  & & 5 & $12.99_{-4.14}^{+6.62}$ & $1.58_{-.12}^{+.13}$ & $3.16_{-.08}^{+.08}$ & 26.3 & 19  \\

  \ion{Fe}{17} & 15.014 & 0 & $1.77$ & $1.57$ & $6.39$ & 129.5 & 52  \\
  & & 0.5 & $2.02$ & $1.58$ & $6.40$ & 131.4 & 52  \\
  & & 1 & $2.33$ & $1.59$ & $6.40$ & 133.6 & 52  \\
  & & 2 & $3.23$ & $1.60$ & $6.40$ & 139.3 & 52  \\
  & & 5 & $9.96$ & $1.61$ & $6.44$ & 175.6 & 52  \\

   \ion{Fe}{17} & 16.780 & 0 & $3.38_{-.45}^{+.31}$ & $1.54_{-.39}^{+.33}$ & $3.01_{-.06}^{+.07}$ & 30.3 & 32  \\
   & & 0.5 & $3.94_{-.63}^{+.56}$ & $1.65_{-.23}^{+.26}$ & $3.01_{-.06}^{+.07}$ & 31.1 & 32  \\
   & & 1 & $4.50_{-.83}^{+.89}$ & $1.73_{-.21}^{+.23}$ & $3.03_{-.07}^{+.06}$ & 32.2 & 32  \\
   & & 2 & $5.61_{-1.17}^{+1.53}$ & $1.86_{-.18}^{+.20}$ & $3.06_{-.07}^{+.07}$ & 34.4 & 32  \\
   & & 5 & $8.61_{-2.76}^{+4.30}$ & $2.13_{-.18}^{+.19}$ & $3.19_{-.08}^{+.07}$ & 40.4 & 32  \\

  \ion{O}{8} \lya & 18.969 & 0 & $3.14$ & $1.01$ & $4.66$ & 204.3 & 72  \\
  &  & 0.5 & $3.64$ & $1.57$ & $4.64$ & 198.0 & 72  \\
  &  & 1 & $4.35$ & $1.51$ & $4.64$ & 193.4 & 72  \\
  &  & 2 & $6.38$ & $1.57$ & $4.63$ & 188.9 & 72  \\
  &  & 5 & $20.20$ & $1.65$ & $4.67$ & 222.3 & 72  \\

  \ion{N}{7} \lyb & 20.910 & 0 & $4.93_{-1.03}^{+.66}$ & $1.41_{-.40}^{+.62}$ & $1.66_{-.10}^{+.07}$ & 40.5 & 36  \\
  & & 0.5 & $6.18_{-1.46}^{+1.89}$ & $1.65_{-.51}^{+.46}$ & $1.66_{-.10}^{+.10}$ & 40.6 & 36  \\
  & & 1 & $7.42_{-1.91}^{+2.67}$ & $1.79_{-.36}^{+.36}$ & $1.66_{-.10}^{+.10}$ & 41.0 & 36  \\
  & & 2 & $10.16_{-2.76}^{+4.26}$ & $1.95_{-.30}^{+.33}$ & $1.66_{-.10}^{+.10}$ & 42.4 & 36  \\
  & & 5 & $26.09_{-7.96}^{+12.44}$ & $2.14_{-.29}^{+.38}$ & $1.69_{-.09}^{+.10}$ & 49.0 & 36  \\
  
  \enddata
  \label{tab:rgs_iso}
\end{deluxetable*}

\subsection{Effects of porosity on inferred mass-loss rates}

In Section~\ref{sec:individual_porosity} we have shown for several
lines in the spectrum of \zp\ that anisotropic porosity models do not
fit the data for any significant porosity length, and that isotropic
porosity models do not fit for large porosity lengths. However,
isotropic porosity models with moderate porosity lengths are formally
allowed for some individual lines.

\citet{Cohen2010} have fit the ensemble of measured \taustar values
for a non-porous model of \zp by using the mass-loss rate as an
adjustable derived parameter, together with a model wind opacity. We
can apply the same technique to the \taustar values measured for
different porosity lengths to obtain an estimate of X-ray derived
mass-loss rate as a function of assumed porosity length. Together with
constraints on the porosity length and geometry, whether from X-ray
profile modeling or from other observations, this relation gives a
quantitative constraint on the contribution of porosity to
uncertainties in mass-loss rates derived from fitting X-ray line profiles.

To this end, it is most instructive to consider not the mass-loss rate
itself, but the fractional change in mass-loss rate relative to a
non-porous model. The derived mass-loss rate scales linearly with
measured \taustar, and thus the fractional change in derived mass-loss
rate must equal the fractional change in measured \taustar. Thus, as a
simple way to estimate the fractional change in mass-loss rate, we
calculate the fractional change in measured \taustar for the ensemble
of lines, for a given assumed porosity length. The estimated mass-loss
rate increase for isotropic porosity models is roughly linear in
\hinf, with $\Delta \dot{M} / \dot{M} \sim 0.4$ for $\hinf =
\rstar$. Large porosity lengths indeed produce large changes in
inferred mass-loss rate compared to non-porous models, and anisotropic
porosity models have a stronger effect than isotropic models. However,
all of the models with strong effects are ruled out by our fit
results.


\section{Comparison with previous results}
\label{sec:comparison}

One of the main results of Section~\ref{sec:results} is that anisotropic
porosity models are disfavored, as well as isotropic porosity models
with large porosity lengths. This is at odds with the conclusions of
\citet{OFH06}, who report that anisotropic porosity models provide
good fits to the line profiles of several bright O stars observed
with \cxo, while non-porous models do not fit their data as well.

We show below that the difference in our findings has its root in a
less obvious assumption of both our models and those of Oskinova et
al.: the radial upper bound for X-ray emission, \rmax. Oskinova et
al. assume $\rmax = 5 \star$ for \zp, and as high as $\rmax = 9
\rstar$ for other stars in their sample. On the other hand, in this
paper we have until now assumed $\rmax = \infty$.

\subsection{The effect of the radial upper bound}

\begin{figure}[ht]
  \includegraphics[angle=0, width=3.0in]{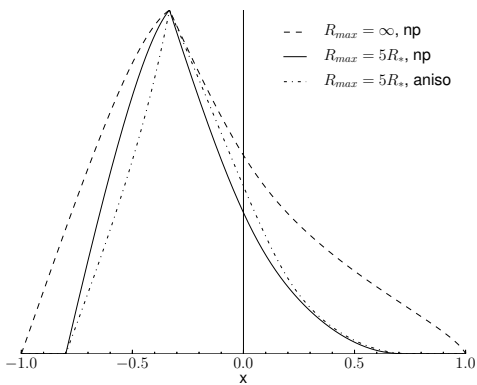}
  \caption{Comparison of three model profiles: a model with no
    porosity and no upper radial cutoff; a model with no porosity and
    upper radial cutoff of 5 \rstar; and a model with anisotropic
    porosity, $\hinf = \rstar$, and an upper radial cutoff of 5
    \rstar.  All of the models assume $\taustar = 3$. Note that both of
    the models with a cutoff have corresponding cutoffs to X-ray
    emission in the wings of the profile.
\label{fig:rmaxmodel}}
\end{figure}

In Figure~\ref{fig:rmaxmodel} we show a comparison of models
illustrating the effect of a finite upper radial cutoff to the X-ray
emission in line profile models. For a given value of \taustar, a
model with a finite radial cutoff can produce a much more asymmetric
line profile. Of course, porosity can mitigate this to some
extent. However, note that the models with a finite radial cutoff to
X-ray emission have no flux at all in the wings of the profile, since
there is no X-ray emitting plasma far out in the wind, where the
highest velocities are. This important difference applies to porous
models just as much as non-porous models.


\subsection{Comparison with observed line profiles}

To attempt to reconcile our results with those of \citet{OFH06}, we
have performed fits to the \ion{O}{8} line in the RGS spectrum of \zp
assuming the same upper and lower radial bounds to X-ray emission that
they used: $\ro = 1.5 \rstar$ and $\rmax = 5 \rstar$. We have also fit
the same model to the \cxo spectra and get comparable results to the
RGS fits, but with lower statistical significance, and we therefore
focus on the RGS results in this section.

\begin{figure}[ht]
  \includegraphics[angle=90, scale=\profilePlotPanelSize]{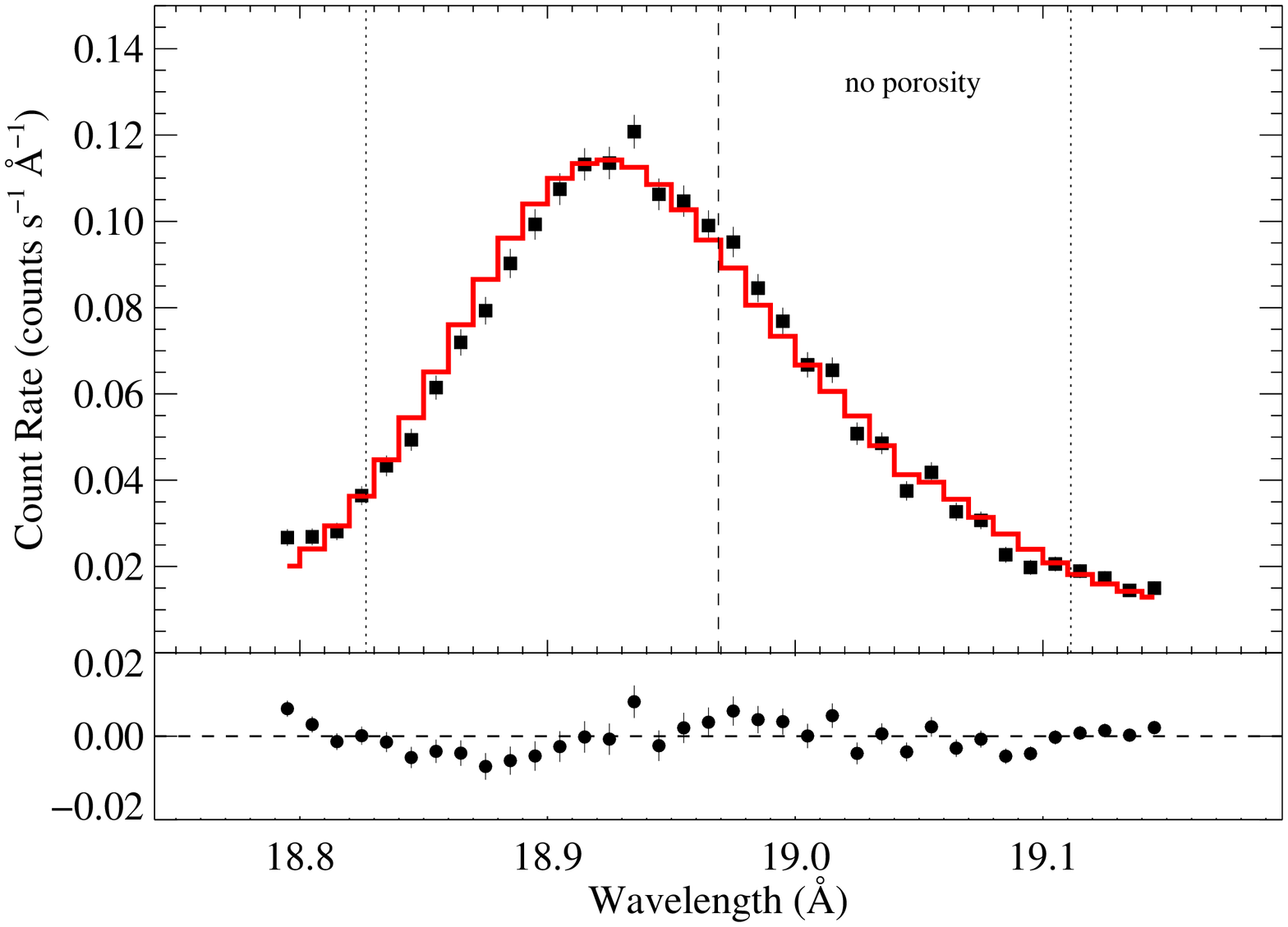}
  \includegraphics[angle=90, scale=\profilePlotPanelSize]{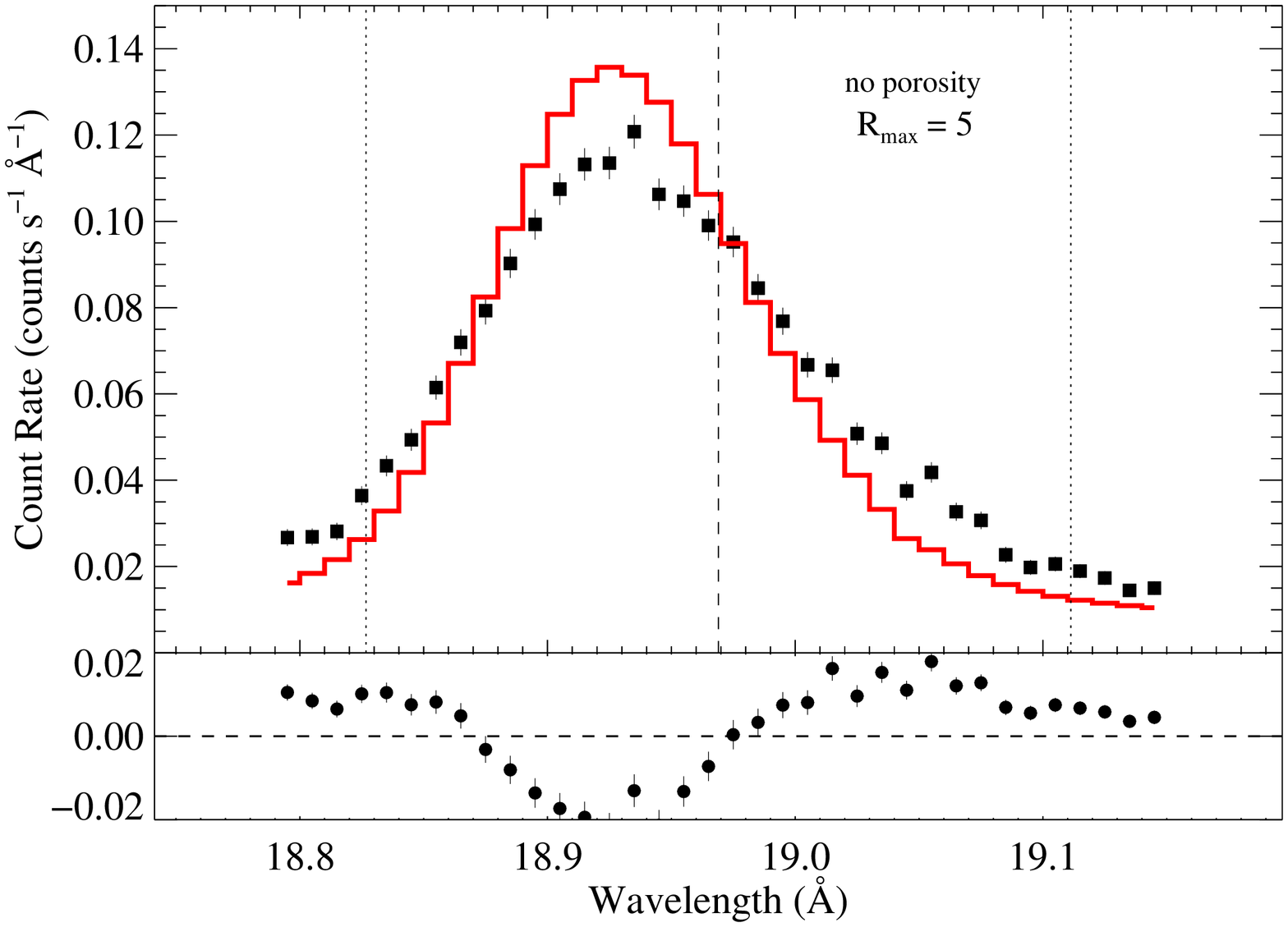}
  \includegraphics[angle=90, scale=\profilePlotPanelSize]{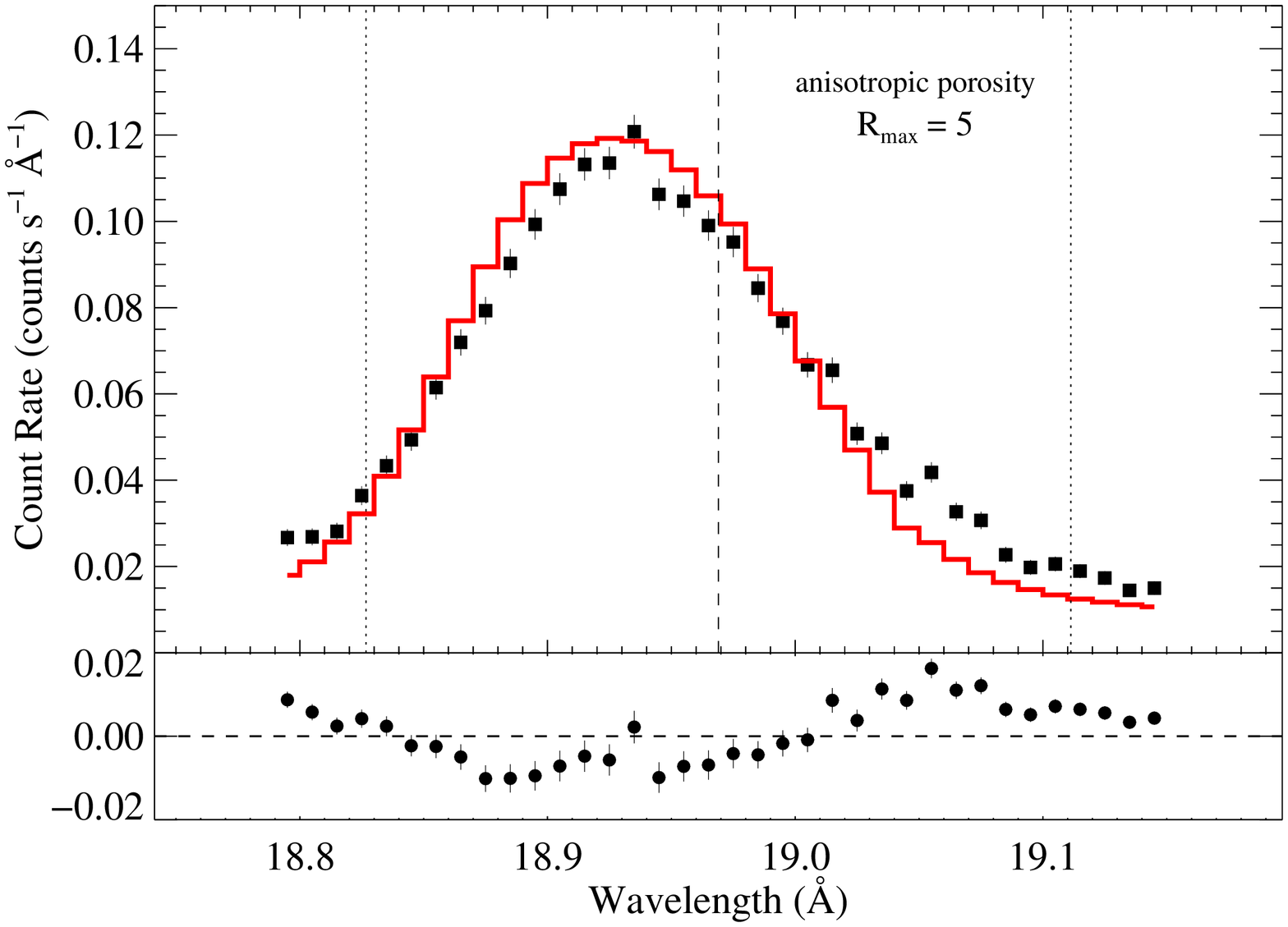}
  \caption{Three model fits to the \ion{O}{8} Ly$\alpha$ line in the
    \xmm RGS spectrum of \zp. RGS2 data are shown, but the fits are
    performed simultaneously to data from both RGS. The first panel
    shows the nonporous fit reported in Section~\ref{sec:results}. The
    second panel shows a non-porous model with $\ro = 1.5 \rstar$
    and $\rmax = 5 \rstar$. The third panel shows a model with
    anisotropic clumps, and with the same radial cutoffs. The last two
    panels correspond approximately to the models shown in Figure 8 of
    \citet{OFH06}. (A color version of this figure is available in the
    online journal.)
\label{fig:rmaxfits}}
\end{figure}

We fit a non-porous model as well as models with both isotropic and
anisotropic porosity. The results of our fits are given in
Table~\ref{tab:rmaxfits} and the best fit models are shown in
Figure~\ref{fig:rmaxfits}. Because the fits are formally unacceptable,
we do not report model parameter confidence intervals.  Note that the
model with isotropic clumps preferred no porosity effect in the fit,
so we do not list it separately.

The results of our fits assuming finite \rmax are qualitatively
similar to Oskinova et al. We find that under these assumptions, no
value of \taustar can provide a good fit to the data in the absence of
porosity, while on the other hand we find a much improved fit from an
anisotropic porosity model. Nevertheless, the anisotropic porosity
model with finite upper radial cutoff does not provide nearly as good
a fit as the nonporous model of Table~\ref{tab:rgs_nonporous} with no
upper radial cutoff, which is also shown in the first panel of
Figure~\ref{fig:rmaxfits} for comparison. The poor fit of both models
with $\rmax = 5 \rstar$ to the data is due in large part to the lack of
flux in the wings of the profile, especially the red wing. This lack
of flux in the wings is a necessary consequence of a finite radial
upper cutoff.

\begin{deluxetable}{cccccc}
  \tablecaption{Effect of radial bounds of $\ro = 1.5 R_*$
and $\rmax = 5 R_*$ on fit to \ion{O}{8} Ly$\alpha$ in \xmm RGS spectrum}
  \tablehead{
    \colhead{$h_\infty$} &
    \colhead{\taustar} & 
    \colhead{Type} & 
    \colhead{Normalization} & 
    \colhead{$\chi^2$} & 
    \colhead{$N_{\rm bins}$} \\
    \colhead{(\rstar)} & 
    \colhead{} & 
    \colhead{} &
    \colhead{($10^{-4}$ photons cm$^{-2}$ s$^{-1}$)} & 
    \colhead{} & 
    \colhead{} 
  }
  \startdata 
  0 & 2.34 & np, iso & 4.404 & 1421.0 & 72 \\
  1.03 & 18.46 & aniso & 4.506 & 769.6 & 72 \\
  \enddata
\label{tab:rmaxfits}
\end{deluxetable}

\subsection{Constraints on the radial upper bound from forbidden line strength}

An obvious way to obtain a strong constraint on the presence of hot
plasma at large radii is by modeling the weak forbidden lines of the
He-like triplets, which are sensitive to the radial distribution of
X-ray emitting plasma \citep{Kel01}. This sensitivity comes about
because the upper level of the forbidden line is metastable, and can
be depopulated by photoexcitation from the star's intense UV radiation
field, the strength of which is radially dependent due to geometrical
dilution.

The forbidden-to-intercombination line ratio is predicted for
different He-like ions as a function of radius in Figure 3 of
\citet{LPKC06}. It is evident from this plot that the forbidden lines
of \ion{N}{6} and \ion{O}{7} are strongly suppressed for $r < 20
\rstar$, which already suggests that an upper radial cutoff of $\rmax
= 5 \rstar$ is likely not compatible with the detection of weak
forbidden lines.  However, as shown in \citet{LPKC06}, it is
straightforward to include the radial dependence of the $f/i$ ratio in
the integrand of the line profile calculation in
Equation~\ref{eq:windintegral} and test the effects of an upper radial
cutoff against the measured He-like complex data.

\begin{figure}[ht]
  \includegraphics[angle=90, scale=\profilePlotPanelSize]{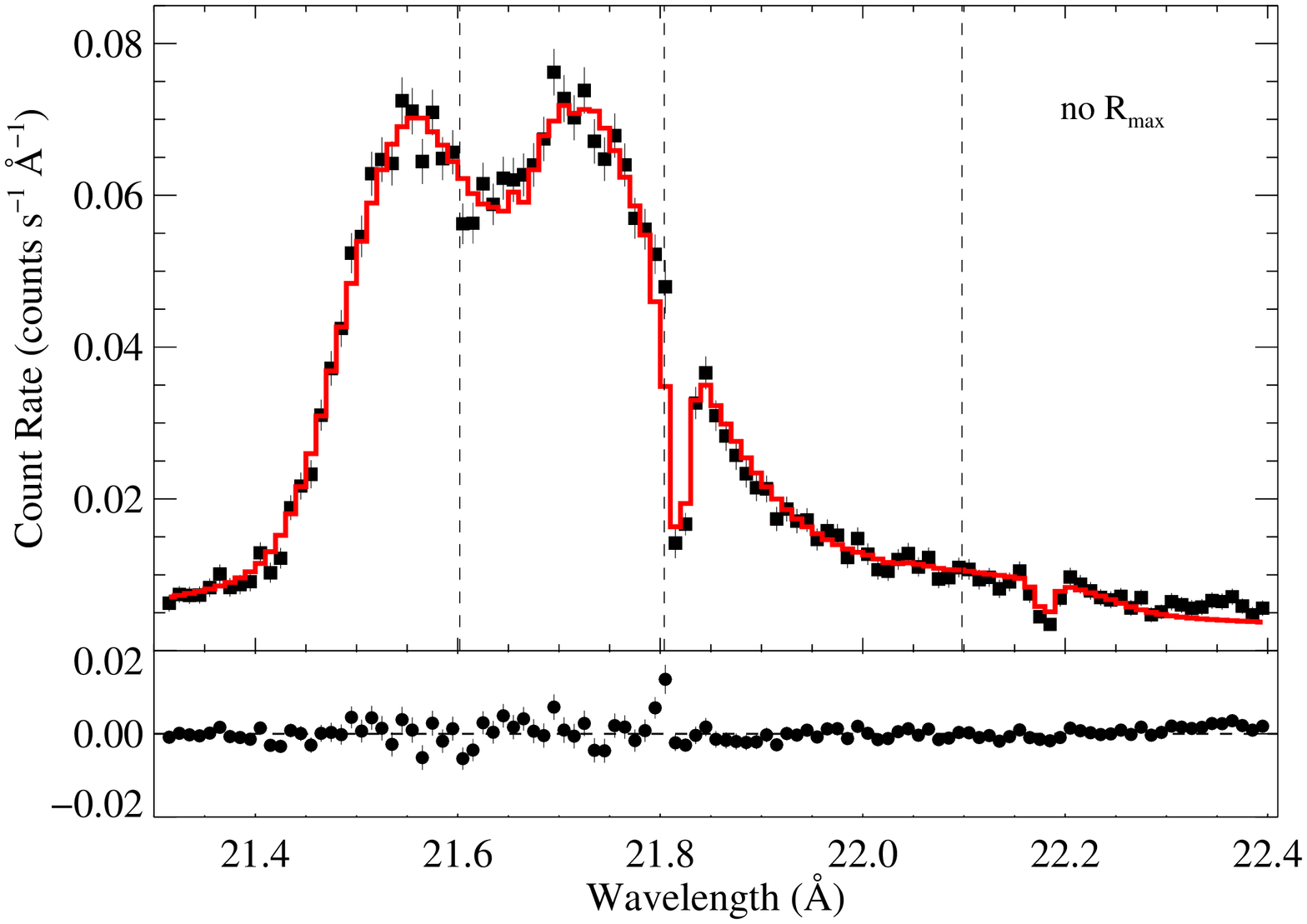}
  \includegraphics[angle=90, scale=\profilePlotPanelSize]{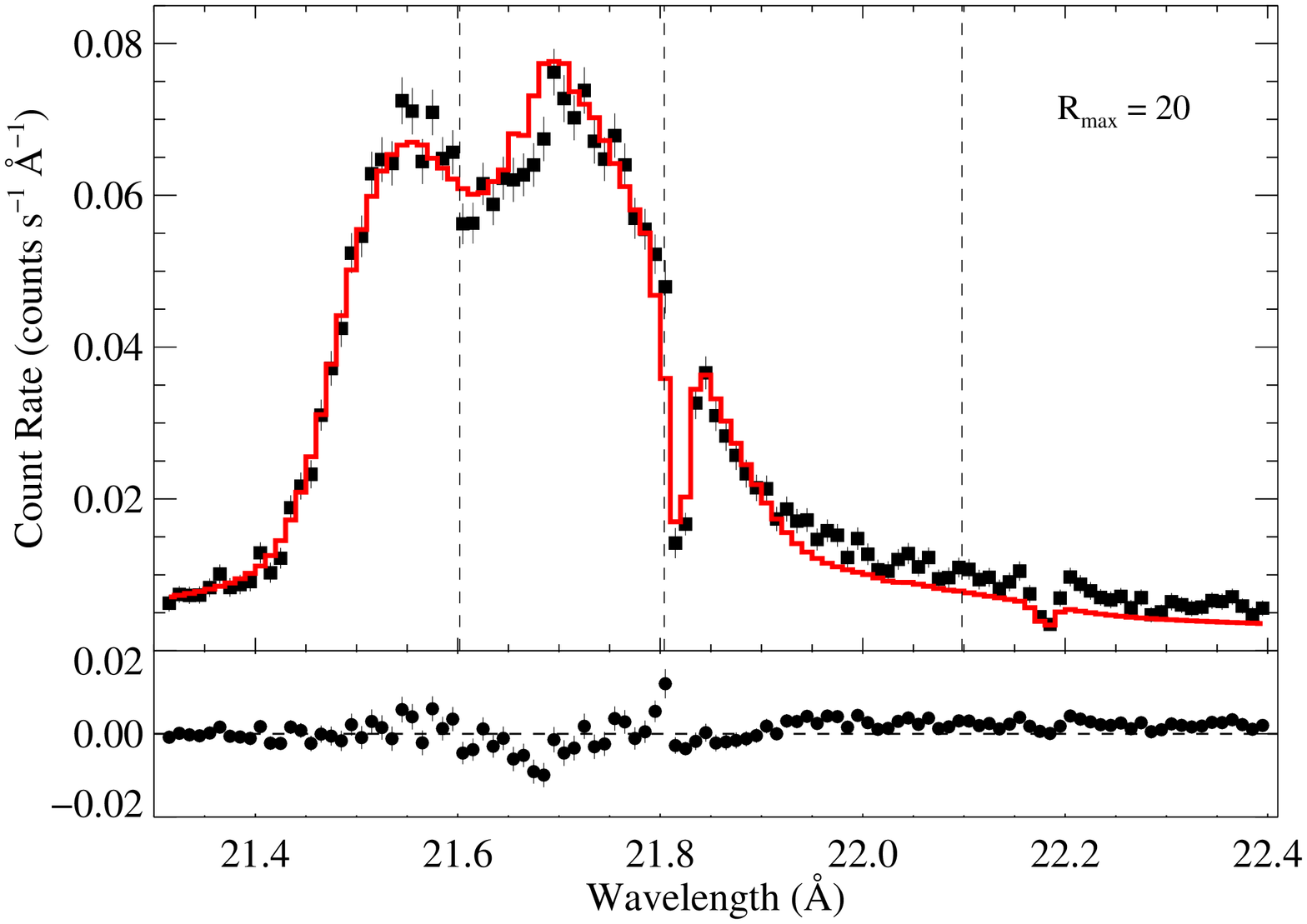}
  \includegraphics[angle=90, scale=\profilePlotPanelSize]{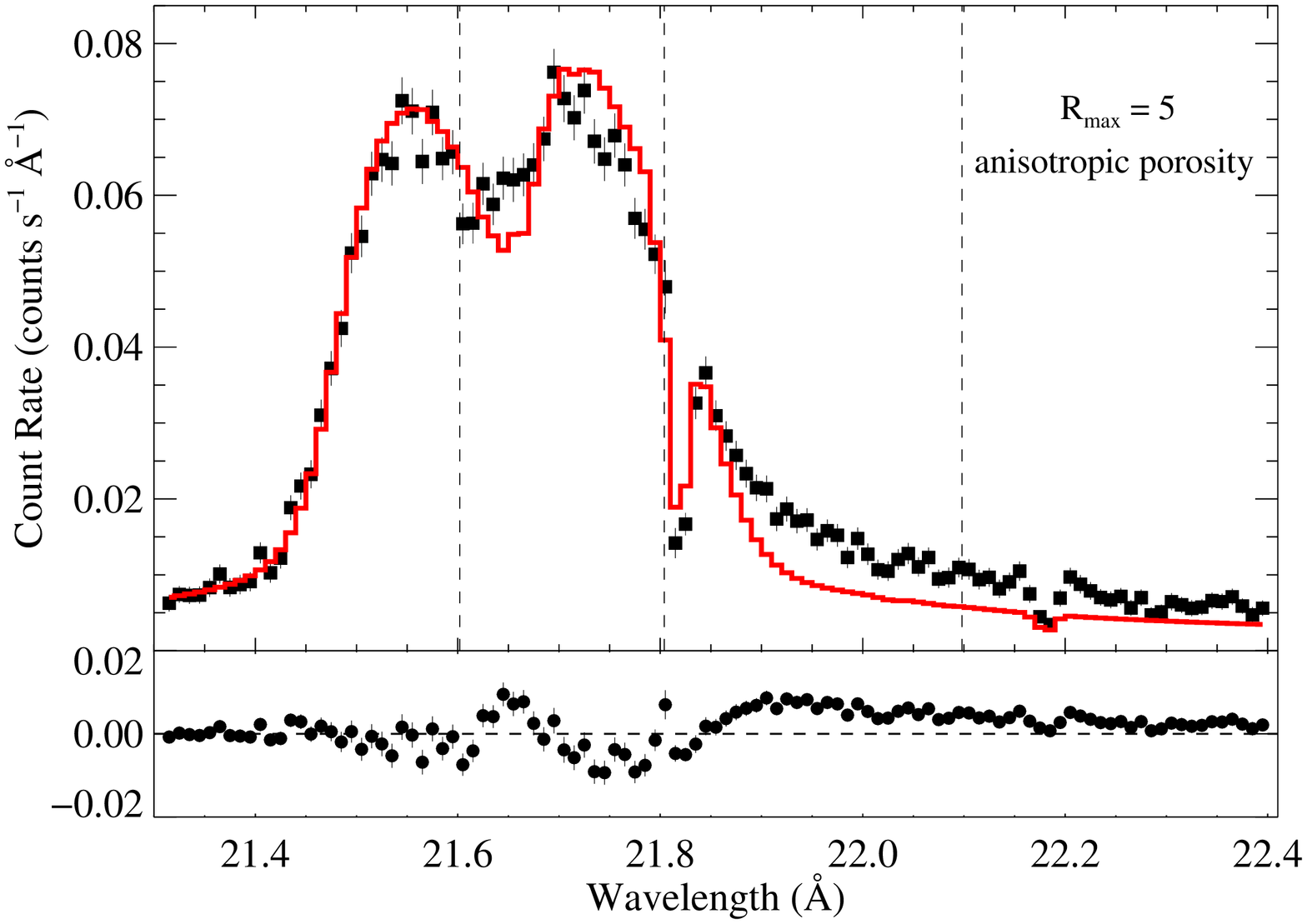}
  \caption{\xmm RGS spectrum of the \ion{O}{7} complex of \zp,
    together with best fit models assuming $\rmax = \infty$ (first
    panel), $\rmax = 20 \rstar$ (second panel), and $\rmax = 5 \rstar$
    (third panel). The three vertical dashed lines indicate the
    positions of the rest wavelengths of the resonance,
    intercombination, and forbidden lines, from left to right. The
    model shown in third panel includes the effects of anisotropic
    porosity, and is comparable to the model favored in
    \citet{OFH06}. Both models with finite radial upper cutoffs
    underpredict emission from the forbidden line, which is primarily
    formed at large radii. (A color version of this figure is
    available in the online journal.) \label{fig:helikeOVII}}
\end{figure}

\begin{figure}[ht]
  \includegraphics[angle=90, scale=\profilePlotPanelSize]{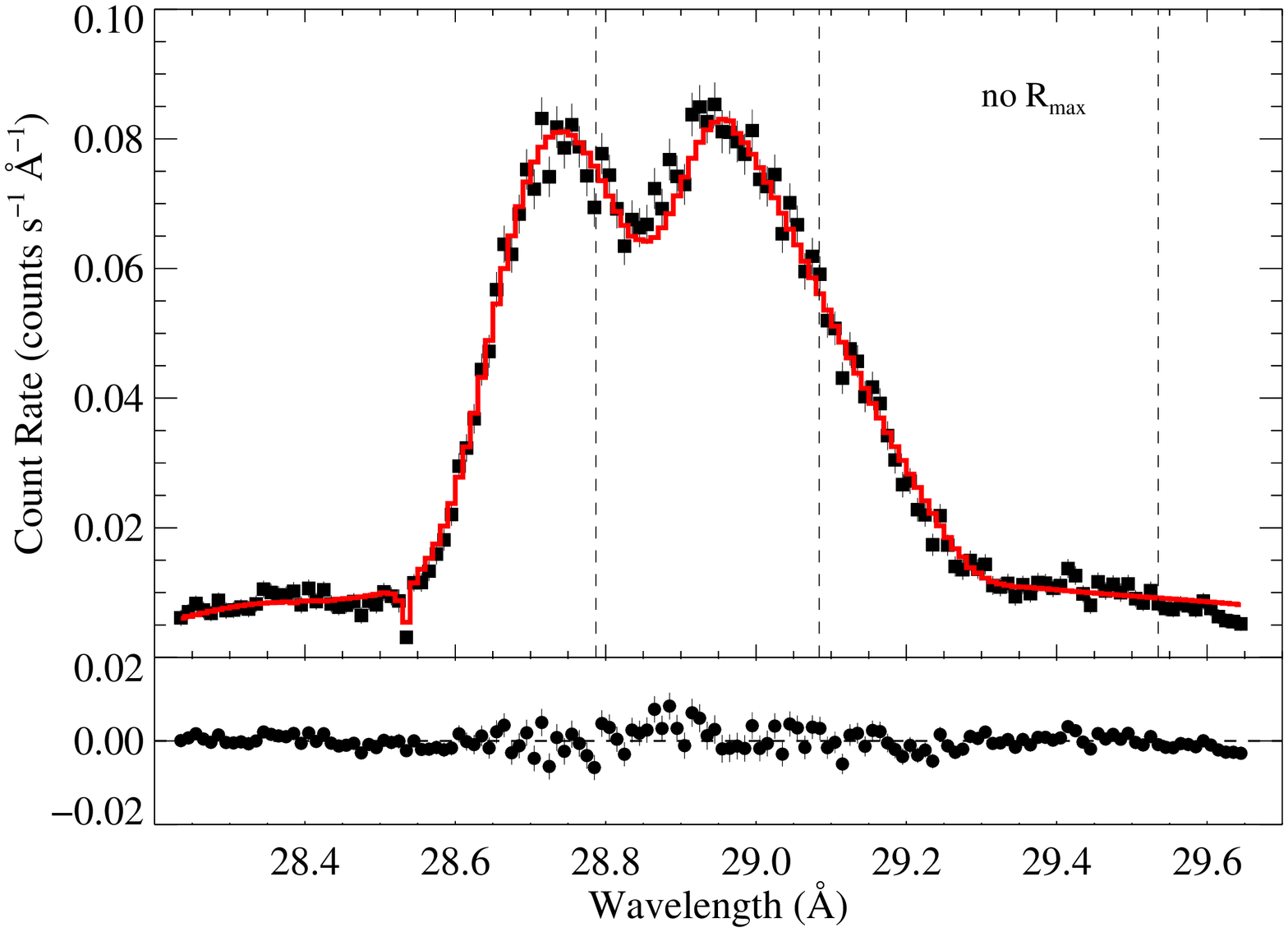}
  \includegraphics[angle=90, scale=\profilePlotPanelSize]{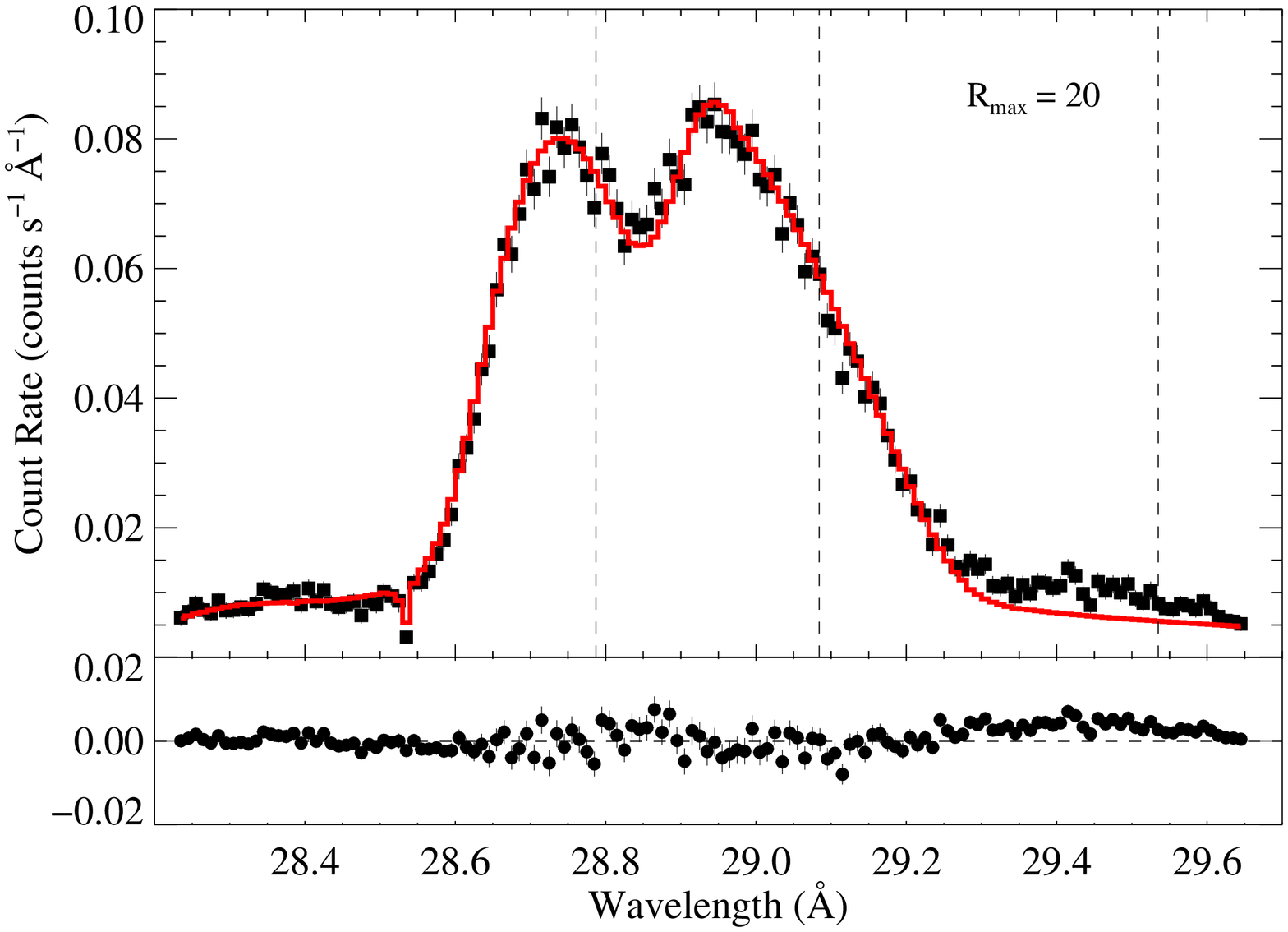}
  \includegraphics[angle=90, scale=\profilePlotPanelSize]{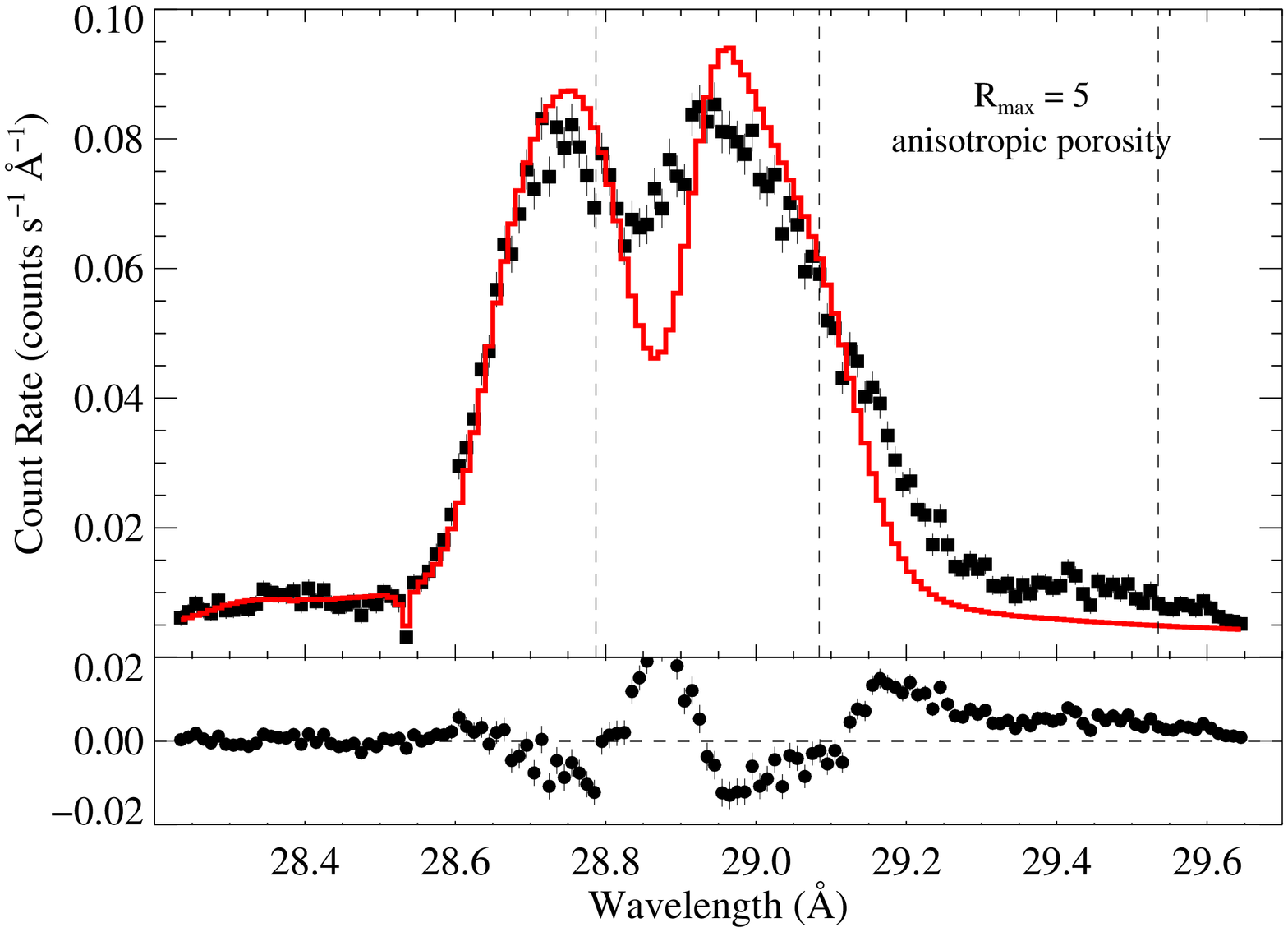}
  \caption{\xmm RGS spectrum of the \ion{N}{6} complex of \zp,
    together with best fit models assuming $\rmax = \infty$ (first
    panel), $\rmax = 20 \rstar$ (second panel) , and $\rmax = 5 \rstar$
    (third panel). The three vertical dashed lines indicate the
    positions of the rest wavelengths of the resonance,
    intercombination, and forbidden lines, from left to right. The
    model shown in third panel includes the effects of anisotropic
    porosity, and is comparable to the model favored in
    \citet{OFH06}. Both models with finite radial upper cutoffs
    underpredict emission from the forbidden line, which is primarily
    formed at large radii. The weak feature at 28.4656 \AA\ is
    \ion{C}{6} Ly$\beta$. (A color version of this figure is available
    in the online journal.) \label{fig:helikeNVI}}
\end{figure}

We fit this model to the \ion{O}{7} and \ion{N}{6} triplets in the RGS
spectrum of \zp. Resonance scattering was included in our models, as
described in \citet{LOKP07}.  The weak \ion{C}{6} Ly$\beta$ line
(which falls on the red wing of the resonance line of \ion{N}{6}) was
included in our fit to the \ion{N}{6} complex, as in
\citet{LOKP07}. Both complexes were fit under the competing
assumptions that $\rmax = \infty$ or $20 \rstar$. In order to make a
more direct comparison with the radial distribution of X-ray emitting
plasma assumed in Oskinova et al., we also tried to fit the data with
models using their values of $\ro = 1.5 \rstar$ and $\rmax = 5
\rstar$, and including the effects of porosity by allowing \hinf to be
a free parameter. The results of our fits are given in
Table~\ref{tab:helike}, and the models are shown in
Figures~\ref{fig:helikeOVII} and \ref{fig:helikeNVI}.

The models with a finite radial cutoff clearly do not produce enough
emission from the forbidden line to fit the data. On the other hand,
the models with no radial cutoff fits the $f/i$ line ratios (and the
line complexes) quite well. We thus conclude that a $\rmax = 5
\rstar$ cutoff is strongly disfavored by the observed forbidden line
strengths as well as the wings of individual lines.

\begin{deluxetable*}{ccccccccccc}
  \tablecaption{He-like triplet models with finite radial upper
    cutoff: fit results to the \xmm RGS spectra}
  \tablehead{
    \colhead{ion} & 
    \colhead{\taustar} & 
    \colhead{\ro} & 
    \colhead{$\rmax$} &
    \colhead{$h_\infty$} &
    \colhead{$\tau_{0,*}$} &
    \colhead{$G$} &
    \colhead{Normalization} & 
    \colhead{$\chi^2$} & 
    \colhead{$N_{\rm bins}$} \\
    \colhead{} & 
    \colhead{} & 
    \colhead{(\rstar)} & 
    \colhead{(\rstar)} & 
    \colhead{(\rstar)} & 
    \colhead{} &
    \colhead{} &
    \colhead{($10^{-4}$ photons cm$^{-2}$ s$^{-1}$)} & 
    \colhead{} & 
    \colhead{} 
  }
  \startdata 
  \ion{N}{6} & 4.79 & 2.28 & $\infty$ & 0 &$\infty$ & 1.162 & 17.52 & 181.5 & 142 \\
             & 5.70  & 4.0  & 20 & 0 & $\infty$  & 1.176  & 16.96 & 381.5 & 142 \\
             & 26.28  & 1.5  & 5 & 0.72 &$\infty$  & 1.085  & 16.05 & 1435.9 & 142 \\
  \ion{O}{7} & 3.96 & 1.59 & $\infty$ & 0 & 2.68   & 1.005 & 7.582 & 146.8 & 109 \\
             & 7.75 & 4.0 & 20     & 0 & 8.32   & 1.174 & 7.408 & 311.1 & 109 \\
             & 25.57 & 1.5 & 5     & 0.77 & 0.85   & 1.095 & 7.171 & 740.8 & 109 \\
  \enddata
\label{tab:helike}
\end{deluxetable*}

\section{Discussion}
\label{sec:discussion}

One of the most important results of our emission line profile
modeling is that anisotropic porosity models are disfavored. This can
be understood in terms of the markedly different profile shapes
predicted by anisotropic porosity models, with a strong bump at line
center due to the ``Venetian blind effect'', as exemplified in
Figure~\ref{fig:15.014_model}. There is no such bump evident in any of
the profiles we have modeled, and this is reflected in our formal
statistical constraints.

On the other hand, isotropic porosity models can provide good fits to
the lines we studied, provided the porosity lengths are no larger than
\rstar.  Porosity length and mass-loss rate are degenerate to a
certain extent, since they both influence the degree of profile
asymmetry. We have quantified this degeneracy for one line in
Figure~\ref{fig:15.014_contour}. Large porosity lengths ($\hinf >
\rstar$) and their associated large adjustments to derived mass-loss
rates are disfavored by the data. Ruling out large porosity lengths is
possible because porosity length and mass-loss rate are not fully
degenerate parameters, leading to measurable differences in profile
shape, as shown in Figure~\ref{fig:15.014_model}.  Smaller non-zero
porosity lengths are allowed, but lead to only modest adjustments to
derived mass-loss rates.

We have compared our results to those of \citet{OFH06}, who also
compare both porous and non-porous models to X-ray spectra of \zp, but
reach the opposite conclusion that models assuming anisotropic porosity
are allowed, and non-porous models disfavored. We showed that this
difference in conclusions is a direct result of a difference in
assumptions regarding the spatial distribution of X-ray emitting
plasma. Specifically, we assumed in our modeling that X-ray emission
extends to effectively infinite radius, while Oskinova et al. assumed
that X-ray emission turns off above 5 $\rstar$.

In Figure~\ref{fig:rmaxmodel} we compared the effect of different
assumptions regarding \rmax, and found that models with a finite
radial cutoff have much less flux in the wings of the line profile
than models with no cutoff, which is a straightforward consequence of
the fact that all emission at large redshift and blueshift originates
at large radii. We fit models with different assumed values for
\rmax to the \ion{O}{8} Ly$\alpha$ line, and found that models
with a finite radial cutoff do not fit the wings of the observed line
profile of \zp.

We further probed the radial distribution of X-ray emission in the
wind of \zp by fitting the He-like triplet lines of \ion{O}{7} and
\ion{N}{6} with models accounting for the radial dependence of the
forbidden-to-intercombination line ratio \citep{LPKC06}. As is evident
from Figures~\ref{fig:helikeOVII} and \ref{fig:helikeNVI}, models with
a finite radial cutoff do not reproduce the observed forbidden line
flux, which indicates that there is significant X-ray emission
originating from $R > 20 \rstar$.

The finding of significant X-ray emission from large radii is one of
the major, if unintended, results of this study. One might be
surprised by this result based on the simple argument that X-rays
should originate where the wind is accelerating and can generate
strong shocks. \citet{RO02} have performed hydrodynamic simulations
specifically aimed at understanding the outer wind structure of O
stars, and they find that X-ray emitting shocks can be generated at
tens of stellar radii. Furthermore, the cooling time for few MK plasma
formed at $r \gax 10 \rstar$ in the wind of $\zeta$ Pup is comparable to
the flow time, so even if new shocks are not generated at very large
radii, X-ray emission can still persist \citep{FKPPP97}.

Under similar modeling assumptions as \citet{OFH06}, in particular
that X-ray emission is cut off above 5 \rstar, we too find that porous
models with anisotropic clumps are favored over non-porous models.
But we also show that such a small cutoff radius for X-ray emission is
not consistent with either the observed flux in the wings of X-ray
emission lines, or the strength of forbidden line in He-like
triplets. These constraints are well matched by models without an
arbitrary X-ray cutoff. With such extended emission, fitting line
profiles favors models with either no or a modest isotropic porosity,
with anisotropic porosity strongly disfavored.

\citet{Cohen2010} have derived a mass-loss rate for \zp from modeling
of X-ray line profiles under the assumption that porosity effects are
negligible. Taking the conclusions of the present work at face value,
the correction to this mass-loss rate is at most 40 \% if moderate
porosity effects are present ($\hinf \sim \rstar$). This conclusion is
further supported by the fact that the uncorrected X-ray mass-loss
rate of \citet{Cohen2010} agrees with mass-loss rates from non-X-ray
observational diagnostics within uncertainties
\citep{2011A&A...535A..32N, Bouret2012}.

We thus conclude that, at least for \zp, X-ray line profiles are a
good independent diagnostic of mass-loss rates, and that they are not
subject to strong systematic errors from clumping on any scale. The
other O stars observed by \cxo and \xmm should ideally also be
subjected to a similar study, although it would perhaps be surprising
to find major differences in porosity effects in the wind of \zp in
comparison with other O stars.  In fact, early O supergiants like \zp
with their high mass-loss-rate winds are the stars for which porosity
effects are expected to be the strongest. Because the essence of
porosity is the optical thickness of individual clumps, O stars with
lower mass-loss rates than \zp require more extreme clump properties
just to produce the same porosity effect.

The relative robustness against wind inhomogeneity effects of X-ray
line profile measurements of mass-loss rates suggests their use as a
primary mass-loss rate diagnostic, on an equal footing with
traditional diagnostics such as H$\alpha$, thermal radio emission, and
UV absorption line profiles.  However, X-ray line profile measurements
are currently only possible for the brightest stars in the
Galaxy. Until the advent of much larger X-ray spectroscopic
observatories, with square meter effective areas and which could
undertake a large-scale survey of O star X-ray spectra, we suggest a
program of benchmarking as many stars as possible with all available
mass-loss rate diagnostics. Only by continuing the tradition of a
multiwavelength, multidiagnostic approach can we hope to disentangle
the observational signatures of mass-loss and wind structure.

\acknowledgements

We acknowledge the comments of the referee, Achim Feldmeier, which
significantly improved the presentation of this article.  Support for
this work was provided by the National Aeronautics and Space
Administration through \cxo award number AR7-8002X and ADAP award
number NNX11AD26G to Swarthmore College. J.O.S. and S.P.O. acknowledge
support from NASA award ATP NNX11AC40G to the University of
Delaware. J.O.S. also acknowledges current support from DFG-grant
Pu117/8-1.

\clearpage

\appendix
\section{Expanded fit results \label{sec:appendix}}
\label{sec:appendix}

\begin{figure}[hb]
  \includegraphics[angle=90,scale=\appendixPlotPanelSize]{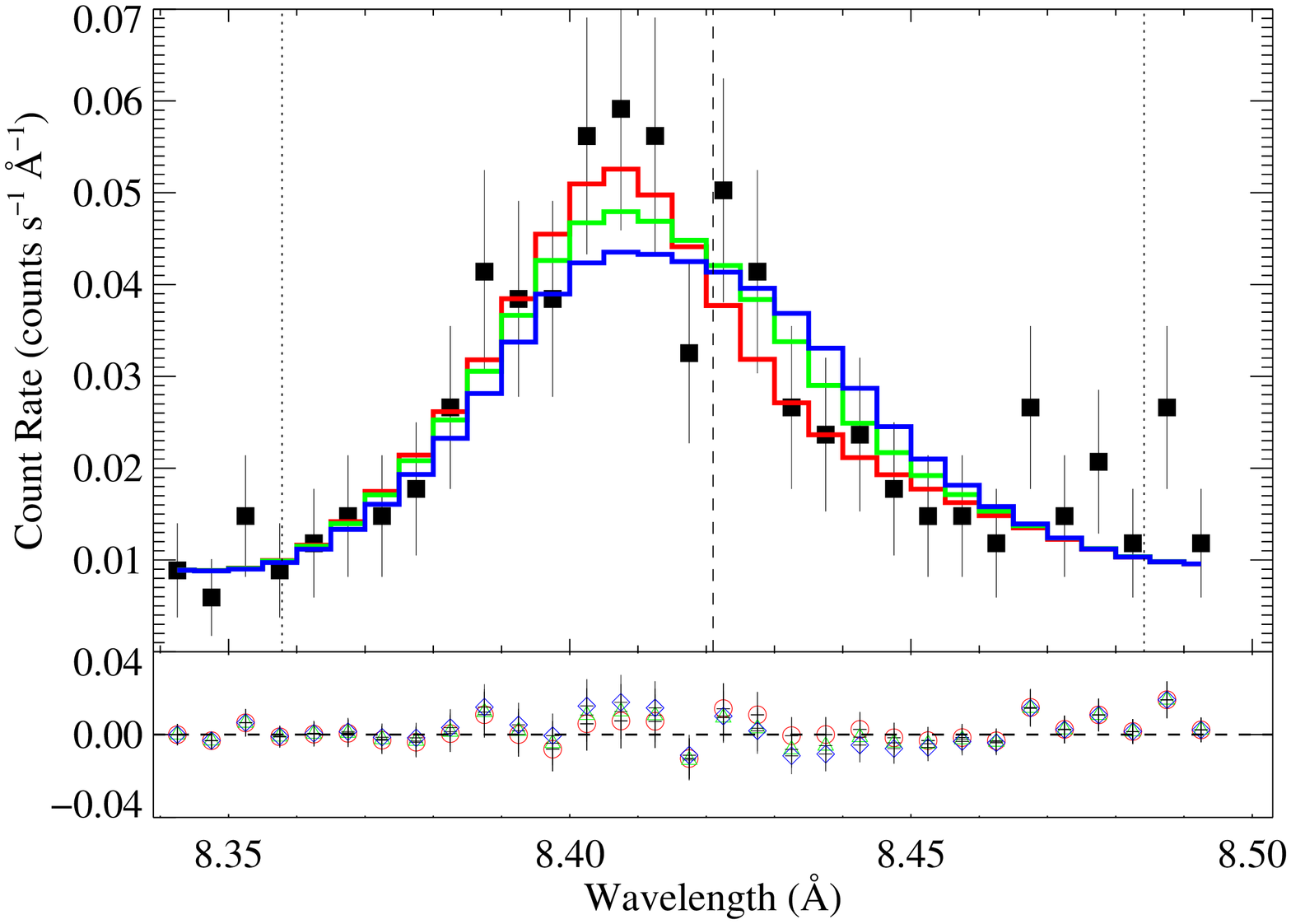}
  \caption{The \cxo MEG measurements of the \ion{Mg}{12} line at 8.421
    \AA, with the best-fit anisotropic porosity models
    superimposed. The red, green, and blue models assume \hinf = 0, 1,
    and 5 \rstar, respectively. (A color version of this figure is
    available in the online journal.) \label{fig:8.421_aniso}}
\end{figure}

\begin{figure}[hb]
  \includegraphics[angle=90,scale=\appendixPlotPanelSize]{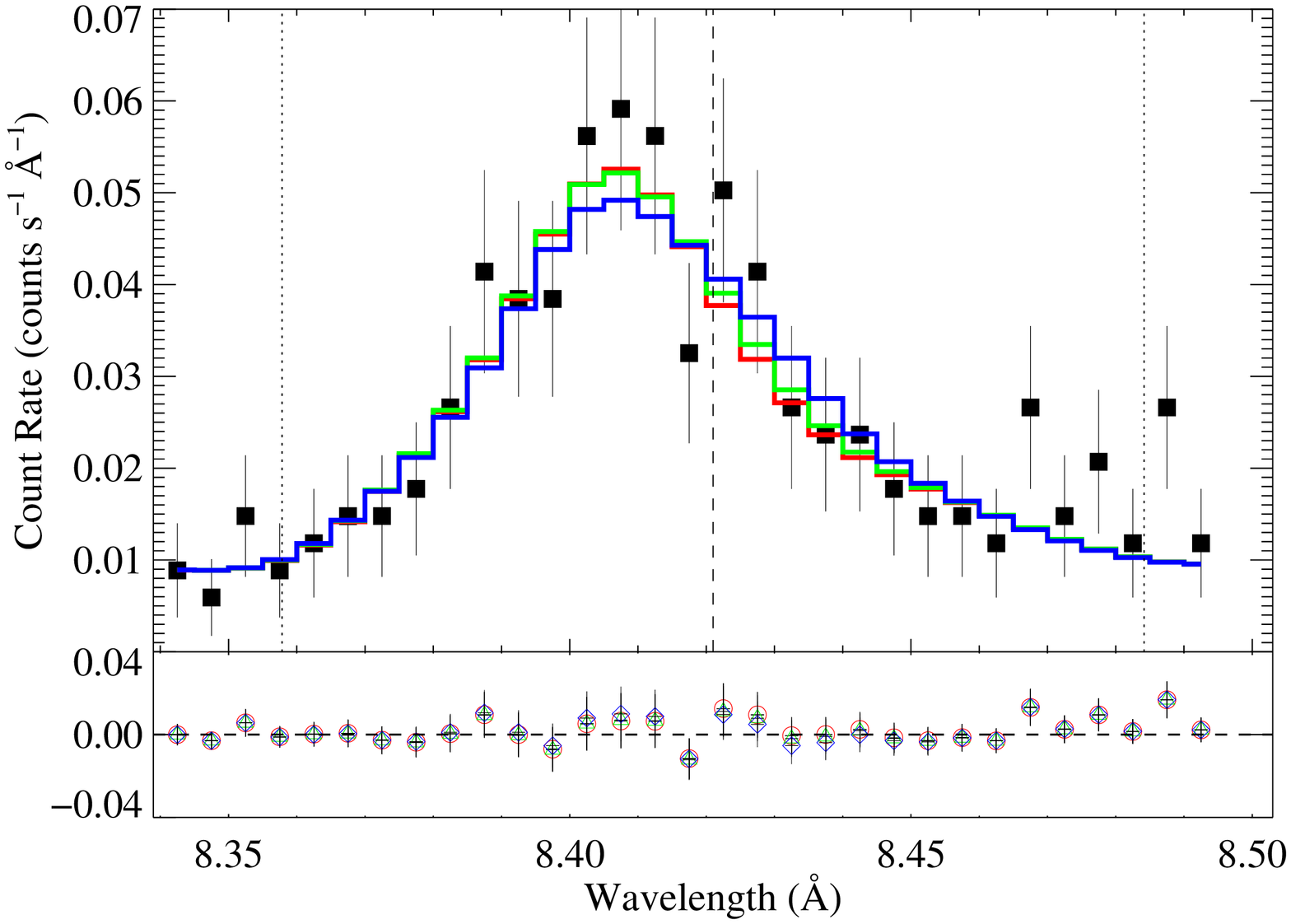}
  \caption{The \cxo MEG measurements of the \ion{Mg}{12} line at 8.421
    \AA, with the best-fit isotropic porosity models superimposed. The
    red, green, and blue models assume \hinf = 0, 1, and 5 \rstar,
    respectively. (A color version of this figure is available in the
    online journal.)\label{fig:8.421_iso}}
\end{figure}

\begin{figure}
  \includegraphics[angle=90,scale=\appendixPlotPanelSize]{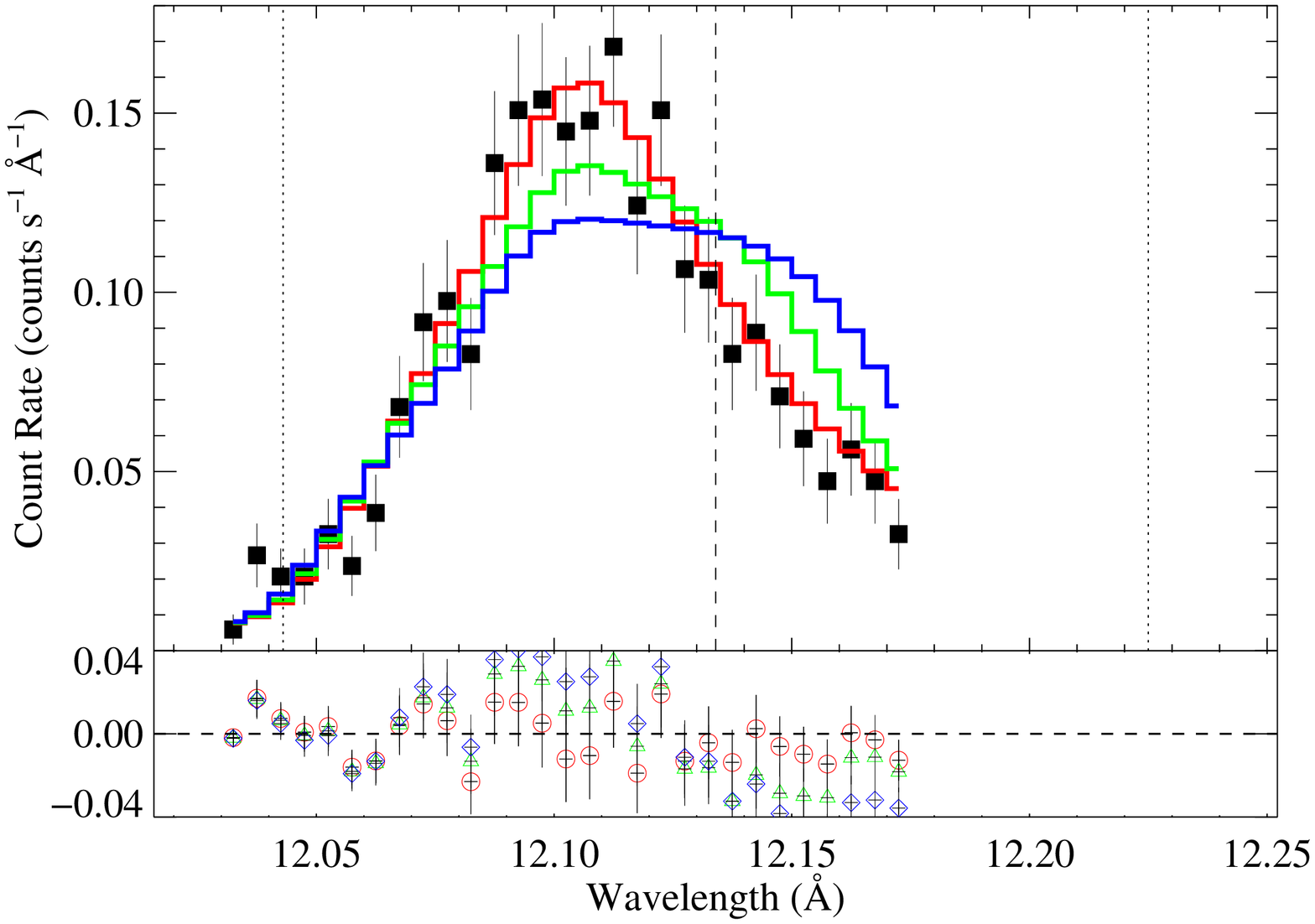}
  \includegraphics[angle=90,scale=\appendixPlotPanelSize]{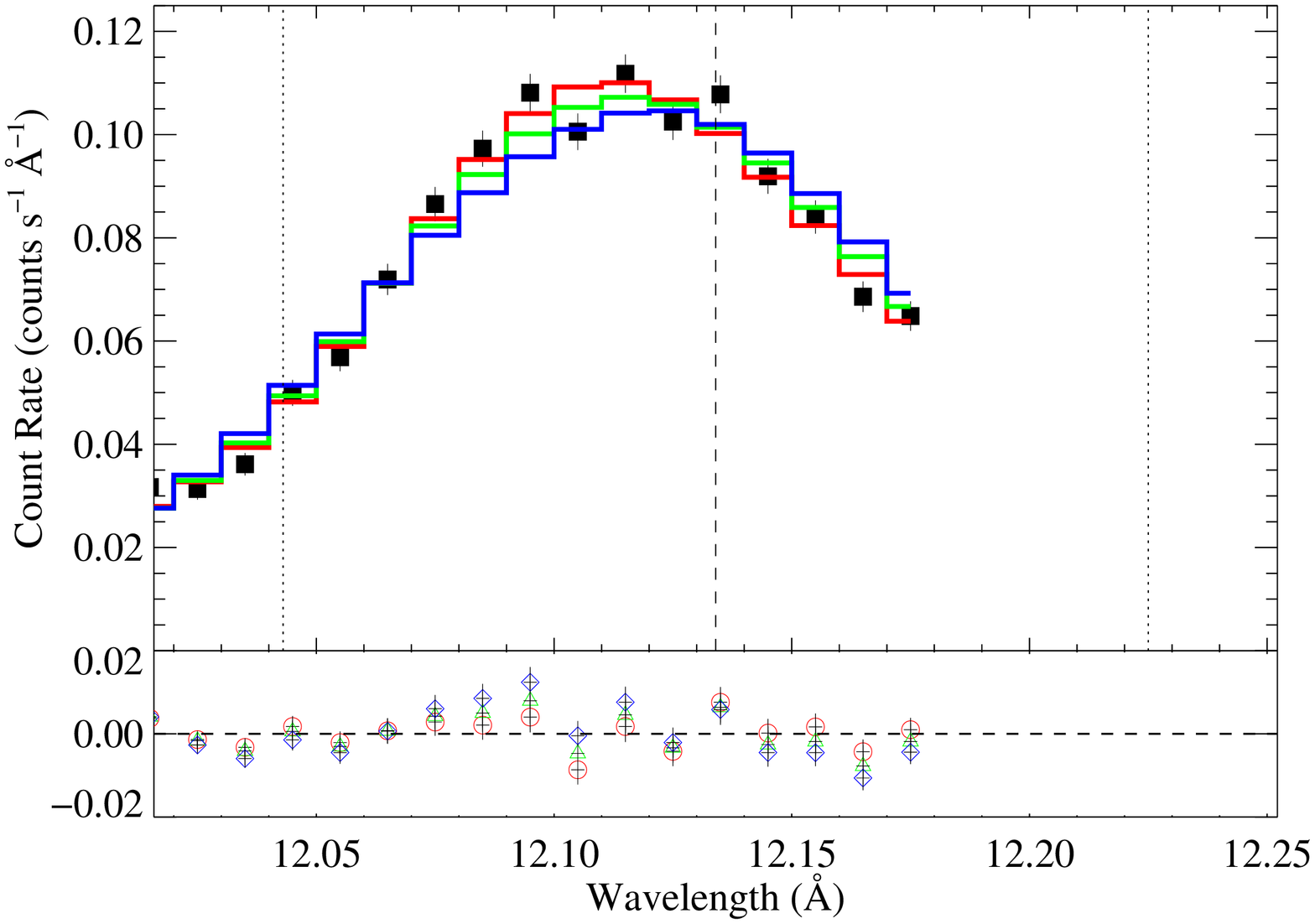}
\caption{The \cxo MEG (upper panel) and \xmm RGS (lower panel)
  measurements of the \ion{Ne}{10} line at 12.134 \AA, with the
  best-fit anisotropic porosity models superimposed. The red, green,
  and blue models assume \hinf = 0, 1, and 5 \rstar, respectively. (A
  color version of this figure is available in the online
  journal.) \label{fig:12.134_aniso}}
\end{figure}

\begin{figure}
  \includegraphics[angle=90,scale=\appendixPlotPanelSize]{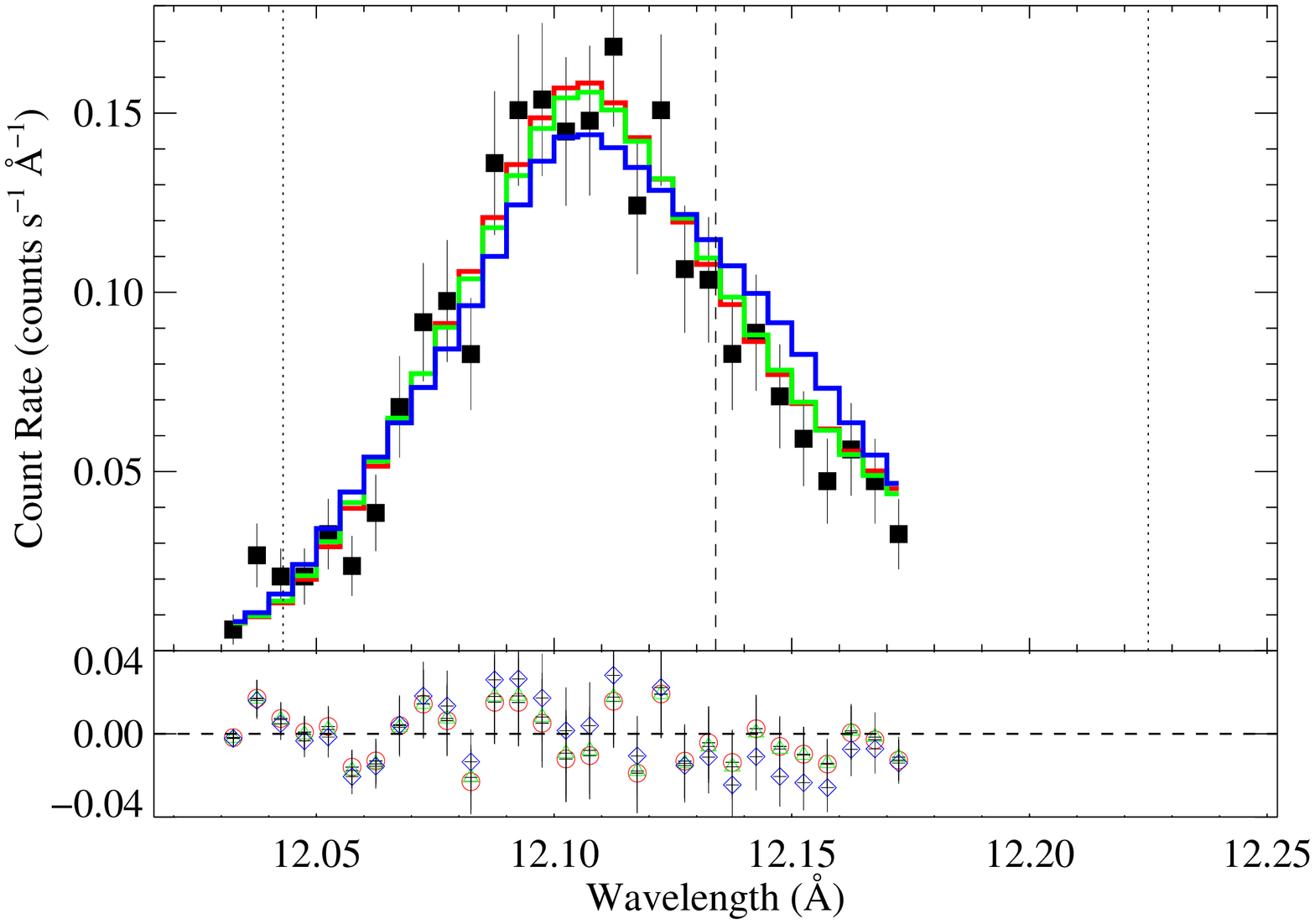}
  \includegraphics[angle=90,scale=\appendixPlotPanelSize]{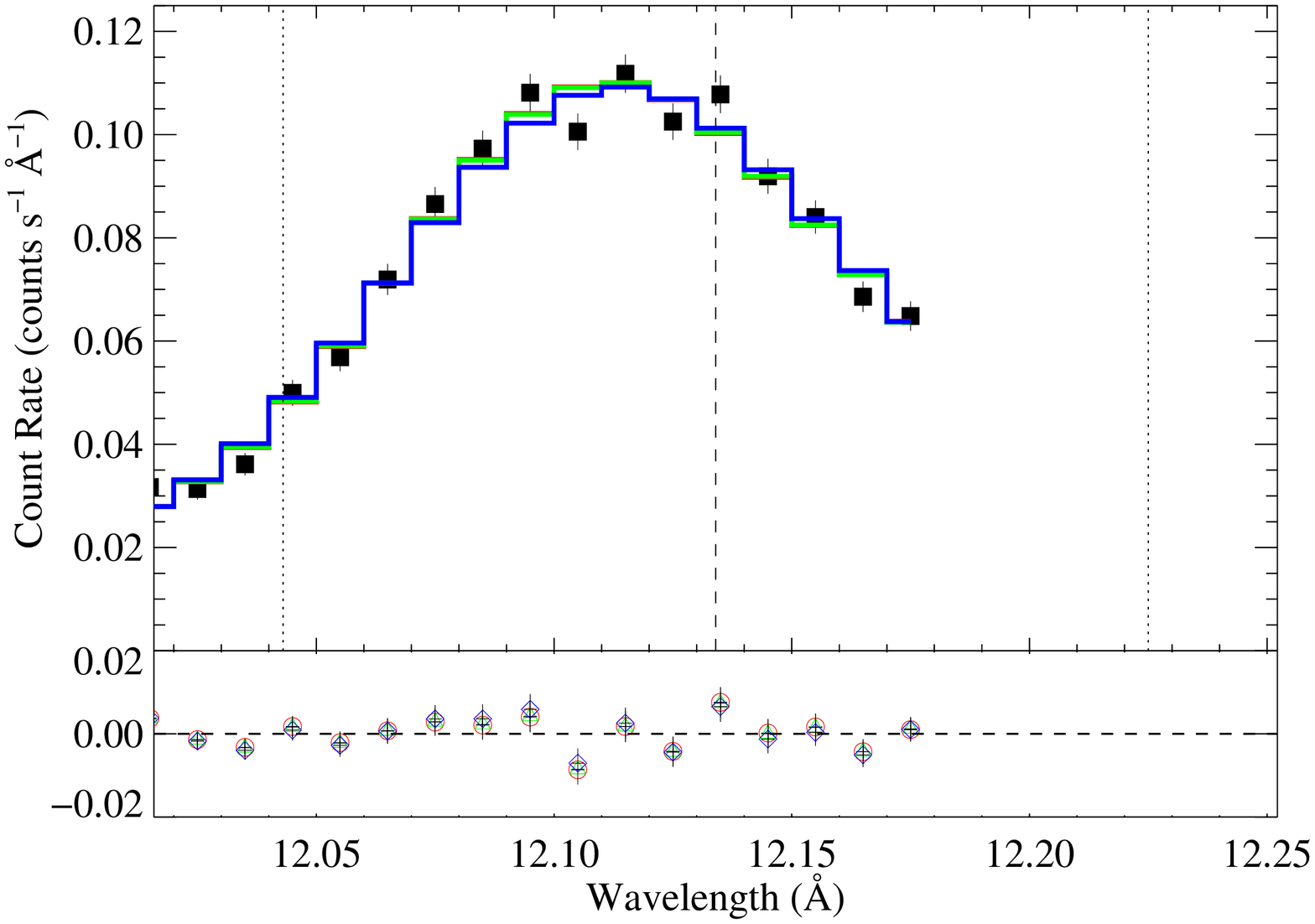}
\caption{The \cxo MEG (upper panel) and \xmm RGS (lower panel)
  measurements of the \ion{Ne}{10} line at 12.134 \AA, with the
  best-fit isotropic porosity models superimposed. The red, green, and
  blue models assume \hinf = 0, 1, and 5 \rstar, respectively. (A
  color version of this figure is available in the online
  journal.) \label{fig:12.134_iso}}
\end{figure}

\begin{figure}
\includegraphics[angle=90,scale=\appendixPlotPanelSize]{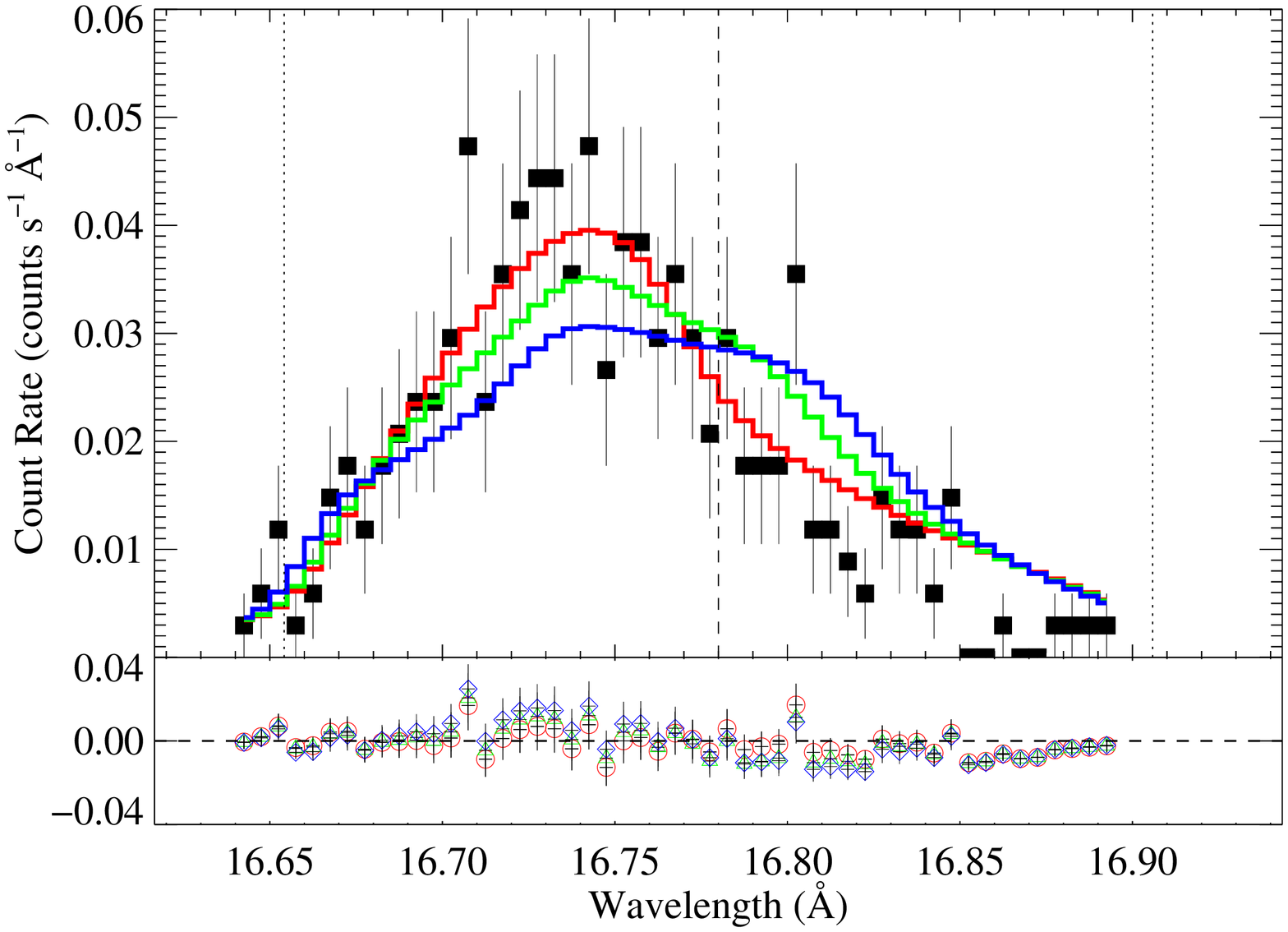}
\caption{The \cxo MEG measurements of the \ion{Fe}{17} line at 16.780
  \AA, with the best-fit anisotropic porosity models superimposed. The
  red, green, and blue models assume \hinf = 0, 1, and 5 \rstar,
  respectively. (A color version of this figure is available in the
  online journal.) \label{fig:16.780_aniso}}
\end{figure}

\begin{figure}
\includegraphics[angle=90,scale=\appendixPlotPanelSize]{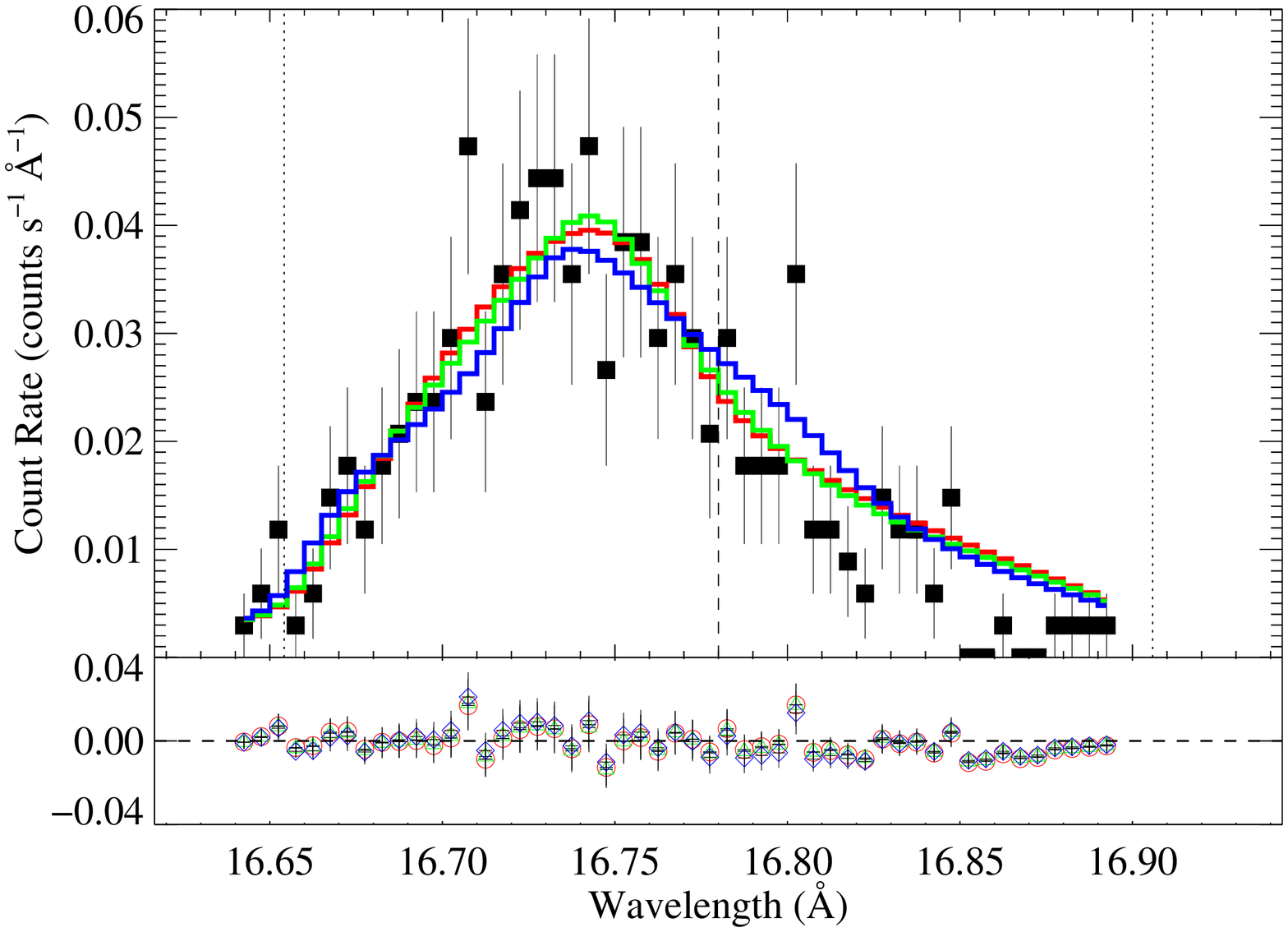}
\caption{The \cxo MEG measurements of the \ion{Fe}{17} line at 16.780
  \AA, with the best-fit isotropic porosity models superimposed. The
  red, green, and blue models assume \hinf = 0, 1, and 5 \rstar,
  respectively. (A color version of this figure is available in the
  online journal.) \label{fig:16.780_iso}}
\end{figure}

\begin{figure}
  \includegraphics[angle=90,scale=\appendixPlotPanelSize]{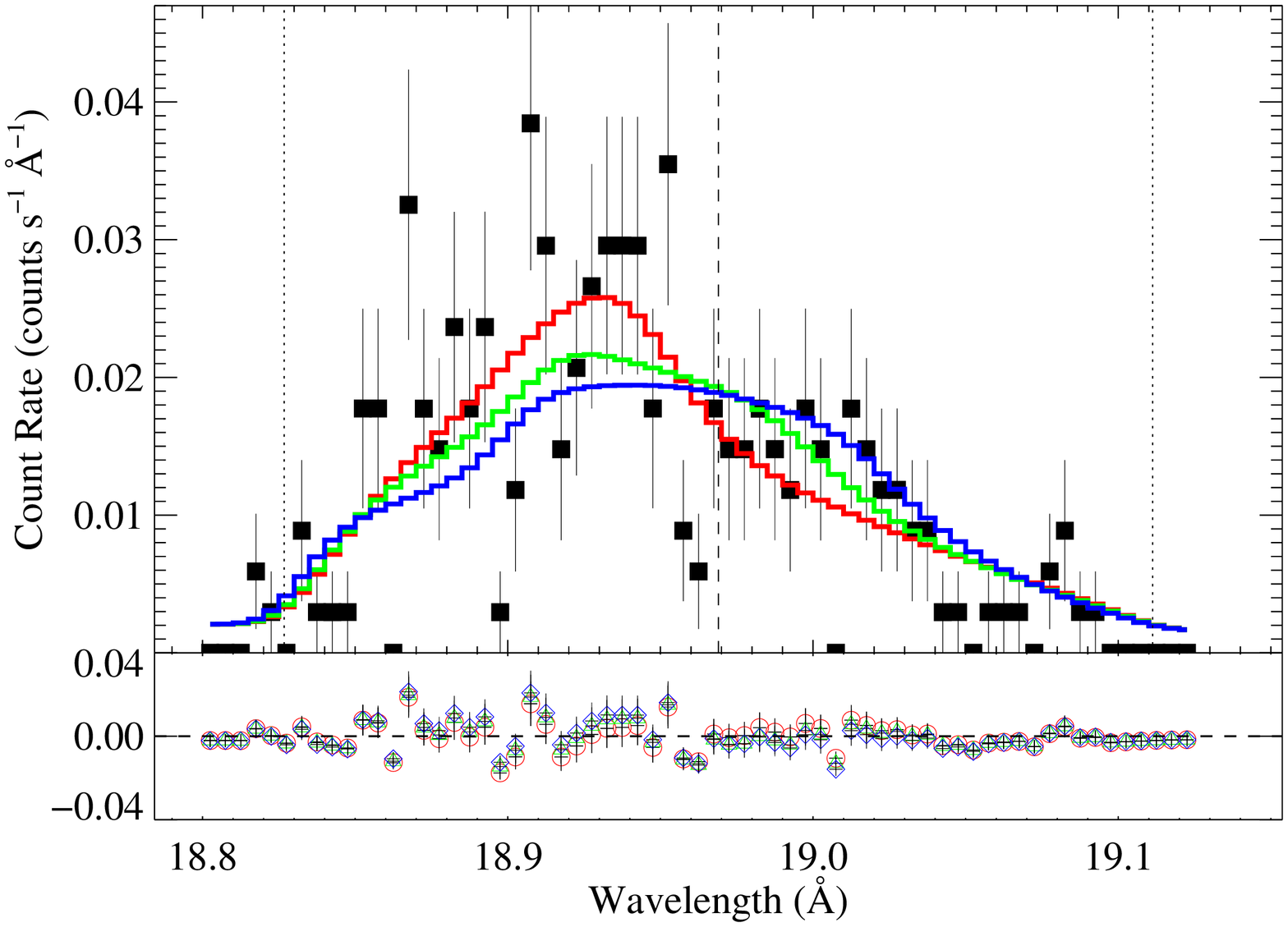}
  \includegraphics[angle=90,scale=\appendixPlotPanelSize]{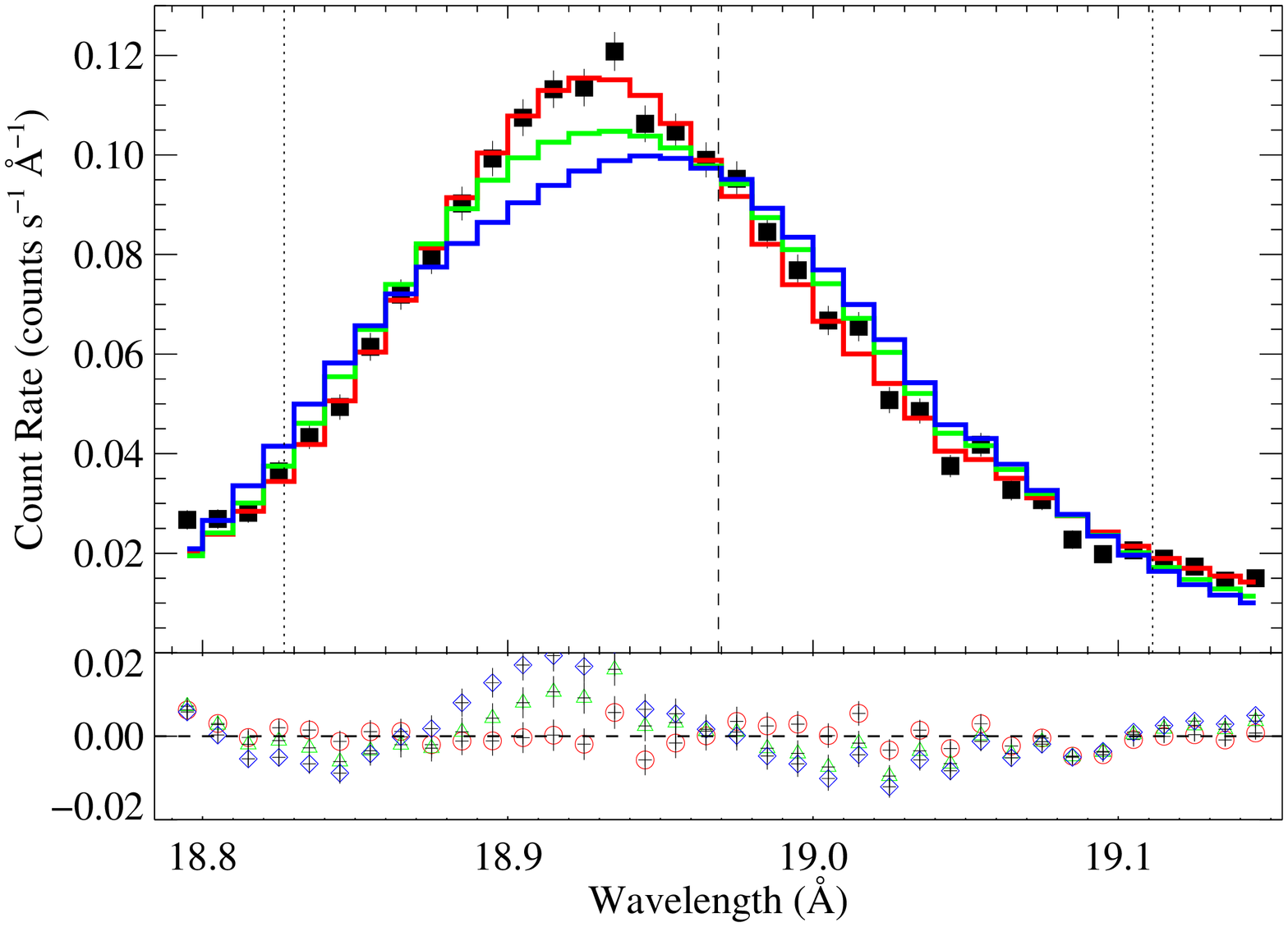}
\caption{The \cxo MEG (upper panel) and \xmm RGS (lower panel)
  measurements of the \ion{O}{8} line at 18.969 \AA, with the best-fit
  anisotropic porosity models superimposed. The red, green, and blue
  models assume \hinf = 0, 1, and 5 \rstar, respectively. (A color
  version of this figure is available in the online
  journal.) \label{fig:18.969_aniso}}
\end{figure}

\begin{figure}
  \includegraphics[angle=90,scale=\appendixPlotPanelSize]{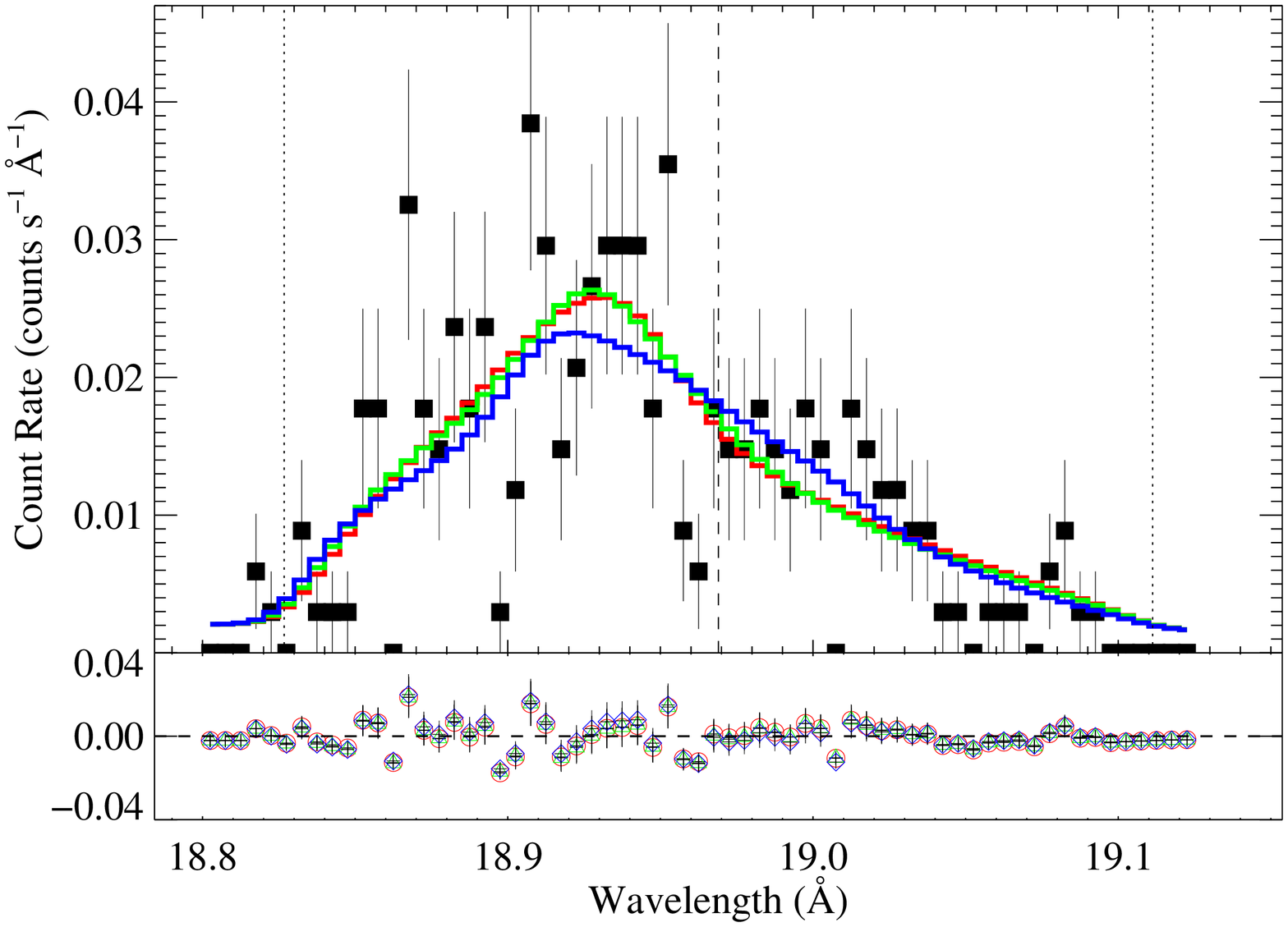}
  \includegraphics[angle=90,scale=\appendixPlotPanelSize]{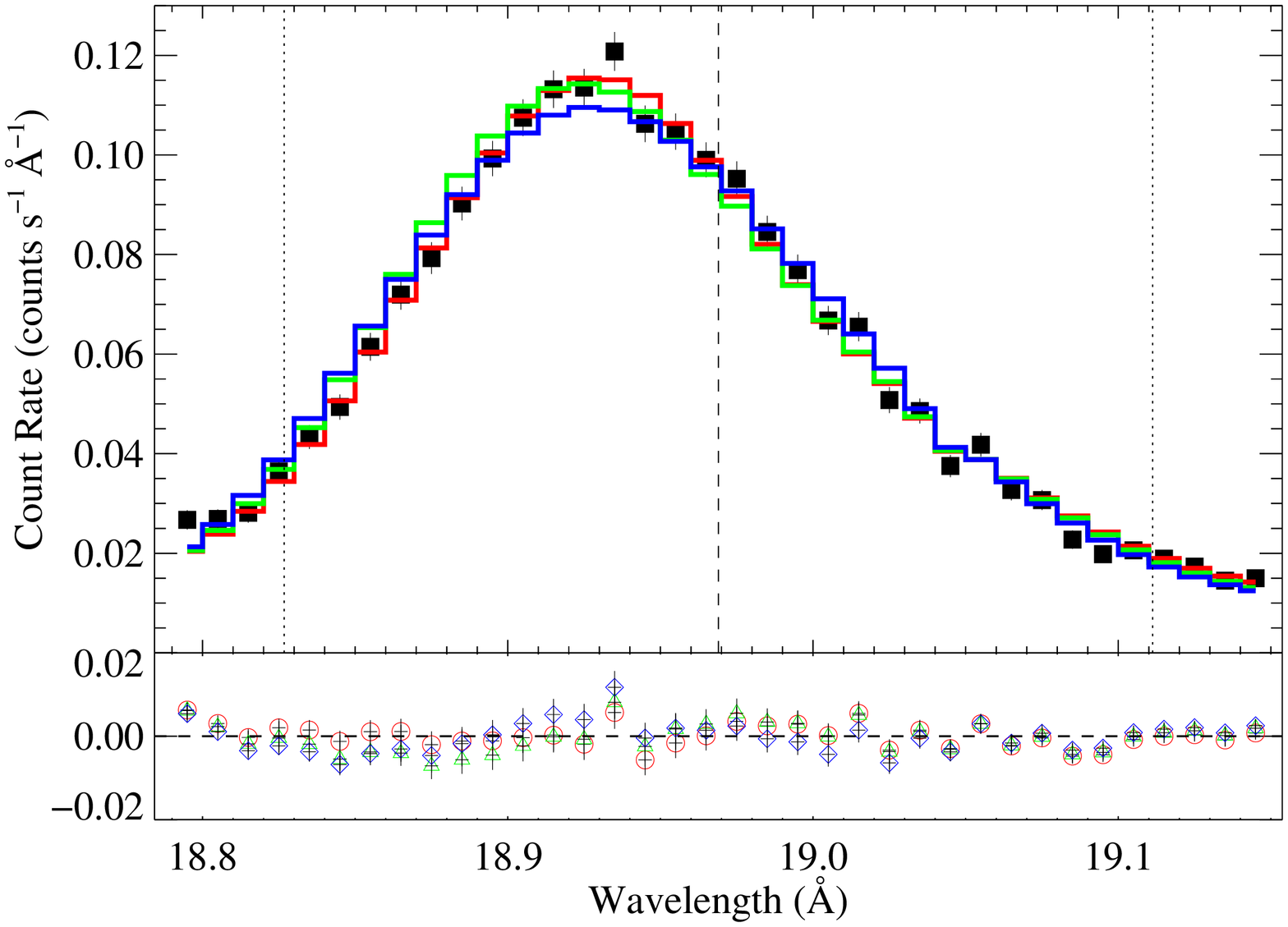}
\caption{The \cxo MEG (upper panel) and \xmm RGS (lower panel)
  measurements of the \ion{O}{8} line at 18.969 \AA, with the best-fit
  isotropic porosity models superimposed. The red, green, and blue
  models assume \hinf = 0, 1, and 5 \rstar, respectively. (A color
  version of this figure is available in the online
  journal.) \label{fig:18.969_iso}}
\end{figure}

\begin{figure}
\includegraphics[angle=90,scale=\appendixPlotPanelSize]{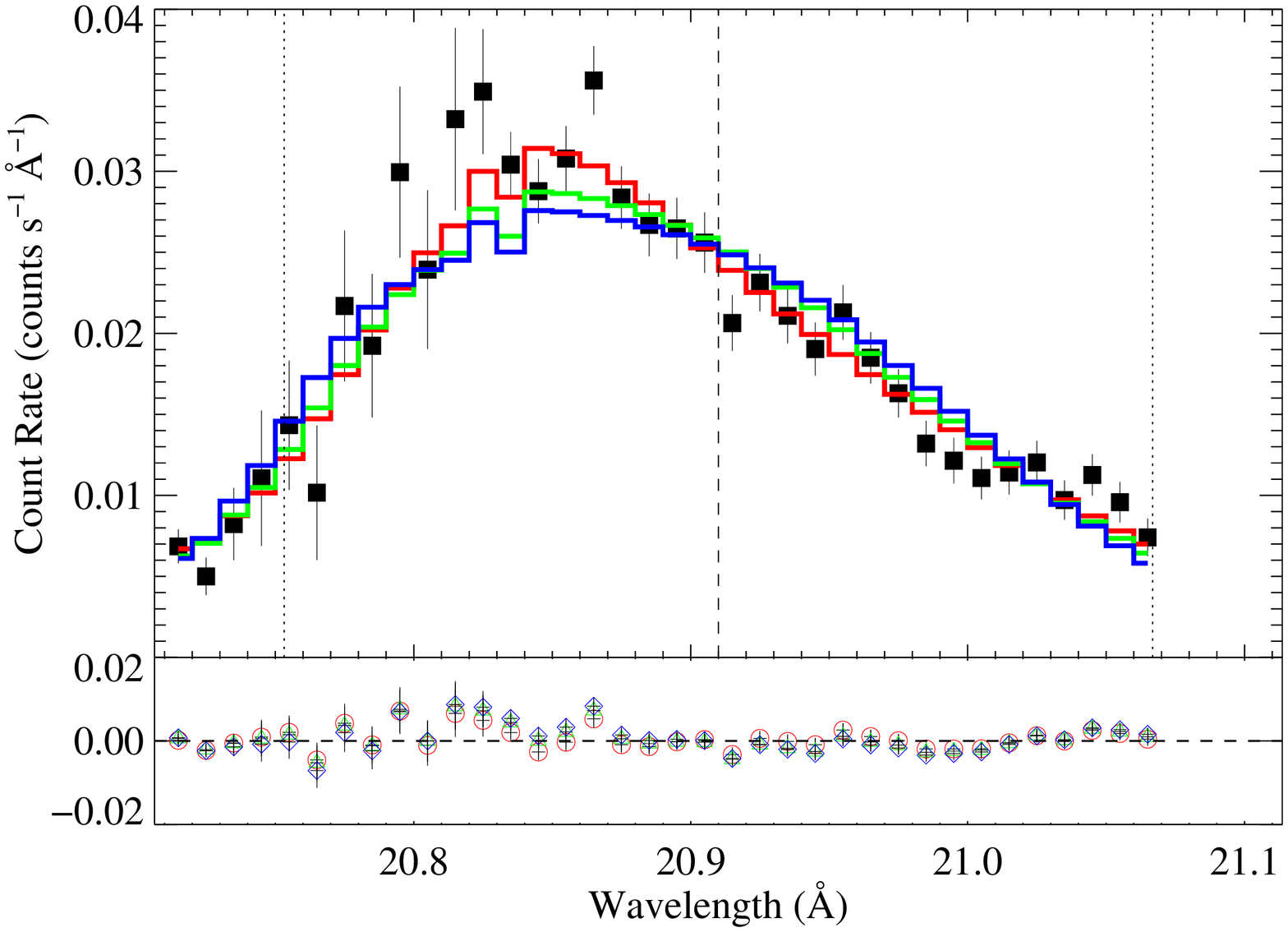}
\caption{The \xmm RGS measurements of the \ion{N}{7} line at 20.910
  \AA, with the best-fit anisotropic porosity models superimposed. The
  red, green, and blue models assume \hinf = 0, 1, and 5 \rstar,
  respectively. (A color version of this figure is available in the
  online journal.) \label{fig:20.910_aniso}}
\end{figure}

\begin{figure}
\includegraphics[angle=90,scale=\appendixPlotPanelSize]{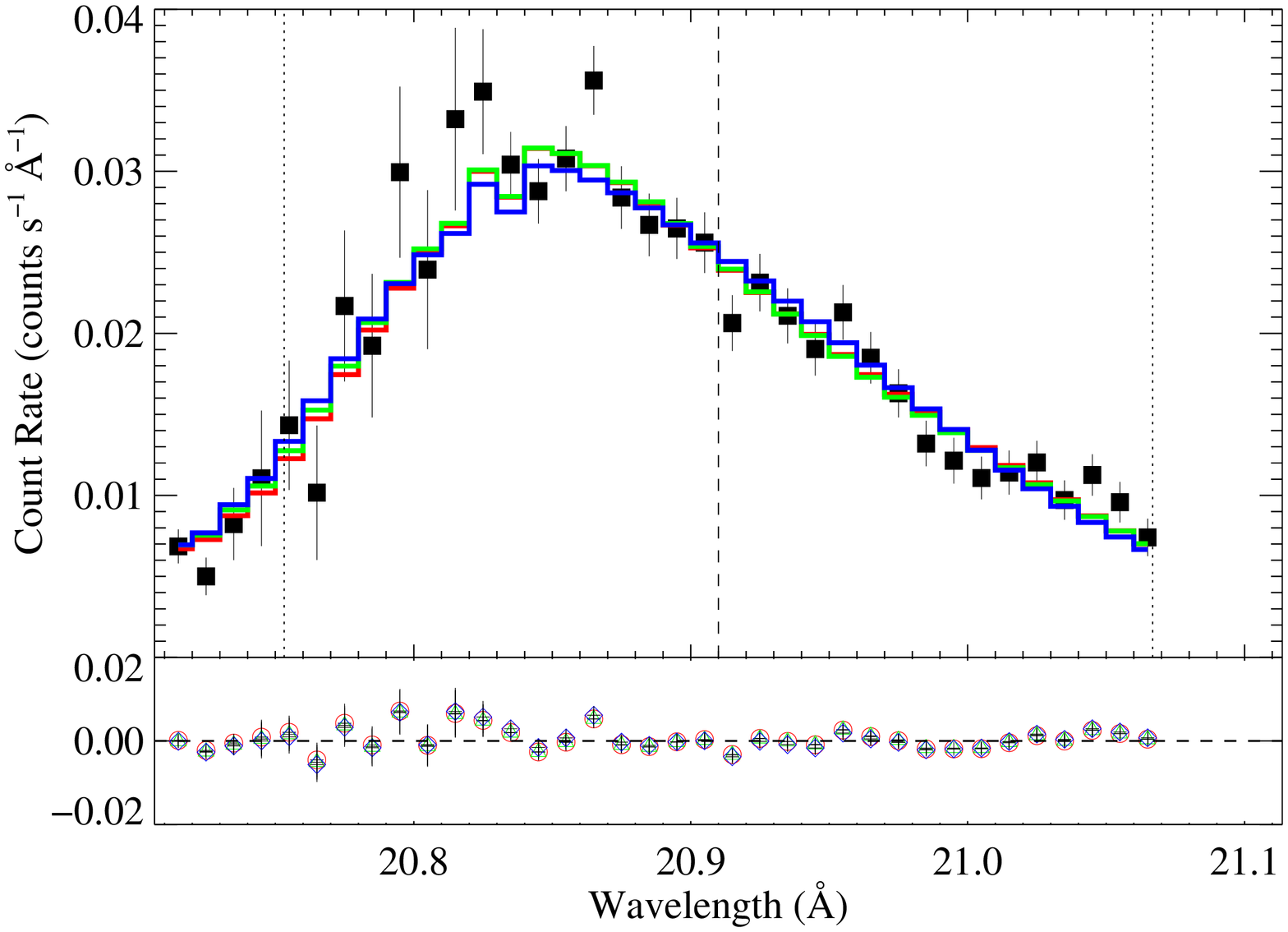}
\caption{The \xmm RGS measurements of the \ion{N}{7} line at 20.910
  \AA, with the best-fit isotropic porosity models superimposed. The
  red, green, and blue models assume \hinf = 0, 1, and 5 \rstar,
  respectively. (A color version of this figure is available in the
  online journal.) \label{fig:20.910_iso}}
\end{figure}

In this section, in Figures~\ref{fig:8.421_aniso}-\ref{fig:20.910_iso}
we show plots of all the fits reported in
Tables~\ref{tab:chandra_aniso}-\ref{tab:rgs_iso}. For each line that
we fit, we show plots of anisotropic porosity models in one figure,
and isotropic porosity models in another figure. We repeat the
non-porous model in both the anisotropic and isotropic porosity fit
figures. To keep the figures legible, we only show models with \hinf =
0, 1, and 5 \rstar.

\clearpage

\bibliography{xray-ostar}

\begin{thebibliography}{35}
\expandafter\ifx\csname natexlab\endcsname\relax\def\natexlab#1{#1}\fi

\bibitem[{{Arnaud}(1996)}]{1996ASPC..101...17A}
{Arnaud}, K.~A. 1996, in ASP Conf. Ser., Vol. 101, Astronomical Data Analysis
  Software and Systems V, ed. G.~H. {Jacoby} \& J.~{Barnes}, (San Francisco,
  CA: ASP), 17

\bibitem[{{Bouret} {et~al.}(2012){Bouret}, {Hillier}, {Lanz}, \&
  {Fullerton}}]{Bouret2012}
{Bouret}, J.-C., {Hillier}, D.~J., {Lanz}, T., \& {Fullerton}, A.~W. 2012,
  \aap, 544, A67

\bibitem[{{Brucato}(1971)}]{1971MNRAS.153..435B}
{Brucato}, R.~J. 1971, \mnras, 153, 435

\bibitem[{{Cash}(1979)}]{C79}
{Cash}, W. 1979, \apj, 228, 939

\bibitem[{{Churazov} {et~al.}(1996){Churazov}, {Gilfanov}, {Forman}, \&
  {Jones}}]{1996ApJ...471..673C}
{Churazov}, E., {Gilfanov}, M., {Forman}, W., \& {Jones}, C. 1996, \apj, 471,
  673

\bibitem[{{Cohen} {et~al.}(2010){Cohen}, {Leutenegger}, {Wollman},
  {Zsarg{\'o}}, {Hillier}, {Townsend}, \& {Owocki}}]{Cohen2010}
{Cohen}, D.~H., {Leutenegger}, M.~A., {Wollman}, E.~E., {Zsarg{\'o}}, J.,
  {Hillier}, D.~J., {Townsend}, R.~H.~D., \& {Owocki}, S.~P. 2010, \mnras, 405,
  2391

\bibitem[{{Coia} \& {Pollock}(2008)}]{tn0080}
{Coia}, D., \& {Pollock}, A.~M.~T. 2008, XMM-{\it Newton} Calibration Document
  CAL-TN-0080

\bibitem[{{den Herder} {et~al.}(2001){den Herder}, {Brinkman}, {Kahn},
  {Branduardi-Raymont}, {Thomsen}, {Aarts}, {Audard}, {Bixler}, {den Boggende},
  {Cottam}, {Decker}, {Dubbeldam}, {Erd}, {Goulooze}, {G{\"u}del}, {Guttridge},
  {Hailey}, {Janabi}, {Kaastra}, {de Korte}, {van Leeuwen}, {Mauche},
  {McCalden}, {Mewe}, {Naber}, {Paerels}, {Peterson}, {Rasmussen}, {Rees},
  {Sakelliou}, {Sako}, {Spodek}, {Stern}, {Tamura}, {Tandy}, {de Vries},
  {Welch}, \& {Zehnder}}]{dHel01}
{den Herder}, J.~W., {et~al.} 2001, \aap, 365, L7

\bibitem[{{Feldmeier} {et~al.}(1997{\natexlab{a}}){Feldmeier}, {Kudritzki},
  {Palsa}, {Pauldrach}, \& {Puls}}]{FKPPP97}
{Feldmeier}, A., {Kudritzki}, R.-P., {Palsa}, R., {Pauldrach}, A.~W.~A., \&
  {Puls}, J. 1997{\natexlab{a}}, \aap, 320, 899

\bibitem[{{Feldmeier} {et~al.}(2003){Feldmeier}, {Oskinova}, \&
  {Hamann}}]{FOH03}
{Feldmeier}, A., {Oskinova}, L., \& {Hamann}, W.-R. 2003, \aap, 403, 217

\bibitem[{{Feldmeier} {et~al.}(1997{\natexlab{b}}){Feldmeier}, {Puls}, \&
  {Pauldrach}}]{FPP97}
{Feldmeier}, A., {Puls}, J., \& {Pauldrach}, A.~W.~A. 1997{\natexlab{b}}, \aap,
  322, 878

\bibitem[{{Fullerton} {et~al.}(2006){Fullerton}, {Massa}, \& {Prinja}}]{FMP06}
{Fullerton}, A.~W., {Massa}, D.~L., \& {Prinja}, R.~K. 2006, \apj, 637, 1025

\bibitem[{{Haser}(1995)}]{Haser95PHD}
{Haser}, S.~M. 1995, PhD thesis, Universit{\"a}ts-Sternwarte der
  Ludwig-Maximillian Universit{\"a}t, M{\"u}nchen, (1995)

\bibitem[{{Kahn} {et~al.}(2001){Kahn}, {Leutenegger}, {Cottam}, {Rauw},
  {Vreux}, {den Boggende}, {Mewe}, \& {G{\"u}del}}]{Kel01}
{Kahn}, S.~M., {Leutenegger}, M.~A., {Cottam}, J., {Rauw}, G., {Vreux}, J.-M.,
  {den Boggende}, A.~J.~F., {Mewe}, R., \& {G{\"u}del}, M. 2001, \aap, 365,
  L312

\bibitem[{{Leutenegger} {et~al.}(2007){Leutenegger}, {Owocki}, {Kahn}, \&
  {Paerels}}]{LOKP07}
{Leutenegger}, M.~A., {Owocki}, S.~P., {Kahn}, S.~M., \& {Paerels}, F.~B.~S.
  2007, \apj, 659, 642

\bibitem[{{Leutenegger} {et~al.}(2006){Leutenegger}, {Paerels}, {Kahn}, \&
  {Cohen}}]{LPKC06}
{Leutenegger}, M.~A., {Paerels}, F.~B.~S., {Kahn}, S.~M., \& {Cohen}, D.~H.
  2006, \apj, 650, 1096

\bibitem[{{Lucy} \& {Solomon}(1970)}]{LS70}
{Lucy}, L.~B., \& {Solomon}, P.~M. 1970, \apj, 159, 879

\bibitem[{{Lucy} \& {White}(1980)}]{LW80}
{Lucy}, L.~B., \& {White}, R.~L. 1980, \apj, 241, 300

\bibitem[{{Massa} {et~al.}(2003){Massa}, {Fullerton}, {Sonneborn}, \&
  {Hutchings}}]{MFSH03}
{Massa}, D., {Fullerton}, A.~W., {Sonneborn}, G., \& {Hutchings}, J.~B. 2003,
  \apj, 586, 996

\bibitem[{{Morton}(1967)}]{Morton67a}
{Morton}, D.~C. 1967, \apj, 147, 1017

\bibitem[{{Najarro} {et~al.}(2011){Najarro}, {Hanson}, \&
  {Puls}}]{2011A&A...535A..32N}
{Najarro}, F., {Hanson}, M.~M., \& {Puls}, J. 2011, \aap, 535, A32

\bibitem[{{Naz{\'e}} {et~al.}(2012){Naz{\'e}}, {Flores}, \& {Rauw}}]{Naze2012}
{Naz{\'e}}, Y., {Flores}, C.~A., \& {Rauw}, G. 2012, \aap, 538, A22

\bibitem[{{Oskinova} {et~al.}(2004){Oskinova}, {Feldmeier}, \&
  {Hamann}}]{OFH04}
{Oskinova}, L.~M., {Feldmeier}, A., \& {Hamann}, W.-R. 2004, \aap, 422, 675

\bibitem[{{Oskinova} {et~al.}(2006){Oskinova}, {Feldmeier}, \&
  {Hamann}}]{OFH06}
---. 2006, \mnras, 372, 313

\bibitem[{{Oskinova} {et~al.}(2007){Oskinova}, {Hamann}, \&
  {Feldmeier}}]{2007A&A...476.1331O}
{Oskinova}, L.~M., {Hamann}, W.-R., \& {Feldmeier}, A. 2007, \aap, 476, 1331

\bibitem[{{Owocki}(2008)}]{2008cihw.conf..121O}
{Owocki}, S.~P. 2008, in Clumping in Hot-Star Winds, ed. W.-R. {Hamann},
  A.~{Feldmeier}, \& L.~M. {Oskinova} (Universit\"{a}tsverlag Potsdam), 121

\bibitem[{{Owocki} {et~al.}(1988){Owocki}, {Castor}, \& {Rybicki}}]{OCR88}
{Owocki}, S.~P., {Castor}, J.~I., \& {Rybicki}, G.~B. 1988, \apj, 335, 914

\bibitem[{{Owocki} \& {Cohen}(2001)}]{OC01}
{Owocki}, S.~P., \& {Cohen}, D.~H. 2001, \apj, 559, 1108

\bibitem[{{Owocki} \& {Cohen}(2006)}]{OC06}
---. 2006, \apj, 648, 565

\bibitem[{{Press} {et~al.}(2007){Press}, {Teukolsky}, {Vetterling}, \&
  {Flannery}}]{NR2007}
{Press}, W.~H., {Teukolsky}, S.~A., {Vetterling}, W.~T., \& {Flannery}, B.~P.
  2007, {Numerical Recipes. The Art of Scientific Computing} (3rd ed.;
  Cambridge: Cambridge Univ. Press)

\bibitem[{{Puls} {et~al.}(2006){Puls}, {Markova}, {Scuderi}, {Stanghellini},
  {Taranova}, {Burnley}, \& {Howarth}}]{Pel06}
{Puls}, J., {Markova}, N., {Scuderi}, S., {Stanghellini}, C., {Taranova},
  O.~G., {Burnley}, A.~W., \& {Howarth}, I.~D. 2006, \aap, 454, 625

\bibitem[{{Runacres} \& {Owocki}(2002)}]{RO02}
{Runacres}, M.~C., \& {Owocki}, S.~P. 2002, \aap, 381, 1015

\bibitem[{{Sundqvist} {et~al.}(2012){Sundqvist}, {Owocki}, {Cohen},
  {Leutenegger}, \& {Townsend}}]{2012MNRAS.420.1553S}
{Sundqvist}, J.~O., {Owocki}, S.~P., {Cohen}, D.~H., {Leutenegger}, M.~A., \&
  {Townsend}, R.~H.~D. 2012, \mnras, 420, 1553

\bibitem[{{Sundqvist} {et~al.}(2010){Sundqvist}, {Puls}, \&
  {Feldmeier}}]{2010A&A...510A..11S}
{Sundqvist}, J.~O., {Puls}, J., \& {Feldmeier}, A. 2010, \aap, 510, A11+

\bibitem[{{Sundqvist} {et~al.}(2011){Sundqvist}, {Puls}, {Feldmeier}, \&
  {Owocki}}]{2011A&A...528A..64S}
{Sundqvist}, J.~O., {Puls}, J., {Feldmeier}, A., \& {Owocki}, S.~P. 2011, \aap,
  528, A64+

\end{thebibliography}

\end{document}